\documentclass[11pt]{article}
\usepackage{epsf}
\usepackage{bm,physics}
\usepackage{cite}
\usepackage{color}
\usepackage{amsmath,amssymb}
\allowdisplaybreaks
\usepackage{graphicx}
\usepackage[colorlinks,citecolor=blue,linkcolor=blue]{hyperref}
\usepackage{comment}
\usepackage{geometry}
\renewcommand{\thefootnote}{\fnsymbol{footnote}}
\def\thefootnote{\fnsymbol{footnote}}

\makeatletter

\@addtoreset{equation}{section}
\makeatother

\begin{document}

\begin{titlepage}

\begin{center}

\vskip .45in

\bigskip\bigskip
{\Large \bf Inflaton Regeneration via Scalar Couplings: \vspace{2mm} \\
Generic Models and the Higgs Portal}

\vskip .65in

{\large 
Kunio~Kaneta$^{1}$,
Tomo~Takahashi$^{2}$ and
Natsumi~Watanabe$^{3}$
\vspace{2mm} \\
}
\vskip 0.2in

{\em 
$^{1}$Faculty of Education, Niigata University, Niigata 950-2181, Japan
\vspace{2mm}\\
$^{2}$Department of Physics, Saga University, Saga 840-8502, Japan
\vspace{2mm}\\
$^{3}$Graduate School of Science and Engineering, Saga University, Saga 840-8502, Japan
}

\end{center}
\vskip .5in

\begin{abstract}
The standard cosmological paradigm assumes that the inflaton field becomes dynamically negligible during the post-reheating evolution of the Universe. We demonstrate that this assumption fails for a broad class of inflationary models where the potential behaves as a monomial form $V(\phi) \propto \phi^k$ (with $k \ge 4$) around the minimum. 
In such scenarios, the effective inflaton mass depends on the field amplitude and vanishes asymptotically as the Universe expands. 
This vanishing-mass mechanism renders the inflaton kinematically accessible to the thermal plasma long after reheating, facilitating the regeneration of inflaton quanta through 1-to-2 decays and 2-to-2 scatterings of bath particles. 
This mechanism is quite generic and the coupling responsible for reheating can be constrained if the inflaton is overproduced, while the inflaton quanta can constitute dark matter in specific scenarios. 
Furthermore, if reheating occurs via the Standard Model Higgs portal, the  process can be further constrained by big bang nucleosynthesis, cosmic microwave background, and colliders such as the LHC. This mechanism provides a new framework for probing post-inflationary reheating.

\end{abstract}

\end{titlepage}

\tableofcontents

\renewcommand{\thepage}{\arabic{page}}
\setcounter{page}{1}
\renewcommand{\thefootnote}{\#\arabic{footnote}}
\setcounter{footnote}{0}

\section{Introduction} 
\label{sec:introduction}

The prevailing cosmological model, $\Lambda$CDM, places an early epoch of cosmic inflation to establish the initial conditions of the observable Universe. 
This period of exponential expansion successfully addresses the horizon and flatness problems that plagued classical Big Bang cosmology~\cite{Guth:1980zm,Sato:1981qmu,Linde:1981mu,Albrecht:1982wi}, while simultaneously generating the primordial density perturbations necessary for the formation of large-scale structure~\cite{Bardeen:1980kt,Mukhanov:1981xt,Hawking:1982cz,Bardeen:1983qw}.
However, the transition from the cold, low-entropy state of inflation to the hot, thermalized plasma of the radiation-dominated era, the epoch known as reheating, remains one of the most significant theoretical uncertainties in early universe cosmology.

The dynamics of this post-inflationary epoch are conventionally modeled as a one-way street.
In this standard paradigm, the energy density of the coherent inflaton condensate, $\rho_\phi$, is transferred irreversibly to the Standard Model (SM) degrees of freedom or to an intermediate sector that subsequently decays into the SM particles~\cite{Abbott:1982hn,Dolgov:1982th,Albrecht:1982mp}. 
This energy transfer initiates the Hot Big Bang, generating the primordial soup of relativistic particles that eventually undergo nucleosynthesis (BBN)~\cite{Alpher:1948ve,Alpher:1948srz,Wagoner:1966pv,Walker:1991ap}.

Once this energy transfer is complete and the Universe has thermalized at the reheating temperature $T_{\rm reh}$, the inflaton field is typically assumed to play no further role in the cosmological history and vanish during the post-reheating evolution.
It is treated as a relic of the past, effectively vanishing from the thermal history of the Universe. 
This assumption is foundational to most simplified cosmological histories; it allows cosmologists to treat the radiation era and subsequent matter era as independent of the specific microphysics of the inflationary sector, provided the correct initial perturbation spectrum is generated.

This standard assumption of the vanishing inflaton relies on a specific, often implicit, premise: that the inflaton remains heavy or decouples completely from the thermal bath after reheating ends. 
In the simplest chaotic inflation model~\cite{Linde:1983gd}, such as the one governed by a quadratic potential $V(\phi) = \frac{1}{2}m_\phi^2 \phi^2$, the inflaton mass $m_\phi$ is a constant fundamental parameter. 
To satisfy the normalization of the scalar perturbation amplitude ($A_s \sim 2 \times 10^{-9}$~\cite{Planck:2018jri}), this mass is typically fixed at a high scale of $10^{13}$ GeV, close to the Grand Unified Theory (GUT) scale.

The production of a particle with such a large mass would indeed be kinematically suppressed from the thermal bath at temperatures $T \ll m_\phi$. 
As the Universe cools below the inflaton mass, the number density of any remaining inflaton quanta would be exponentially suppressed by a Boltzmann factor $e^{-m_\phi/T}$. 
Consequently, in these quadratic models, the assumption that the inflaton disappears from the thermal history is kinematically justified. 
However, modern observational data, particularly from the Planck satellite and BICEP/Keck arrays~\cite{Planck:2018jri,BICEP:2021xfz}, disfavor simple quadratic potentials due to their prediction of a large tensor-to-scalar ratio.
Instead, the data favor models with flatter potentials at large field values, such as Starobinsky-like inflation~\cite{Starobinsky:1980te,Bezrukov:2007ep} or $\alpha$-attractors (T-models and E-models)~\cite{Kallosh:2013yoa,Kallosh:2013hoa}, which often exhibit different behavior near their minima.

This paper focuses on a broad class of observationally consistent models where the inflaton potential behaves as a monomial 
\begin{align}
    V(\phi) \simeq \lambda M_P^4 \left(\frac{\phi}{M_P}\right)^k
    \label{eq:power_law_potential}
\end{align}
with a potential coupling $\lambda$ and $k \ge 4$ near the minimum.
Note that $k$ is assume to be an even integer.
These potentials arise naturally in various theoretical frameworks, including supergravity~\cite{Kallosh:2013yoa,Roest:2013fha,Kallosh:2014rga}.

For these potentials, the effective mass of the inflaton is defined by the curvature of the potential, $m_{\phi}^2(\phi) = V''(\phi)$.
Crucially, for $k \ge 4$, this second derivative is not constant but depends on the field amplitude $\phi$:
\begin{align}
    m_{\phi}^2(\phi) = \lambda^{2/k}k(k-1)M_P^2\left(\frac{\phi}{M_P}\right)^{k-2}.
\end{align}
During the reheating phase and subsequent expansion, the inflaton condensate oscillates around the minimum of its potential.
As the Universe expands, the amplitude of these oscillations decreases due to Hubble friction and particle decay. 
Consequently, for $k \ge 4$, the effective mass $m_{\phi}$ drops rapidly as the field amplitude decays~\cite{Turner:1983he,Garcia:2020wiy,Kaneta:2025xuo}\footnote{
While fragmentation creates a large initial abundance $Y_{\text{frag}}$, the field-dependent mass $m_{\phi}(\phi)$ ensures these particles are unstable and decay, and their abundance is strongly depleted at the onset of radiation domination~\cite{Garcia:2023dyf}.
This depletion requires that the decay process be sufficiently rapid compared to the rate at which the mass drops; otherwise, incomplete decay would result in inefficient reheating and a non-negligible initial relic abundance.
}.
This dependency creates a profound departure from the standard quadratic case. 
As the amplitude drops, the inflaton becomes asymptotically massless (or very light) at late times. 
Far from remaining a heavy, decoupled spectator, the inflaton evolves into a light degree of freedom. 
This ``vanishing mass mechanism" implies that the standard assumption, that the inflaton is kinematically forbidden from the thermal bath, is invalid for this class of models. 
This transition reopens its kinematic accessibility to the thermal bath, allowing the inflaton regeneration after reheating, as pointed out in our previous paper \cite{Kaneta:2025xuo}.
Indeed, the same couplings responsible for reheating the Universe inevitably facilitate the thermal production of inflaton particles from the primordial plasma. 
In such a case, far from disappearing, the inflaton particles may be repopulated, potentially constituting the observed Dark Matter (DM) abundance today, and if they are overproduced, such a scenario is excluded. This indicates that, via the inflaton regeneration, one can probe reheating models. 
In this work, by extending our previous work \cite{Kaneta:2025xuo}, we rigorously evaluate this abundance of the inflaton particle for a general portal coupling to matter. 
Furthermore, we also discuss the Higgs portal scenario, where the requirement of successful reheating and the constraints from BBN and the Large Hadron Collider (LHC) define a tight corridor of viability for the inflaton as a dark matter candidate.

Although our study does not specifically focus on a scenario in which the inflaton particle is identified as dark matter, since for particular choice of parameters the inflaton can behave as dark matter, it is worth contrasting this mechanism with previous studies where an inflaton-like scalar plays the role of dark matter, sometimes referred to as ``WIMPflation"~\cite{Lerner:2009xg,Okada:2010jd,Mukaida:2014kpa,Hooper:2018buz,Garcia:2021gsy}. In these works, the inflaton (or an inflaton-like scalar) acquires a fixed mass at the weak (or TeV) scale after inflation, and the observed relic abundance results from a conventional freeze-out or freeze-in process with a time-independent mass. By contrast, in the class of models considered in this paper the inflaton mass is not fixed and vanishes dynamically as the condensate amplitude decays during reheating.
The inflaton particle is then only regenerated once a scalar field $\chi$ undergoes a phase transition and acquires its vacuum expectation value. This two-step structure, in which the mass vanishes during reheating and is regenerated only after the $\chi$ phase transition, has no counterpart in Refs.~\cite{Lerner:2009xg,Okada:2010jd,Mukaida:2014kpa,Hooper:2018buz,Garcia:2021gsy}, and is the central new ingredient of the present analysis.

The paper is organized as follows.
In Section~\ref{sec:framework}, we establish the theoretical framework for an inflaton field interacting with a generic singlet scalar field. We then discuss the production of inflaton particles from the thermal bath and their relic abundance in Section~\ref{sec:production}. In Section~\ref{sec:higgs_portal}, we apply our framework to the specific case of the Standard Model Higgs portal, discussing various cosmological and experimental constraints. Finally, we conclude in Section~\ref{sec:conclusion}. 
In Appendices, the details of some computations are provided.

\section{Framework: Inflaton and Singlet Scalar Dynamics}
\label{sec:framework}

In this section, we describe a theoretical framework for an inflaton field $\phi$ interacting with a generic singlet scalar field $\chi$ in a scenario where the inflaton potential at the bottom is dominated by a monomial term of a higher power, and the reheating is driven by portal couplings to $\chi$. 
We additionally note that our scenario may have interesting implications for recent measurements of the spectral index of primordial fluctuations, potentially offering nontrivial constraints in light of current observational data.

\subsection{Inflaton Potential and Evolution}
\label{subsec:inflaton}

We consider the inflaton field $\phi$ with a power-law potential near its minimum, given by Eq.~\eqref{eq:power_law_potential}.
The inflaton field can be divided as $\phi(t, \vec{x}) = \bar{\phi}(t) + \varphi (t, \vec{x})$ where $\bar{\phi}$ and $\varphi$ respectively correspond to the condensate (zero mode) and the inflaton quanta (non-zero momentum modes). In what follows, we write the condensate as $\phi$, omitting the bar unless ambiguity arises. 
For $k \ge 4$, the effective mass of the inflaton 
decreases as the Universe expands, becoming negligible at late times. This property is crucial for the post-reheating regeneration of inflaton quanta from the thermal bath, as the production is kinematically allowed when the bath temperature $T$ exceeds the effective inflaton mass.

During inflation, inflaton potential may be different from the one given in Eq.~\eqref{eq:power_law_potential}. Specifying the shape of the potential during inflation serves to determine the potential coupling $\lambda$ from the normalization of primordial fluctuaions
and the amplitude of the inflaton field at the end of inflation, $\phi_e$. For concreteness, we consider the T-model inflation \cite{Kallosh:2013hoa,Kallosh:2013pga}, in which the potential is given by
\begin{align}
\label{eq:Vinf}
   V(\phi) 
   &= 
   \lambda M_P^4\left[
      \sqrt{6}\tanh\left(\frac{\phi}{\sqrt{6}M_P}\right)
   \right]^k.
\end{align}
Note that the choice of the inflaton potential during inflation does not affect the dynamics of the inflaton after reheating, which is the primary focus of this work. 
However, this model can be consistent with recent results on primordial power spectrum from ACT \cite{AtacamaCosmologyTelescope:2025nti} and SPT \cite{SPT-3G:2025bzu} in combination with Planck \cite{Planck:2018jri} and BICEP/Keck \cite{BICEP:2021xfz}, which can also constrain the strength of interaction responsible for reheating once the model is fixed.

The slow-roll parameters for the potential~\eqref{eq:Vinf} are given by 
\begin{align}
  \epsilon_V = \frac12 M_P^2 \left( \frac{V'}{V}\right)^2
   = \frac{k^2}{3} \frac{1}{\sinh^2 (2\phi /(\sqrt{6} M_P)} 
   \,, \\[12pt]
   \eta_V  = M_P^2\frac{V^{''}}{V}
   = 2 \epsilon_V - \frac{2k}{3} 
   \frac{\cosh (2\phi / (\sqrt{6} M_P))}{\sinh^2 (2\phi / (\sqrt{6} M_P))}  \,,
\end{align}
where a prime represents a derivative with respect to $\phi$. The number of $e$-folds at the horizon exit of a reference scale is calculated as 
\begin{align}
    N_\ast = \frac{1}{M_P^2} \int_{\phi_{\rm end}}^{\phi_\ast} \frac{V}{V'}  d\phi  
    = 
    \frac{3}{2k} \left( \cosh (\frac{2\phi_\ast}{\sqrt{6} M_P}) - \cosh (\frac{2\phi_e}{\sqrt{6} M_P}) \right) \,,
\end{align}
with $\phi_\ast$ and $\phi_e$ being the values of $\phi$ at the horizon exit and the end of inflation, respectively. By using the above expressions, the spectral index $n_s$ and the tensor-to-scalar ratio $r$ are given as \cite{Kallosh:2013yoa}
\begin{align}
    n_s - 1 &
    = \frac{- 4k N_\ast - 3k - 2 g(k)}{2kN_\ast^2 + 2 N g(k) + 3k/2} \,, \label{eq:ns}\\[8pt]
    r &
    = \frac{24k }{2kN_\ast^2 + 2 N_\ast g(k) + 3 k/2} \,,\label{eq:r}
\end{align}
where 
$g(k)=3\cosh(2\phi_e/(\sqrt{6}M_P))$.
When the end of inflation is defined by $\epsilon_V=1$, $\phi_e$ is evaluated by 
\begin{align}
\phi_e = \frac{\sqrt{6}M_P}{2} \cosh^{-1} 
\left( \frac{\sqrt{3 (k^2 + 3)}}{3} \right) 
\end{align}
On the other hand, the end of inflation can also be defined by $\ddot a=0$ or equivalently, $\dot\phi_e^2=V(\phi_e)$~\cite{Ellis:2015pla}, in which $\phi_e$ can be evaluated as~\cite{Garcia:2020wiy}
\begin{align}
   \phi_e 
   &=
   \sqrt{
      \frac{3}{8}
   }M_P \ln\left[
      \frac{1}{2}+\frac{k}{3}\left(k+\sqrt{k^2+3}\right)
   \right] \,.
   \label{eq:phi_e from ddot-a=0}
\end{align}
Once $\phi_e$ is specified,
the inflaton energy density at the end of inflation is given by $\rho_e\equiv \dot\phi_e^2/2+V(\phi_e) = 3V(\phi_e)/2$,
which will be used to compute the reheating temperature.
In the following analysis, we use the definition \eqref{eq:phi_e from ddot-a=0}.
In either case, the $k$-dependence of $n_s$ and $r$ in the large $N_*$ limit is subdominant, and we find $n_s\simeq 1-2/N_*$ and $r\simeq 12/N_*^2$ at leading order.

After the end of inflation, inflaton condensate oscillates
around the minimum of the potential.
During this period, the equation of motion for the inflaton condensate $\phi$ is given by
\begin{align}
\ddot\phi +3H\dot\phi +  V'(\phi) 
   &= 0,
\end{align}
which can be recast into the evolution equation of the energy density of inflaton condensate as
\ 
\begin{align}
   \dot\rho_\phi +3(1+w_\phi)H\rho_\phi = 0,
   \label{eq:Boltzmann_rho_phi}
\end{align}
where the equation-of-state parameter $w_\phi$ is given by
\begin{align}
   w_\phi &= \frac{k-2}{k+2},
\end{align}
and the Hubble parameter $H$ is governed by the inflaton energy density, i.e., $3M_P^2H^2 = \rho_\phi$.
To derive this relationship, we average over a single period of inflaton oscillation, defining $\rho_\phi$ as $\rho_\phi = \langle\dot\phi^2/2+V(\phi)\rangle$, supplemented by the virial theorem, $\langle\dot\phi^2\rangle = k\langle V(\phi)\rangle$, for a potential $V(\phi) \propto \phi^k$.
Solving Eq.~\eqref{eq:Boltzmann_rho_phi}, we find 
\begin{align}
   \rho_\phi(a) &= \rho_{e} \left(\frac{a}{a_e}\right)^{-3(1+w_\phi)},
\end{align}
where $a_e$ is the scale factor at the end of inflation.

After the completion of reheating, $\rho_{\phi}$ is negligible compared to the radiation energy density $\rho_{\rm rad}$ which scales as
\begin{align}
   \rho_{\rm rad}(a) &= \rho_{\rm reh}\left(
      \frac{a}{a_{\rm reh}}
   \right)^{-4},\\[12pt]
   \rho_{\rm reh} &= \frac{g_{\rm reh}\pi^2}{30}T_{\rm reh}^4,
\end{align}
where $T_{\rm reh}$ is the reheating temperature.
Inflaton quanta may be generated from the thermal bath if the inflaton mass $m_\phi$ is smaller than the temperature $T$, i.e., $m_\phi/T<1$ is satisfied.
It is important to note that the effective inflaton mass during reheating is given by
\begin{align}
    m_\phi^2 
    &=
    V''(\phi\ll M_P)
    =
    \lambda^{2/k}k(k-1)M_P^2\left(\frac{\rho_\phi}{M_P^4}\right)^{\frac{k-2}{k}},
\end{align}
and thus, $m_\phi$ is vanishing as $\rho_\phi\sim 0$ after reheating is achieved by the decay of the inflaton condensate.
Therefore, the inflaton quanta $\varphi$ may be treated as massless particles during reheating until reaching the phase transition at which a scalar field coupled to inflaton develops a non-zero vacuum expectation value (VEV), as will be discussed shortly.

\subsection{Coexisting Portal Couplings and Reheating}
\label{subsec:portal}

We extend the minimal scenario~\cite{Kaneta:2025xuo} by assuming the inflaton interacts with a generic singlet scalar field $\chi$ through both cubic and quartic renormalizable couplings. The interaction Lagrangian is given by
\begin{align}
   -\mathcal{L}_{\rm int} 
   &= 
   \frac{1}{2}\sigma \phi^2 \chi^2 + \mu \phi \chi^2,
\end{align}
where $\sigma$ and $\mu$ are the coupling constants. We assume that $\chi$ is a singlet scalar that develops a VEV, denoted as $\langle \chi \rangle = v$. This VEV may be established at an arbitrary scale (e.g., $10\text{ GeV}$ to $10\text{ TeV}$).
However, we do not specify the scale in this work to make our analysis general.

Each of these two couplings may reheat the Universe, but their relative importance depends on the specific parameters.
Assuming the perturbative reheating scenario, we obtain
\begin{align}
   T^{(\mu)}_{\rm reh}
   &=
   \left(
      \frac{30}{g_{\rm reh}\pi^2}
   \right)^{1/4}
   \left[
      \frac{1}{7\pi^{3/2}}\sqrt{\frac{3}{k(k-1)^3}}\frac{\Gamma(\frac{1}{k})}{\Gamma(\frac{1}{2}+\frac{1}{k})}
      \left(
         \frac{\mu}{\lambda^{1/k}M_P}
      \right)^{5/2}
      \left(
         \frac{\mu_{\rm eff}}{\mu}
      \right)^2
   \right]^{\frac{k}{6k-10}}M_P,
   \label{eq:T_mu}\\[12pt]
   T^{(\sigma)}_{\rm reh}
   &=
   \left(
      \frac{30}{g_{\rm reh}\pi^2}
   \right)^{1/4}
   \left[
      \frac{1}{32(2k-5)\pi}\sqrt{\frac{3}{k(k-1)^3}}\frac{\sigma_{\rm eff}^2}{\lambda^{3/k}}
   \right]^{\frac{k}{4k-12}}M_P \,,
   \label{eq:T_sigma}
\end{align}
which are evaluated under the assumption that reheating is solely caused by $\mu$-decay and $\sigma$-scattering, respectively~\cite{Garcia:2020wiy}. Here, $\sigma_{\rm eff}$ and $\mu_{\rm eff}$ are the effective couplings that account for averaging fast oscillations of $\phi$ as well as the non-trivial field dependence of the decay and scattering rates.
It is important to note that during the inflaton oscillation, $\chi$ acquires an effective mass, i.e., either $m_{\chi,{\rm eff}}^2\sim \sigma \phi^2$ or $\mu\phi$, which affects the inflaton decay rate through kinetic blocking if the effective mass exceeds the inflaton mass.
In such a regime, non-perturbative effect may also be in operation, known as preheating.
Although incorporating such a non-perturbative contribution is beyond our scope, we here take into account the kinetic blocking effect in the perturbative contribution. See Ref.~\cite{Garcia:2020wiy} for more details.
Whereas this treatment does not capture the full dynamics of the energy transfer from inflaton to radiation, we can still grab the non-trivial shift in estimating the reheating temperature.
The number of effective relativistic degrees of freedom, $g_*$, at $T=T_{\rm reh}$ is denoted by $g_*(T_{\rm reh})\equiv g_{\rm reh}$, and we take the temperature dependence of $g_*$ into account in our analysis, assuming the thermal particle content being the SM.

\begin{figure}[htbp]
   \centering
   \includegraphics[width=0.45\textwidth]{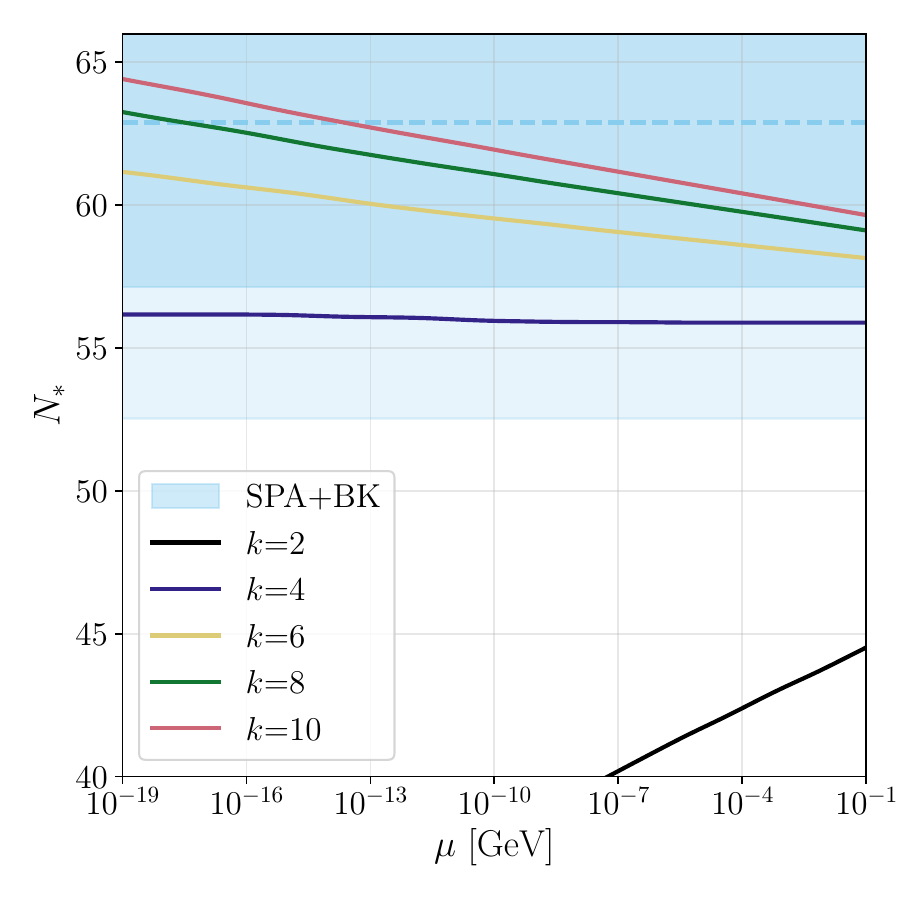}~~~~~
   \includegraphics[width=0.45\textwidth]{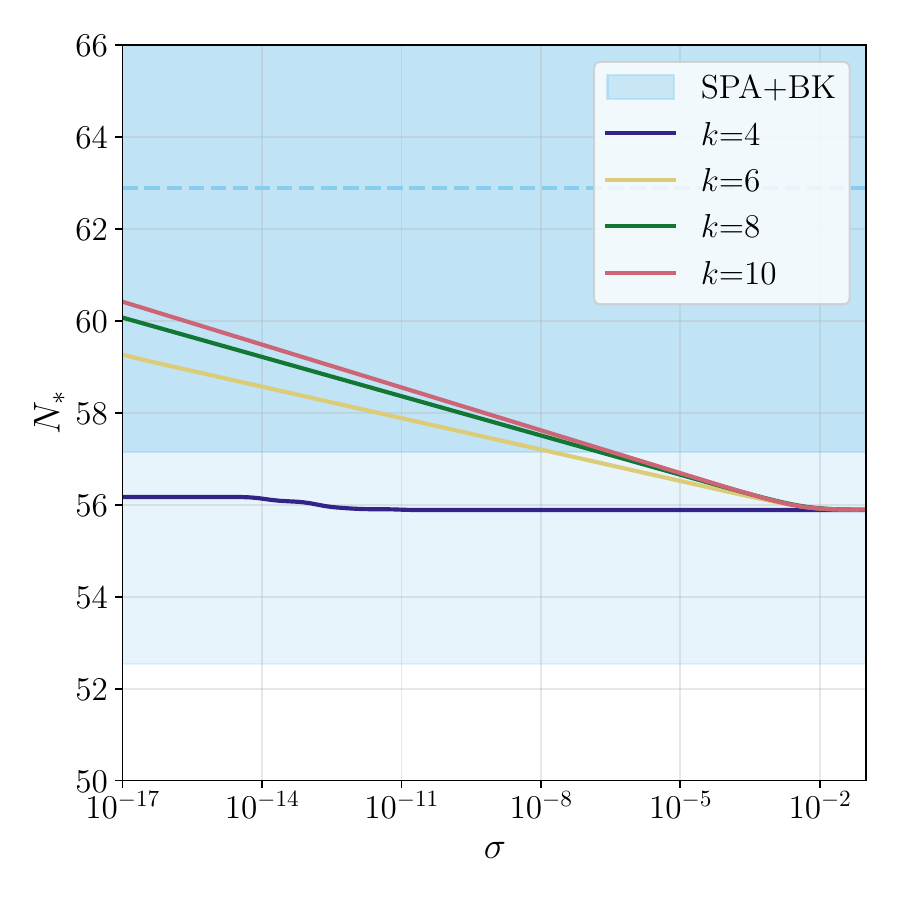}
   \caption{
   The value of $N_*$ as a function of $\mu$ (left) or $\sigma$ (right) consistent with observations.
   Dark and light blue regions correspond to 68\% CL and 95\% CL allowed range of $n_s$ from the combined data~\cite{Balkenhol:2025wms} of Planck~\cite{Planck:2018jri}, SPT~\cite{SPT-3G:2025bzu},  ACT~\cite{AtacamaCosmologyTelescope:2025blo}, and BICEP/Keck~\cite{BICEP:2021xfz}. Reheating with $k=2$ is possible for the $\mu$-coupling case, while it is not for the $\sigma$-coupling case. Therefore, the black solid line ($k=2$) appears only in the left panel.}
   \label{fig:N_coupling}
\end{figure}

Having the analytic estimate of $T_{\rm reh}$ given by Eqs.~\eqref{eq:T_mu} and \eqref{eq:T_sigma}, we may determine the number of $e$-folds $N_*$ at the horizon exit in a self-consistent way.
Taking into account the equation-of-state parameter during reheating $w_{\rm int}=(k-2)/(k+2)$, $N_*$ is given by~\cite{Liddle:2003as,Martin:2013tda}
\begin{align}
   N_* 
   &= \ln \left[ \frac{1}{\sqrt{3}} \left( \frac{\pi^2}{30} \right)^{1/4} \left( \frac{43}{11} \right)^{1/3} \frac{T_0}{H_0} \right] 
   - \ln \left( \frac{k_*}{a_0 H_0} \right) 
   \nonumber\\    
   & \qquad \qquad \qquad \qquad \qquad
   + \frac{1}{4} \ln \left( \frac{V_*^2}{M_P^4 \rho_e} \right)
   + \frac{1 - 3w_{\text{int}}}{12(1 + w_{\text{int}})} \ln \left( \frac{\rho_{\text{reh}}}{\rho_e} \right) 
   - \frac{1}{12} \ln g_{\text{reh}},
\end{align}
where $T_0 = 2.7255$ K and $H_0 = 67.36~ {\rm km}/ {\rm s}/ {\rm Mpc}$ \cite{Planck:2018vyg}. We take the pivot scale $k_* = 0.05~{\rm Mpc}^{-1}$~\cite{Planck:2018jri}.
The potential energy at $N_*$ is denoted as $V_*$.
Since $V_*$ and $\rho_{\rm reh}$ depend on $N_*$, we need an iterative approach to determine $N_*$. See Ref.~\cite{Garcia:2020wiy} for more details.
Figure~\ref{fig:N_coupling} shows the resultant $N_*$ as a function of $\mu$ or $\sigma$.
Adopting the combined analysis of SPT, Planck, ACT, and BICEP/Keck (SPA-BK)~\cite{Balkenhol:2025wms}, the favored region of $n_s = 0.9682\pm 0.0032~(68\%~{\rm CL})$ is also shown in the figure.
Note that the constraint on $N_*$ from $r$ is much weaker than that from $n_s$.
From the figure, it is evident that $k\ge 6$ is favored within 68\% CL in both $\mu$ and $\sigma$ coupling cases, while $k=4$ is still allowed within 95\% CL.
Notice that the case of $k=2$ appears only in the left panel of the figure since reheating can be achieved by $\phi\to\chi\chi$ via the $\mu$ coupling, whereas the energy transfer from $\phi$ to $\chi$ through $\phi\phi\to\chi\chi$ via $\sigma$ coupling is inefficient to realize reheating. Nevertheless, $k=2$ is not favored by the current CMB data.

\begin{figure}
    \centering
    \includegraphics[width=0.48\textwidth]{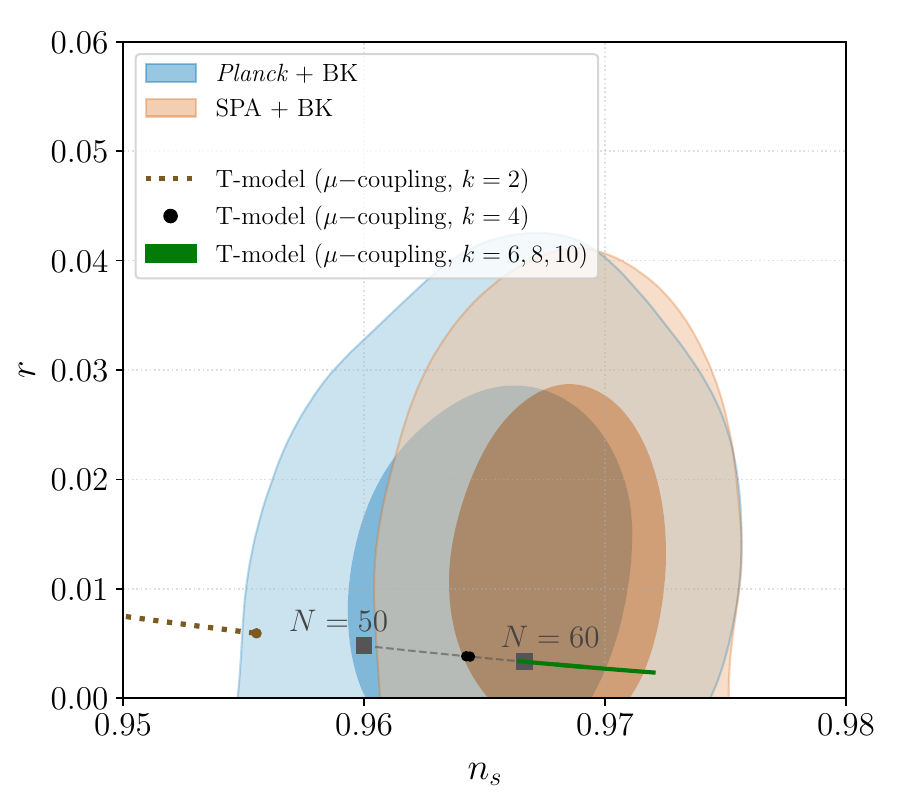}~~~~~
    \includegraphics[width=0.48\textwidth]{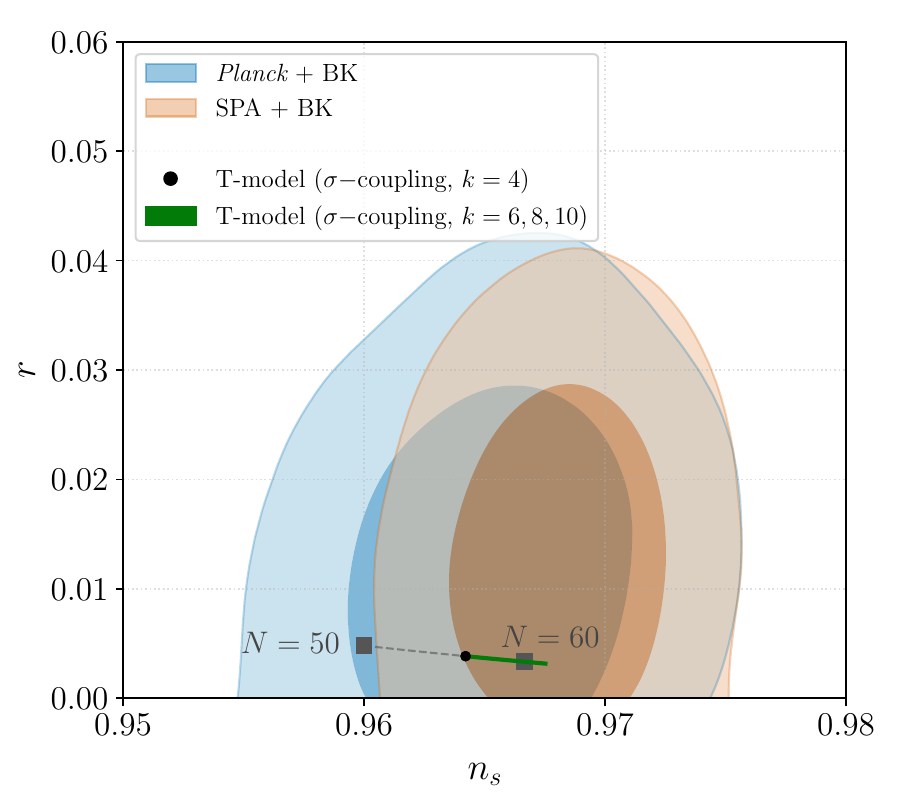}
    \caption{Reheating prediction mapped onto the $n_s$--$r$ plane for the cases where the reheating is realized via $\mu$-couping (left) and $\sigma$-coupling (right). Observational constraints from Planck+BK and SPT (Planck+SPT+ACT)+BK are also shown.
    The case of $k=4$ corresponds to the black dot in both panels since $N_*$ is almost independent from couplings as shown in Fig.~\ref{fig:N_coupling}.}
    \label{fig:ns_r}
\end{figure}

Figure~\ref{fig:ns_r} shows the theoretical prediction and observational constraints in the $n_s$--$r$ plane.
As explained above, $N_*$ is calculated as a function of $\mu$ or $\sigma$, depending on $k$, from which we obtain $n_s$ and $r$.
With the same parameter space used in Fig.~\ref{fig:N_coupling}, we obtain the theoretical predictions for $k=4$ indicated by black dot and for $k\ge 6$ shown in green for both $\mu$- and $\sigma$-couplings in Fig.~\ref{fig:ns_r}.
Note that the predictions for $k\ge 6$ are nearly degenerate, and thus those regions are collectively depicted in the figure.
The gray dashed line, with labels of $N=50$ and $N=60$, corresponds to the T-model prediction without specifying the reheating mechanism\footnote{
For the constraints on the generic $\alpha$-attractor models in light of the recent CMB measurements, see, e.g., Ref.~\cite{Ellis:2025zrf}.
}.
To draw the favored contours by Planck + BK (BICEP/Keck) and SPA (Planck+SPT+ACT) + BK shown in blue and orange in the figure, respectively, we have used the MCMC chain data from Refs.~\cite{Planck:2018jri,BICEP:2021xfz,AtacamaCosmologyTelescope:2025nti} and \cite{Balkenhol:2025wms}.
Notice that $k=2$ for the $\mu$-coupling case is disfavored by SPA+BK data, which adds the motivation to consider $k\ge4$.

\begin{figure}[htbp]
   \centering
   \includegraphics[width=.45\textwidth]{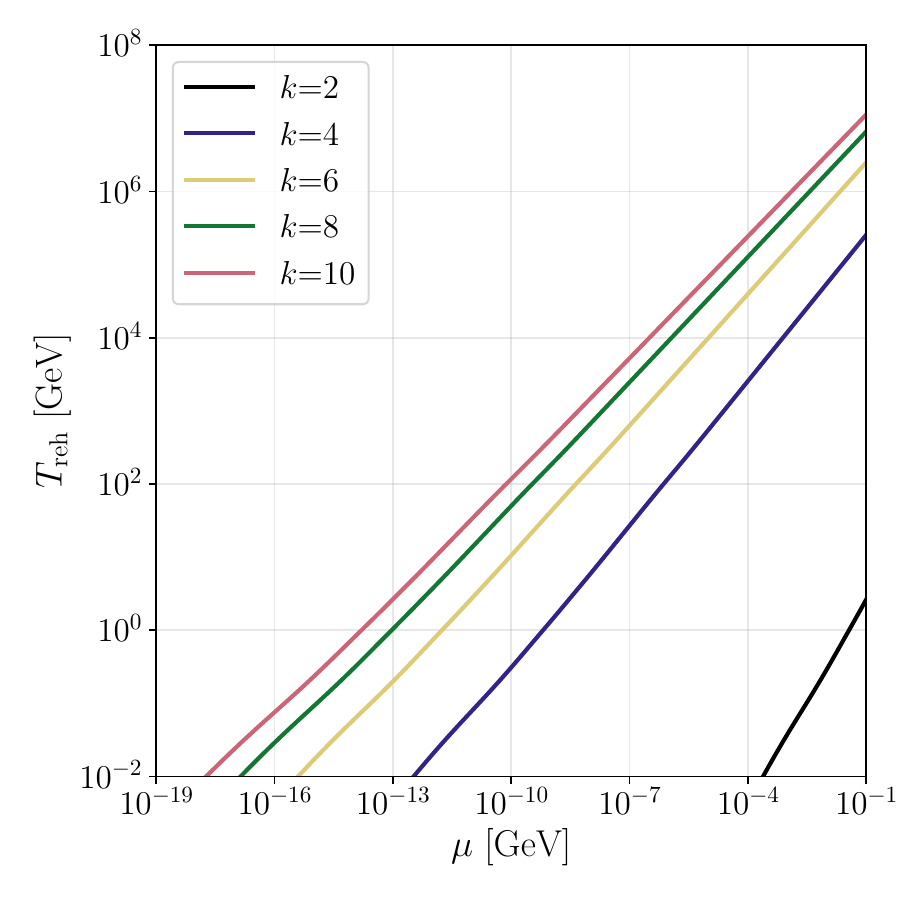}~~~~~
   \includegraphics[width=.45\textwidth]{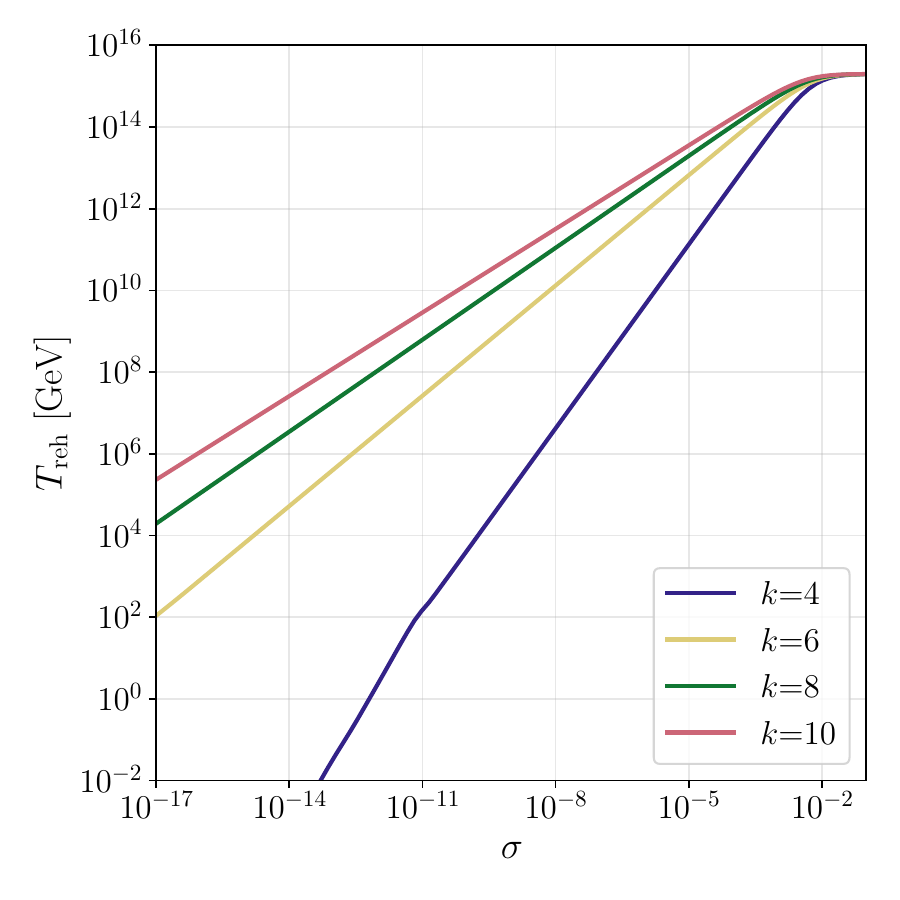}
   \caption{Reheating temperature achieved by perturbative decay via $\mu$-coupling (left) and scattering via $\sigma$-coupling (right).
   }
   \label{fig:Treh_perturbative}
\end{figure}

\begin{figure}[htbp]
   \centering
   \includegraphics[width=.45\textwidth]{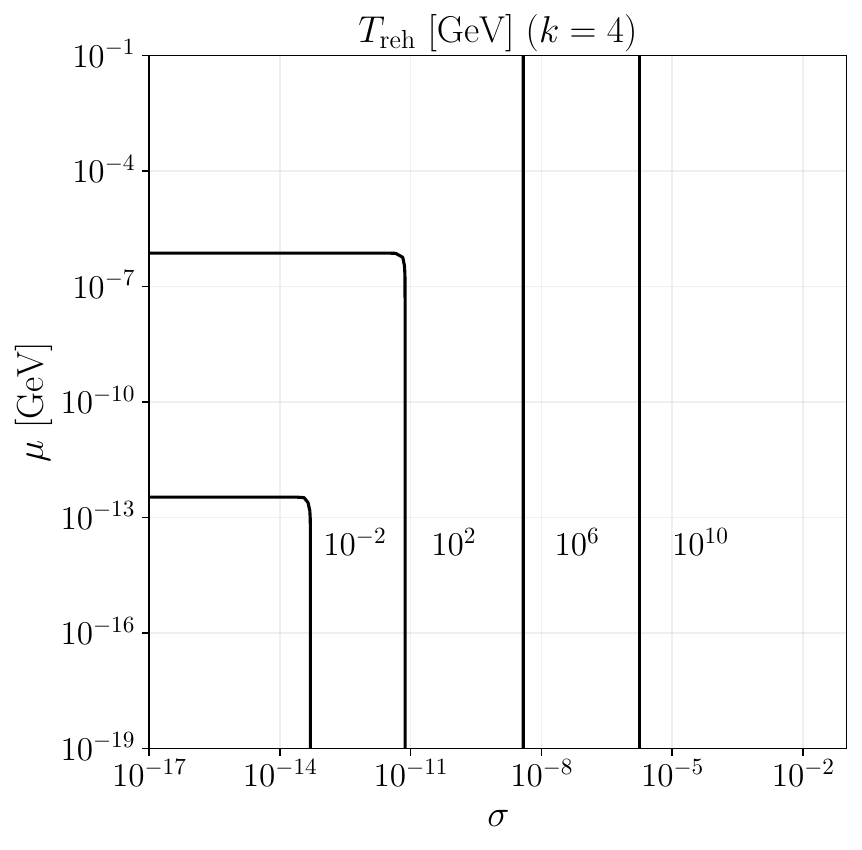}~~~~~
   \includegraphics[width=.45\textwidth]{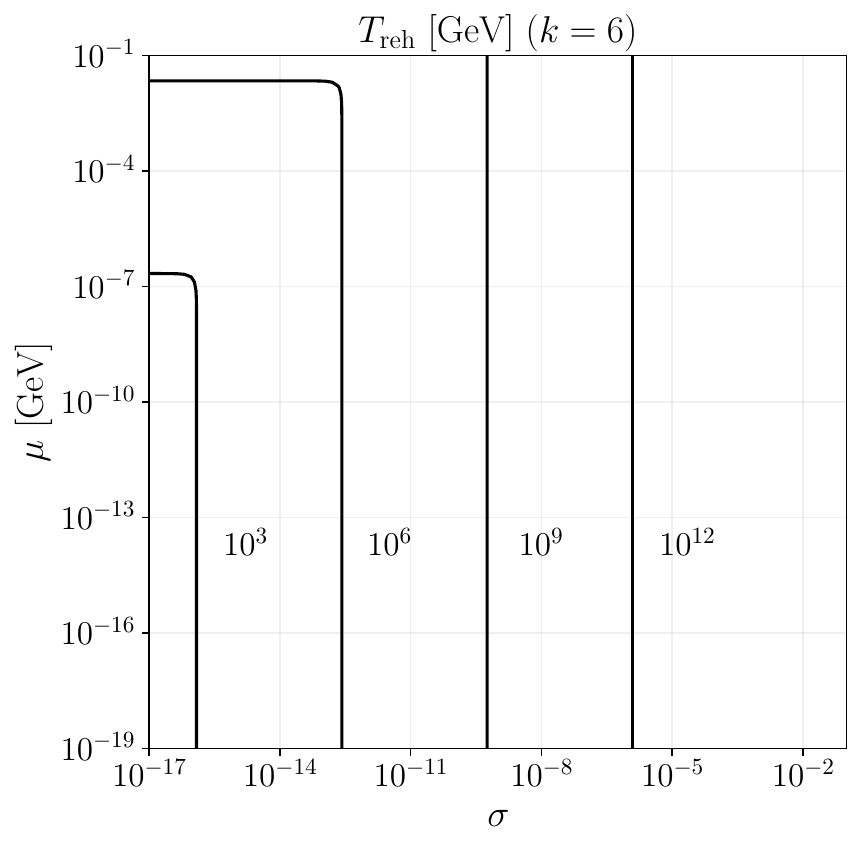}\\
   \includegraphics[width=.45\textwidth]{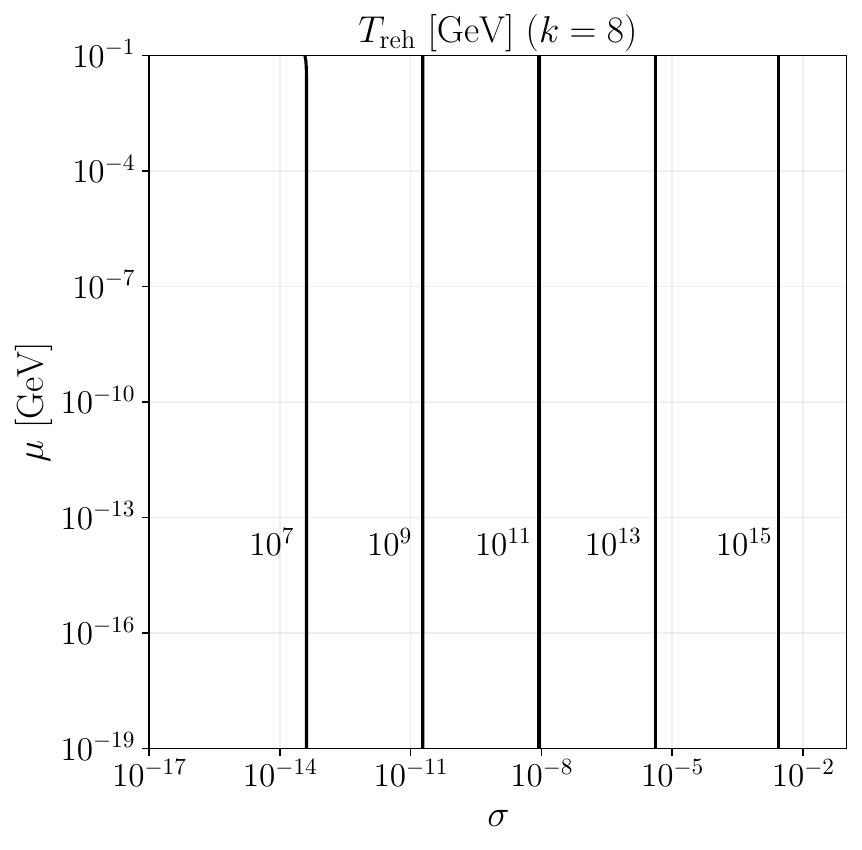}~~~~~
   \includegraphics[width=.45\textwidth]{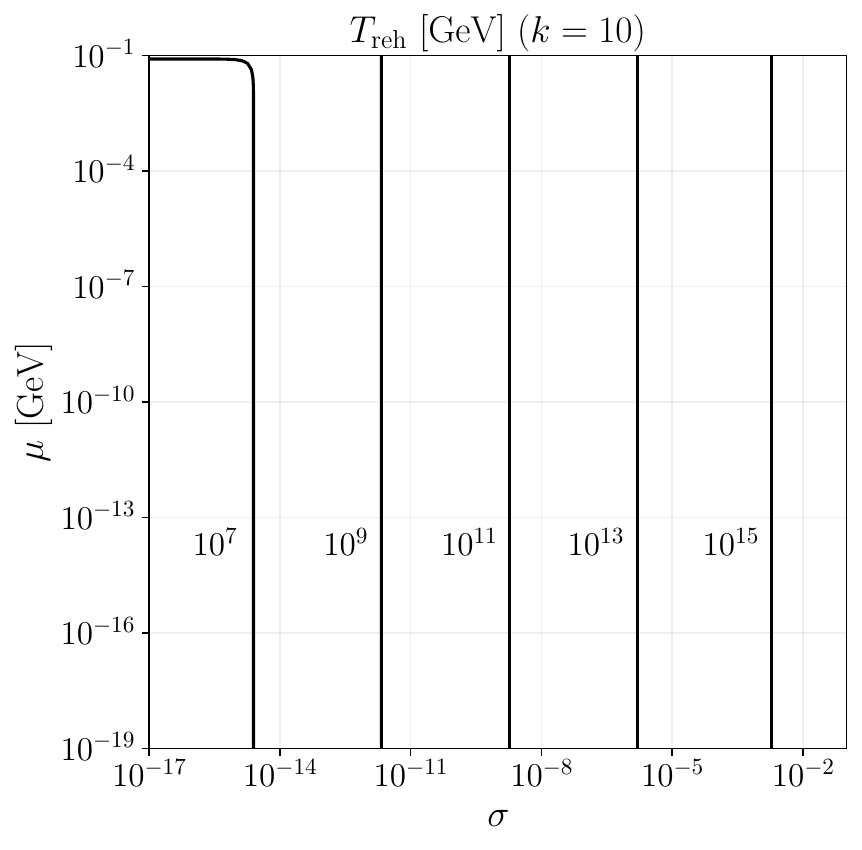}
   \caption{Reheating temperature achieved by the coexisting $\mu$- and $\sigma$-couplings for the cases of $k=4$ (top left), $k=6$ (top right), $k=8$ (bottom left) and $k=10$ (bottom right).}
   \label{fig:Treh_coexisting}
\end{figure}

In determining $N_*$, we also find consistent $T_{\rm reh}$ which is shown as a function of either $\mu$ or $\sigma$ coupling in Fig.~\ref{fig:Treh_perturbative}.
Notably, in the $\sigma$-coupling case, the reheating temperature at $\sigma\gtrsim 10^{-4}$ for $k\ge6$ is suppressed by the kinetic blocking, so $T_{\rm reh}$ does not monotonically increase with~$\sigma$.
If these two couplings coexist, 
the most efficient process dominates the reheating. The reheating temperature is then given by $T_{\rm reh} = \max(T^{(\mu)}_{\rm reh}, T^{(\sigma)}_{\rm reh})$.
Figure~\ref{fig:Treh_coexisting} shows the reheating temperature in such a case by evaluating $T_{\rm reh} = \left[ (T^{(\mu)}_{\rm reh})^4 + (T^{(\sigma)}_{\rm reh})^4 \right]^{1/4}$.

So far we have discussed the reheating processes in detail. Once reheating completes, the Universe is dominated by radiation, during which the inflaton quanta is effectively massless as the inflaton condensate decays away.
Then the inflaton can be regenerated from the thermal bath as its production becomes kinematically accessible.
The phase transition is supposed to occur during radiation-dominated era, and hence $\chi$ develops a non-zero VEV, making the inflaton quanta massive via the coupling between $\varphi$ and $\chi$.
These events occur in the following sequence:
\begin{align}
    \begin{tabular}{c}
         Rehating \\
         ($m_\phi\to 0$)
    \end{tabular}
    \;\;\;\to\;\;\;
    \begin{tabular}{c}
         Radiation Domination \\
         ($\langle \chi \rangle = 0,\, m_\phi\simeq 0$)
    \end{tabular}
    \;\;\;\to\;\;\;
    \begin{tabular}{c}
         Phase Transition \\
         ($\langle\chi\rangle = v,\, m_\phi\neq 0$)
    \end{tabular}
    .\nonumber
\end{align}
In the next subsection, we will focus on the era when the phase transition has already taken place, and thus $\varphi$ has become massive.

\subsection{Mass Matrix and Mixing}
\label{subsec:mass}

Once the reheating completes, the effective mass of inflaton induced by the inflaton potential becomes vanishing.
This is a key feature for inflaton quanta $\varphi$
to be produced from the thermal bath, which will be discussed in the rest of the paper.
While the effective inflaton mass vanishes, $\varphi$ acquires a tree-level mass from the $\sigma$-coupling after $\chi$ develops a VEV.
Furthermore, the $\mu$-coupling induces a mixing between $\varphi$ and $\chi$. Neglecting the small tadpole contribution for $\mu \ll v$, the mass term for the $\varphi$-$\chi$ system in the broken phase is:
\begin{align}
   \mathcal{L}_{m} 
   &= 
   \frac{1}{2} 
   \begin{pmatrix} 
      \varphi & \chi 
   \end{pmatrix} 
   \begin{pmatrix} 
      m_{\varphi\varphi}^2 & m_{\varphi\chi}^2 \\[6pt]
      m_{\varphi\chi}^2 & m_{\chi\chi}^2 
   \end{pmatrix} 
   \begin{pmatrix} 
      \varphi \\
      \chi 
   \end{pmatrix},
   \label{eq:mass_matrix}
\end{align}
where the matrix elements are defined as:
\begin{align}
m_{\varphi\varphi}^2 &= \sigma v^2, \\[8pt]
m_{\varphi\chi}^2 &= 2\mu v.
\end{align}
Here, $m_{\varphi\varphi}$ represents the tree-level mass of the inflaton induced by the $\sigma$-coupling. In the limit $m_\varphi \ll m_\chi$, the mixing angle $\theta$ between the two states is approximately:
\begin{align}
   \theta 
   &\simeq 
   \frac{2\mu v}{m_{\chi}^2}.
\end{align}
The physical states are obtained by rotating the original fields by $\theta$. By replacing $\varphi\to \varphi+\theta \chi$ and $\chi\to\chi-\theta\varphi$ in the Lagrangian, we can move from the interaction basis to the mass basis, which is given by 
\begin{align}
   {\cal L}_{\rm int}
   &=\frac{1}{2}\sigma \varphi^2\chi^2 + g_{\chi\varphi\varphi} \chi\varphi^2 + g_{\chi\chi\varphi} \chi^2\varphi + {\cal O}(\theta^2),
   \label{eq:L_int_chi}\\[8pt]
   g_{\chi\varphi\varphi} 
   &\equiv 
   \sigma v -2\theta\mu\simeq \sigma v-4\mu^2 v/m_\chi^2,\\[8pt]
   g_{\chi\chi\varphi} 
   &\equiv 
   \mu+2\theta\sigma v \simeq \mu+4\sigma\mu v^2/m_\chi^2.
\end{align}
This mixing is a key feature of the hybrid scenario, as it allows for inflaton production via decay channels (e.g., $\chi \to \varphi\varphi$) and additional scattering processes (e.g., $\chi\chi\to\varphi\varphi$ via $\chi$ or $\varphi$ exchange) that are absent in the $\sigma$-only case.

\subsection{Radiative Contributions to Inflaton Mass}
\label{subsec:radiative}

In addition to the tree-level mass induced by the VEV, the inflaton may acquire an effective mass through radiative corrections or self-interactions. For the $\mu$-coupling specifically, the one-loop Coleman-Weinberg potential contributes an effective mass term:
\begin{align}
   m_{\varphi}^2 
   &\simeq 
   \frac{\mu^2}{16\pi^2}.
\end{align}
In our numerical analysis, we adopt a generalized mass for the inflaton that incorporates both tree-level and radiative effects:
\begin{align}
   m_{\varphi}^2 
   &= 
   \sigma v^2 + \frac{\mu^2}{16\pi^2}.
   \label{eq:mass_phi}
\end{align}
This allows us to explore the parameter space where either the $\sigma$-coupling or the $\mu$-coupling dominates the production and mass of the regenerated inflaton quanta.
Note again that the effective mass arising from the inflaton self-coupling, $m_\varphi^2\sim \lambda \phi^{k-2}$, is absent in the post-reheating era as the inflaton condensate $\phi$ ceases to exist.

\subsection{Non-perturbative Production of Inflaton Quanta}
\label{subsec:non-perturbative}

Since the effective mass $m_\phi(\phi)$ varies rapidly while the condensate oscillates, the inflaton perturbations $\varphi$ can be amplified non-perturbatively, and such self-resonance is a priori distinct from the condensate fragmentation invoked above. 
A well-known case in which non-perturbative production dominates the reheating dynamics is Higgs inflation with a quartic potential~\cite{Ema:2016dny,Sfakianakis:2018lzf,DeCross:2015uza,DeCross:2016fdz,DeCross:2016cbs}.
In this case, however, the efficiency of the resonance originates from the non-minimal gravitational coupling $\xi h^2 R$, which is absent in our setup, so this particular mechanism does not directly apply in our setup.

For the class of potentials considered here, self-resonance and fragmentation are not independent channels.
Lattice studies of potentials given by Eq.~\eqref{eq:Vinf} show that the broad resonance band is shrinking as the Universe expands, and it is the first narrow band that slowly drives the condensate to backreaction and fragmentation~\cite{Lozanov:2017hjm}.
No localized configurations such as oscillons are formed in this regime, and the remnant condensate becomes energetically subdominant.

Finally, the portal couplings also render the effective mass of $\chi$ time dependent, so bath particles may likewise be produced non-perturbatively.
This would modify the reheating temperature $T_{\rm reh}$.
However, the effect of the non-perturbative $\chi$ production does not affect the relic abundance of $\varphi$ computed in this work, which is fixed in the infrared regime ($T\sim m_\chi$) well after reheating completes.
The detailed study of the non-perturbative corrections to $T_{\rm reh}$ is left for future work. (See the discussion following Eqs.~\eqref{eq:T_mu} and \eqref{eq:T_sigma}.)

\section{Inflaton Production after Reheating}
\label{sec:production}

This section provides a framework for calculating the abundance of regenerated inflaton quanta. We consider production from the thermal bath consisting of scalar $\chi$ particles (and potentially other SM states) through 1-to-2 decays and 2-to-2 scattering processes. In the following analysis, we assume that the production of inflaton particle proceeds during the radiation-dominated epoch, namely, after the completion of reheating.

\subsection{General Formalism}
\label{subsec:general}

To set our notation, we begin with a generic Boltzmann equation for a number density of a particle~$i$:
\begin{align}
   \dot n_i + 3Hn_i &= \sum_{jk}(-\gamma_{i,jk}+\gamma_{jk,i}) + \sum_{jkl}(-\gamma_{ij,kl}+\gamma_{kl,ij}) + \cdots,
\end{align}
where the ellipsis denotes collision integrals involving more particles.
The collision term for $12\to 34$, i.e., $\gamma_{12,34}$ for instance, is given by
\begin{align}
\gamma_{12,34} 
&= 
\int d\Pi_1 d\Pi_2 d\Pi_3 d\Pi_4 (2\pi)^4 \delta^4(p_1 + p_2 - p_3 - p_4)  \,
|\mathcal{M}_{12,34}|^2 \, f_1 f_2 (1 \pm f_3)(1 \pm f_4),
\end{align}
where $d\Pi_i = \displaystyle\frac{d^3 p_i}{(2 \pi)^3 2E_i}$ is the Lorentz-invariant phase space measure for particle $i$, and $|\mathcal{M}_{ij,kl}|^2$ is the squared matrix element for the process of $ij\to kl$, including the polarization sum and symmetry factors as well as number of particle~$i$ produced/annihilated per reaction. 
$f_i$ denotes a distribution function for particle~$i$. Although the collision term generally includes the factor $1 \pm f_i$, we adopt the dilute gas approximation where 
this factor appears as $1 \pm f_i \simeq 1$ in the following analysis.
For later convenience, we introduce a Lorentz invariant function
\begin{align}
   w_{12,34}(s)
   &\equiv
   \int d{\rm LIPS}_{34} |{\cal M}_{12,34}|^2,
\end{align}
where $d{\rm LIPS}_{34}\equiv d\Pi_3 d\Pi_4 (2\pi)^4 \delta^4(p_1 + p_2 - p_3 - p_4)$, to write the collision integral as
\begin{align}
   \gamma_{12,34} &=
   \int d\Pi_1 d\Pi_2 w_{12,34}(s) f_1 f_2.
\end{align}

In our case, the evolution of the inflaton number density $n_{\varphi}$ is governed by the Boltzmann equation:
\begin{align}
   \frac{dY}{dT} 
   &= 
   -\frac{\langle\sigma_{\varphi\varphi} v_{\rm rel}\rangle \mathfrak{s}}{HT} (Y_{\rm eq}^2 - Y^2),
   \label{eq:Boltzmann_Y}
\end{align}
where $Y = n_{\varphi}/\mathfrak{s}$ is the yield, $\mathfrak{s}$ is the entropy density, and $H$ is the Hubble parameter.
The thermally averaged cross section $\langle\sigma_{\varphi\varphi} v_{\rm rel}\rangle$ is defined as
\begin{align}
   (n_\varphi^{\rm eq})^2\langle\sigma_{\varphi\varphi} v_{\rm rel}\rangle 
   &\equiv 
   \gamma_{\varphi\varphi,\chi}+\gamma_{\varphi\varphi,\chi\chi}.
\end{align}

\subsection{Thermal Averaging and the Gondolo-Gelmini Formula}
\label{subsec:thermal}

In numerically computing the reaction rates across both relativistic and non-relativistic regimes, we utilize the Gondolo-Gelmini formula~\cite{Gondolo:1990dk} to express the reaction rate $\gamma_{12,34}$ as a single integral over the Mandelstam variable $s$ (see Appendices~\ref{appendix:thermally_averaged_cross_sections}, \ref{appendix:lab_to_cm_and_back} and \ref{appendix:gondolo_gelmini_formula} for details):
\begin{align}
   \gamma_{12,34} 
   &= 
   \frac{T}{64\pi^4} \int_{(m_1+m_2)^2}^{\infty} ds \sqrt{s} \beta\left(\frac{m_1}{\sqrt{s}}, \frac{m_2}{\sqrt{s}}\right) w(s) K_1\left(\frac{\sqrt{s}}{T}\right),
   \label{eq:gondolo_gelmini}
\end{align}
where
\begin{align}
   \beta(x,y) &\equiv
   \sqrt{(1-(x-y)^2)(1-(x+y)^2)}.
\end{align}
Note that we have used the Boltzmann distribution $f_i\simeq e^{-E_i/T}$ for the distribution function of the incoming particles, and $n_i^{\rm eq}$ is the equilibrium number density of particle $i$. The function $K_1(x)$ is the modified Bessel function of the second kind.
The non-relativistic limit of the thermally averaged cross section can be readily obtained as
\begin{align}
   \langle\sigma_{12,34} v_{\rm rel}\rangle
   &\simeq
   \frac{1}{4m^2}\left[
      w_0 + \left(6m^2 w_1 - 3w_0\right)\frac{T}{m}
   \right],
\end{align}
where we have taken $m_1=m_2=m$, and
\begin{align}
   w_0 \equiv w(s=4m^2),&\;\;\; w_1\equiv \left.\frac{d w(s)}{ds}\right|_{s=4m^2}.
\end{align}
Notice that $w_0$ and $w_1$ are the s-wave and p-wave contributions, respectively, and only the former is relevant in our case since the latter appears at the next leading order in $T/m$ in the non-relativistic regime.
The relativistic limit will be discussed later.

\subsection{1-to-2 Production: Decay of the Generic Scalar}
\label{subsec:decay}

When the scalar $\chi$ is heavier than the inflaton quanta ($m_\chi > 2m_\varphi$), the decay $\chi \to \varphi\varphi$ becomes one of the primary production channels. 
The partial decay rate in the thermal bath is given by:
\begin{align}
   \gamma_{\chi,\varphi\varphi} 
   &= 
   2\times \frac{m_\chi^2 \Gamma_{\chi\to\varphi\varphi}}{2\pi^2} T K_1\left(\frac{m_\chi}{T}\right),
   \label{eq:decay_rate_generic_scalar}
\end{align}
where a factor of 2 is multiplied as a pair of $\varphi$ is produced per decay.
$\Gamma_{\chi\to\varphi\varphi}$ is the decay width in the rest frame of $\chi$, which is given by
\begin{align}
   \Gamma_{\chi\to\varphi\varphi} 
   &= 
   \frac{g_{\chi\varphi\varphi}^2}{8\pi m_\chi} \beta\left(\frac{m_\varphi}{m_\chi}, \frac{m_\varphi}{m_\chi}\right),
   \label{eq:Gamma_chi-phi-phi}
\end{align}
where the effective coupling $g_{\chi\varphi\varphi}$ incorporates the mixing effects induced by both $\sigma$ and $\mu$.
Note that we have $\gamma_{\chi,\varphi\varphi}=\gamma_{\varphi\varphi,\chi}$ which ensures the detailed balance of the process in the thermal equilibrium.

\subsection{2-to-2 Production: Scattering Kernels}
\label{subsec:scattering}

\begin{figure}[htbp]
   \centering
   \vspace{5mm}
   \includegraphics[width=.8\textwidth]{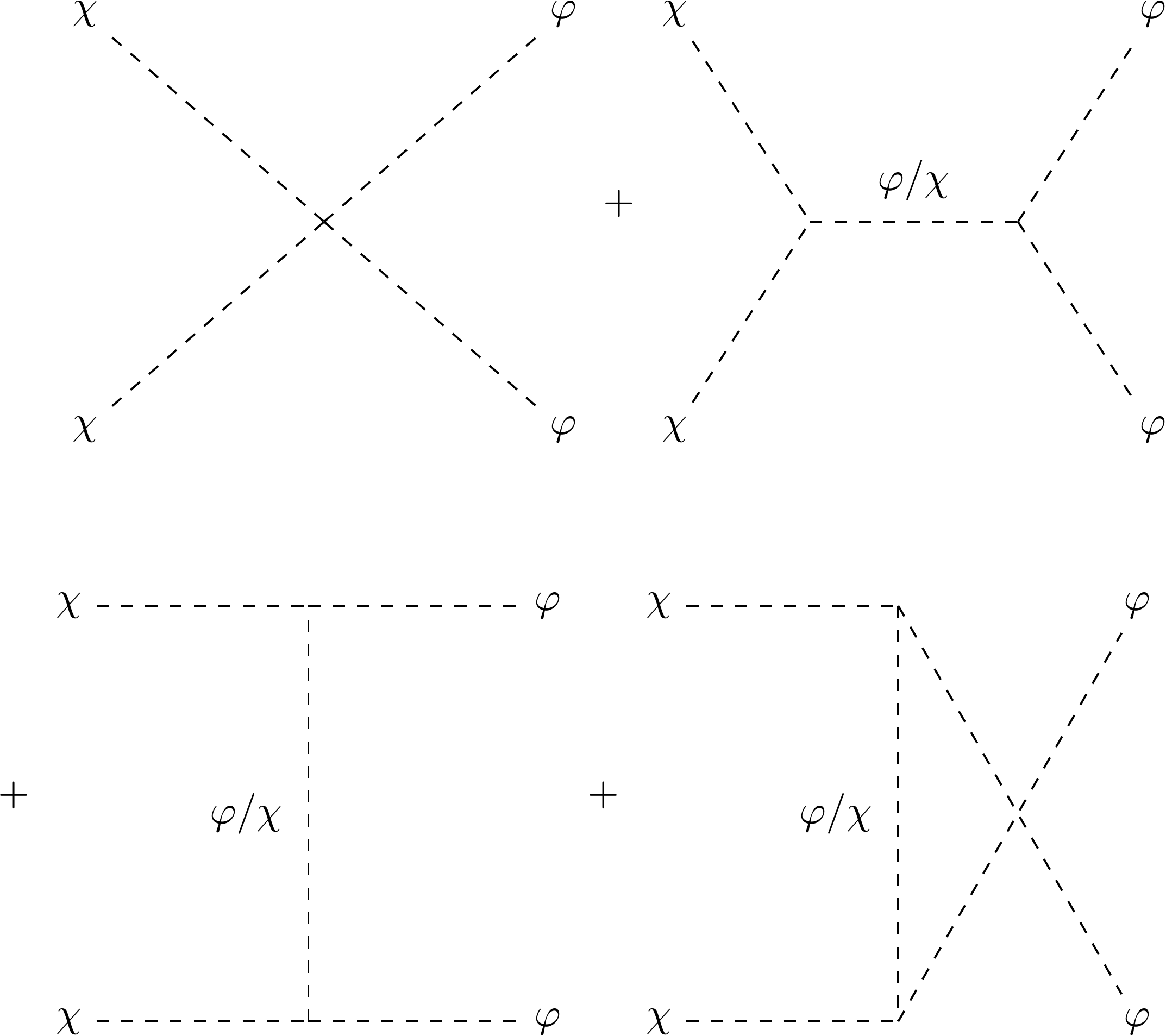}
   \caption{Diagrams of scattering processes. The top-left is the contact interaction, and the top-right shows the $s$-channel process in which $\chi$ or $\varphi$ is exchanged.
   The bottom left and right diagram are respectively} $t$- and $u$-channels, where the exchanged particles are either $\varphi$ or $\chi$.
   \label{fig:scattering}
\end{figure}

The 2-to-2 scattering processes contribute to the production and annihilation of inflaton particle.
As shown in Fig.~\ref{fig:scattering}, these processes include the contact-interaction, $s$-, $t$-, and $u$-channel diagrams.
Since our generic scalar $\chi$ is assumed to develop a VEV, it must have a self coupling.
Perturbing $\chi$ about the minimum of the potential given by
\begin{align}
   V(\chi) &=\frac{\lambda_\chi}{4}(\chi^2-v^2)^2,
\end{align}
we find a $\chi^3$ term as $V\supset (m_\chi^2/2v)\chi^3$, which induces the $s$-channel process.
The $t$- and $u$-channels are induced by exchanging either $\chi$ or $\varphi$, together with $\chi\varphi^2$ or $\chi^2\varphi$ couplings.

The detailed computation of the function $w(s)$ in the collision integral is given in Appendix~\ref{appendix:kernels}.
The relevant contributions are given in Eqs.~\eqref{eq:w_scalar_scattering} and Eqs.~\eqref{eq:w_scalar_scattering_from_A_to_J} with identifying the incoming particle $h$ as $\chi$.
From Eqs.~\eqref{eq:decay_rate_generic_scalar} and \eqref{eq:gondolo_gelmini}, we obtain $\langle\sigma_{\varphi\varphi}v_{\rm rel}\rangle$ that is to be substituted into the collision term of the Boltzmann equation~\eqref{eq:Boltzmann_Y}.

\subsection{Relic abundance}
\label{subsec:relic}

\begin{figure}[htbp]
   \centering
   \includegraphics[width=.45\textwidth]{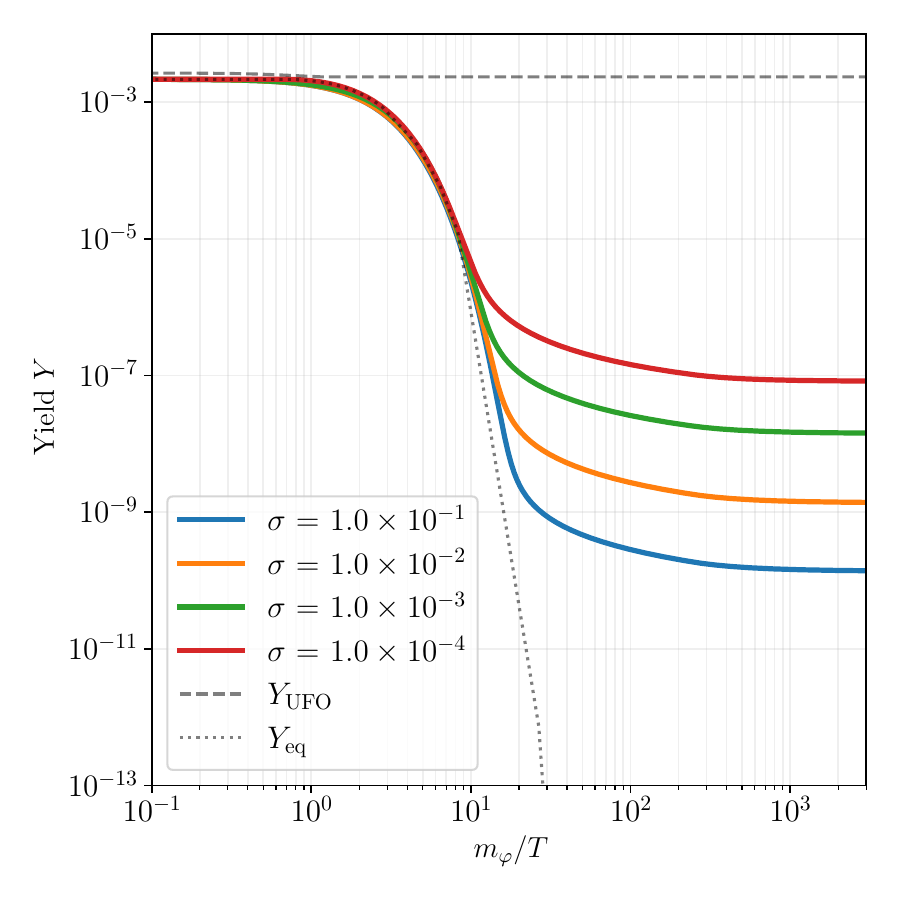}~~~~~
   \includegraphics[width=.45\textwidth]{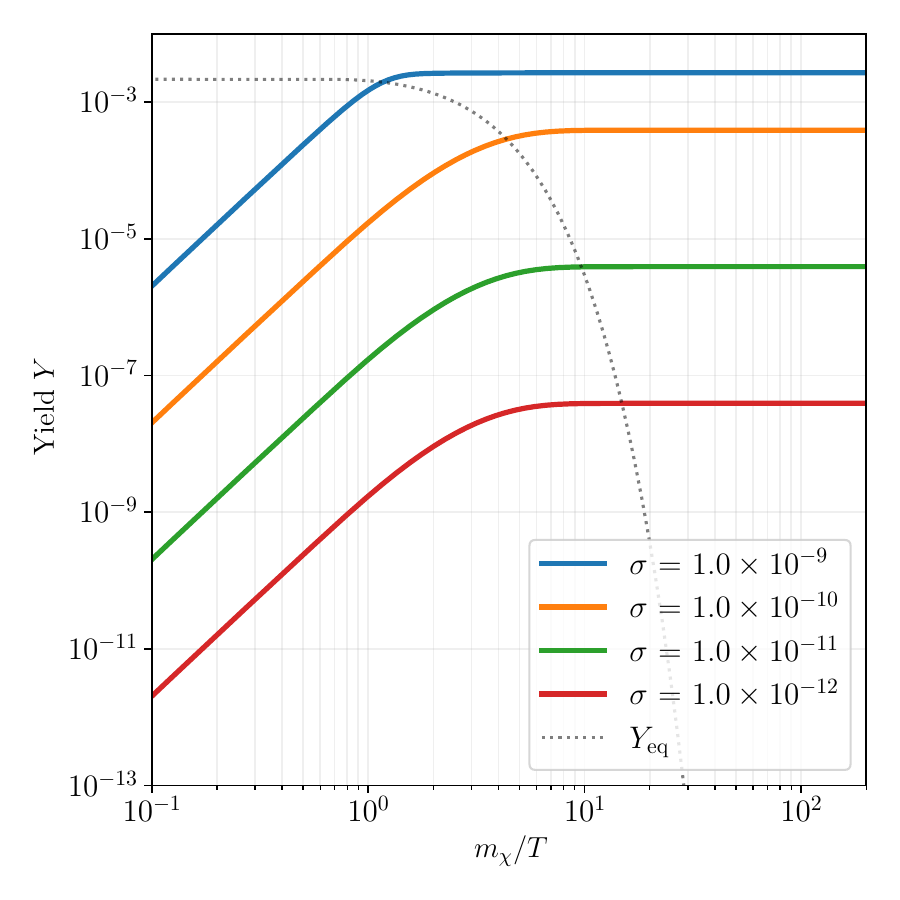}
   \caption{Evolution of $Y$ as a function of $m_\varphi/T$ (left panel) and $m_\chi/T$ (right panel), respectively, for the case of $\mu=0$, $m_\chi = 100$ GeV and $v=10$ TeV.
   The solid lines correspond to different values of $\sigma$ with $Y_{\rm ini}=0$. The black dashed line in the left panel represents the UFO scenario. The black dotted line shows the equilibrium yield $Y_{\rm eq}$.}
   \label{fig:Y_evolution}
\end{figure}

Numerically solving the Boltzmann equation~\eqref{eq:Boltzmann_Y}, we find the comoving number density today,
$Y_0\equiv Y(a=a_0)$, which is related to the present day density parameter $\Omega_\varphi$ as
\begin{align}
   \Omega_\varphi h^2
   &=
   \frac{m_\varphi n_\varphi(a_0)}{\rho_c/h^2}
   \simeq
   2.89\times10^8 \frac{m_\varphi}{\rm GeV}Y_0,
\end{align}
where $\rho_c$ is the critical density given by $\rho_c = 3H_0^2 M_P^2 \simeq 10^{-5} \,h^2~{\rm GeV/cm^3}$ and $H_0 \, (= 100 \, h \, {\rm km/s/Mpc}) $ is the present Hubble parameter.
Figure~\ref{fig:Y_evolution} shows the evolution of $Y$, where we take $\mu=0$, $m_\chi = 100$~GeV and $v=10$~TeV as an example.
Depending on the coupling strength, the production mechanism can be identified as the freeze-out or the freeze-in regimes.
We also call these regimes as the Weakly Interacting Massive Particle (WIMP) and the Feebly Interacting Massive Particle (FIMP) cases, respectively.
The left panel of the figure shows the WIPM case where $\sigma$ is relatively large, while the right panel is the FIMP case  where $\sigma$ is small.
The figure also includes the evolution of the equilibrium yield $Y_{\rm eq}$ for comparison. As $Y_{\rm eq}$ is a function of $m/T$ with $m$ identified either $m_\varphi$ or $m_\chi$, one should be careful when comparing $Y_\varphi$ with $Y_{\rm eq}$. In the left panel of the figure, we fix $\sigma=10^{-4}$, equivalently $m_\varphi=\sqrt{\sigma}v = 100$ GeV, to draw $Y_{\rm eq}$, while $m$ in $Y_{\rm eq}$ is identified as $m_\chi=100$~GeV in the right panel.
In both cases, the initial abundance is taken to be zero.
Note that while we assume that $\chi$ has the nonzero VEV during the $\varphi$ production, there is another possibility where,  during the production,  $\chi$ still has zero VEV, leading to relativistic freeze-out production, which has been discussed and 
dubbed as ultra-relativistic freeze-out (UFO) in Refs.~\cite{Henrich:2025sli,Henrich:2025pca,Henrich:2025gsd}. 
The dashed line in the left panel of Fig.~\ref{fig:Y_evolution} corresponds to the UFO scenario where the yield is estimated as $Y_{\rm UFO}\simeq 45\zeta(3)/(2\pi^4 g_{\mathfrak s}) \simeq 0.003$ with $g_{\mathfrak s}=427/4$, which is however too large a relic abundance in this particular setup, unless a considerable amount of entropy is produced at late times.

\begin{figure}
    \centering
    \includegraphics[width=0.8\linewidth]{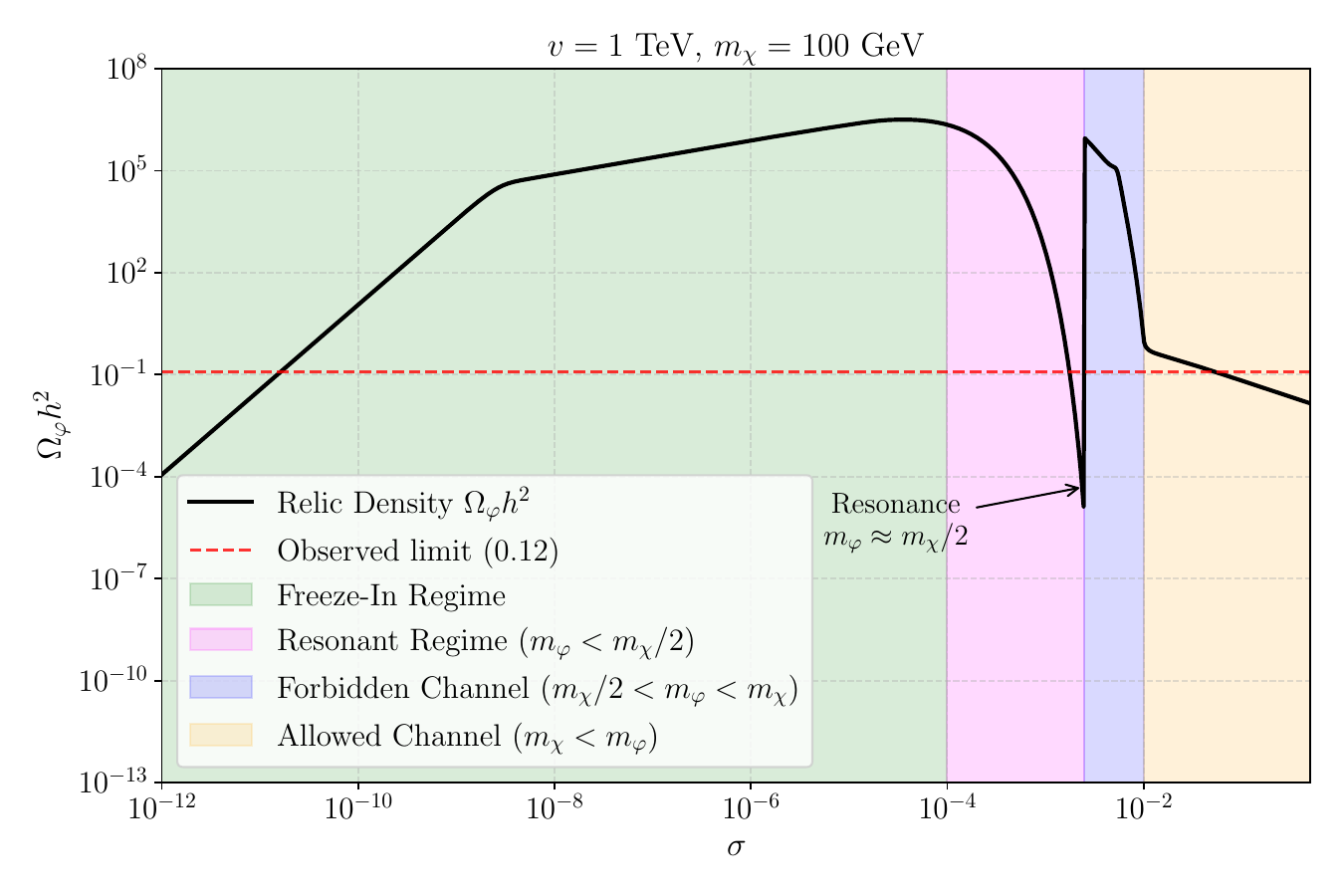}
    \caption{Relic abundance of $\varphi$ as a function of $\sigma$ when fixing $\mu=0$ in the generic scalar case for $m_\chi = 100$ GeV and $v=1$ TeV. The FIMP regime is depicted in green, and the other parameter space ($\sigma \gtrsim 10^{-4}$) is the WIMP regime where the resonant production dominates in the pink region, the forbidden channel determines the abundance in the blue region, and the regular scattering process ($\varphi\varphi\to\chi\chi$) becomes efficient in the orange region.}
    \label{fig:Oh2_sigma_generic_scalar}
\end{figure}

Figure~\ref{fig:Oh2_sigma_generic_scalar} shows $\Omega_\varphi h^2$ as a function of $\sigma$ with fixing $\mu=0$ and the vanishing initial abundance $Y_{\rm ini}=0$.
As can be seen, there are distinctive regimes depending on $\sigma$.
The green region in the figure corresponds to the FIMP regime, while $\varphi$ becomes WIMP in the the rest of the parameter space where the resonant production of $\chi\chi\to \varphi$ dominates in the production in the pink region, the forbidden channel of $\varphi\varphi\to\chi\chi$ determines the abundance in the blue region, and the regular scattering process is the leading contribution in the orange region.

In the WIMP region, $\varphi$ is once in thermal equilibrium and then decouples to freeze-out.
For $m_\varphi\gg m_\chi$, the freeze-out temperature $T_f$ is typically $x_f\equiv m_\varphi/T_f\simeq 25$, at which $\varphi$ is nearly non-relativistic.
The annihilation cross section in the non-relativistic regime becomes
\begin{align}
   \langle\sigma_{\varphi\varphi} v_{\rm rel}\rangle
   &\simeq
   \frac{w_0}{4m_\varphi^2}
   \simeq \frac{\sigma^2}{16\pi m_\varphi^2}
   \simeq 2.2\times10^{-23}{\rm cm^3/s}\times \sigma\left(\frac{v}{100~{\rm GeV}}\right)^{-2}.
\end{align}
Thus, the relic density of $\varphi$ in the WIMP regime is estimated as
\begin{align}
    \Omega_\varphi h^2
    &\simeq 
    10^{-4}\times\left(\frac{g_*(T_f)}{106.75}\right)^{-1/2}\left(\frac{x_f}{25}\right)^{-1}\sigma^{-1}\left(\frac{v}{100~{\rm GeV}}\right)^2,
\end{align}
which explains that, in Fig.~\ref{fig:Oh2_sigma_generic_scalar}, the slope of the black solid line in the orange region is $-1$.
Since $\Omega_\varphi h^2\propto v^2/\sigma$, when we increase $v$, a larger $\sigma$ is needed to obtain $\Omega_\varphi h^2\simeq 0.12$.
Note that such a scenario of ``inflaton" as dark matter has been studied in \cite{Lerner:2009xg,Okada:2010jd,Mukaida:2014kpa,Hooper:2018buz,Garcia:2021gsy}, and is dubbed as WIMPflation in ~\cite{Hooper:2018buz}.
Our particular case can be regarded as such a specific scenario.

When $m_\varphi \ll m_\chi$, the annihilation channel of $\varphi\varphi\to\chi\chi$ is suppressed by the Boltzmann factor $e^{-2(m_\chi-m_\varphi)/T}$, known as the forbidden channel~\cite{Griest:1990kh,DAgnolo:2015ujb}.
In the non-relativistic regime for $\chi$, the annihilation cross section becomes
\begin{align}
    \langle \sigma_{\varphi\varphi}v_{\rm rel}\rangle
    &\simeq
    \frac{w_0}{4m_\chi^2}\left(\frac{n^{\rm eq}_\chi}{n^{\rm eq}_\varphi}\right)^2
    \simeq
    \frac{\sigma^2}{4\pi m_\chi^2}\left(\frac{m_\chi}{m_\varphi}\right)^3 e^{-2(m_\chi-m_\varphi)/T}.
\end{align}
Therefore, as $m_\varphi$ approaches $m_\chi$, the cross section increases rapidly, making the slope of the solid black line in the blue region of Fig.~\ref{fig:Oh2_sigma_generic_scalar} steep.

When $m_\varphi < m_\chi/2$, the dominant contribution in $\langle \sigma_{\varphi\varphi}v_{\rm rel}\rangle$ is the resonant process, i.e., $\varphi \varphi \leftrightarrow \chi$.
The annihilation cross section at the freeze-out temperature in this case becomes
\begin{align}
    \langle \sigma_{\varphi\varphi}v_{\rm rel}\rangle
    &\simeq \frac{\gamma_{\varphi\varphi,\chi}}{(n^{\rm eq}_\varphi)^2}
    \simeq (2\pi)^{3/2}x_f^{3/2}\frac{\Gamma_{\chi\to\varphi\varphi}}{m_\varphi^3}e^{-x_f(m_\chi/m_\varphi -2)},
\end{align}
with
\begin{align}
    \Gamma_{\chi\to\varphi\varphi}
    &=
    \frac{(\sigma v)^2}{8\pi m_\chi}\sqrt{1-\frac{4m_\varphi^2}{m_\chi^2}}.
\end{align}
The reaction rate reaches its maximum value when $m_\chi/(2m_\varphi)-1=1/(4x_f)$, at which we find
\begin{align}
    \langle \sigma_{\varphi\varphi}v_{\rm rel}\rangle
    &\simeq
    2.7\times 10^{-19}~{\rm cm^2/s}\left(\frac{v}{100~{\rm GeV}}\right)^{-2}\left(\frac{x_f}{25}\right)^2.
\end{align}
Thus, the least abundance is achieved as
\begin{align}
    \Omega_\varphi h^2
    &\simeq
    10^{-8}\times \left(\frac{g_*(T_f)}{106.75}\right)^{-1/2}\left(\frac{x_f}{25}\right)^{-1}\left(\frac{v}{100~{\rm GeV}}\right)^2.
\end{align}
This corresponds to the dip appeared at $m_\varphi\simeq m_\chi/2$ in Fig.~\ref{fig:Oh2_sigma_generic_scalar}.

Within the FIMP regime, the kernel of the collision integral simplifies to
\begin{align}
   w_{\chi\chi,\varphi\varphi}(s) &\simeq \frac{\sigma^2}{4\pi},
\end{align}
and thus we obtain
\begin{align}
   \gamma_{\chi\chi,\varphi\varphi}
   &\simeq
   \frac{1}{16\pi^4}\cdot \frac{\sigma^2}{4\pi}\left(
      \int_0^\infty\frac{EdE}{e^{E/T}-1}
   \right)^2
   =
   \frac{\sigma^2 T^4}{2304\pi},
\end{align}
where we have used the Bose-Einstein distribution for $\chi$.
Since $\varphi\varphi\to\chi\chi$ process is negligible in the FIMP regime, the Boltzmann equation becomes
\begin{align}
   \frac{dY}{dT} &\simeq -\frac{\gamma_{\chi\chi,\varphi\varphi}}{\mathfrak{s}TH} \,,
\end{align}
from which we obtain
\begin{align}
    Y_0^{(\rm scattering)}
   \simeq
   \frac{135\sqrt{10}}{2\pi^3g_*^{3/2}(T=m_\chi)}\cdot\frac{\sigma^2}{2304\pi}\frac{M_P}{m_\chi} + Y_{\rm ini} \,,
\end{align}
where $Y_{\rm ini}$ is the initial abundance of $\varphi$.
For the effective degrees of relativistic freedoms in entropy, $g_\mathfrak{s}$, and energy density, $g_\rho$, we have approximated as $g_\mathfrak{s}\simeq g_\rho\simeq g_*$ at $T\simeq m_\chi$ at which the production of $\varphi$ becomes most efficient.

\begin{figure}[htbp]
   \centering
   \includegraphics[width=\textwidth]{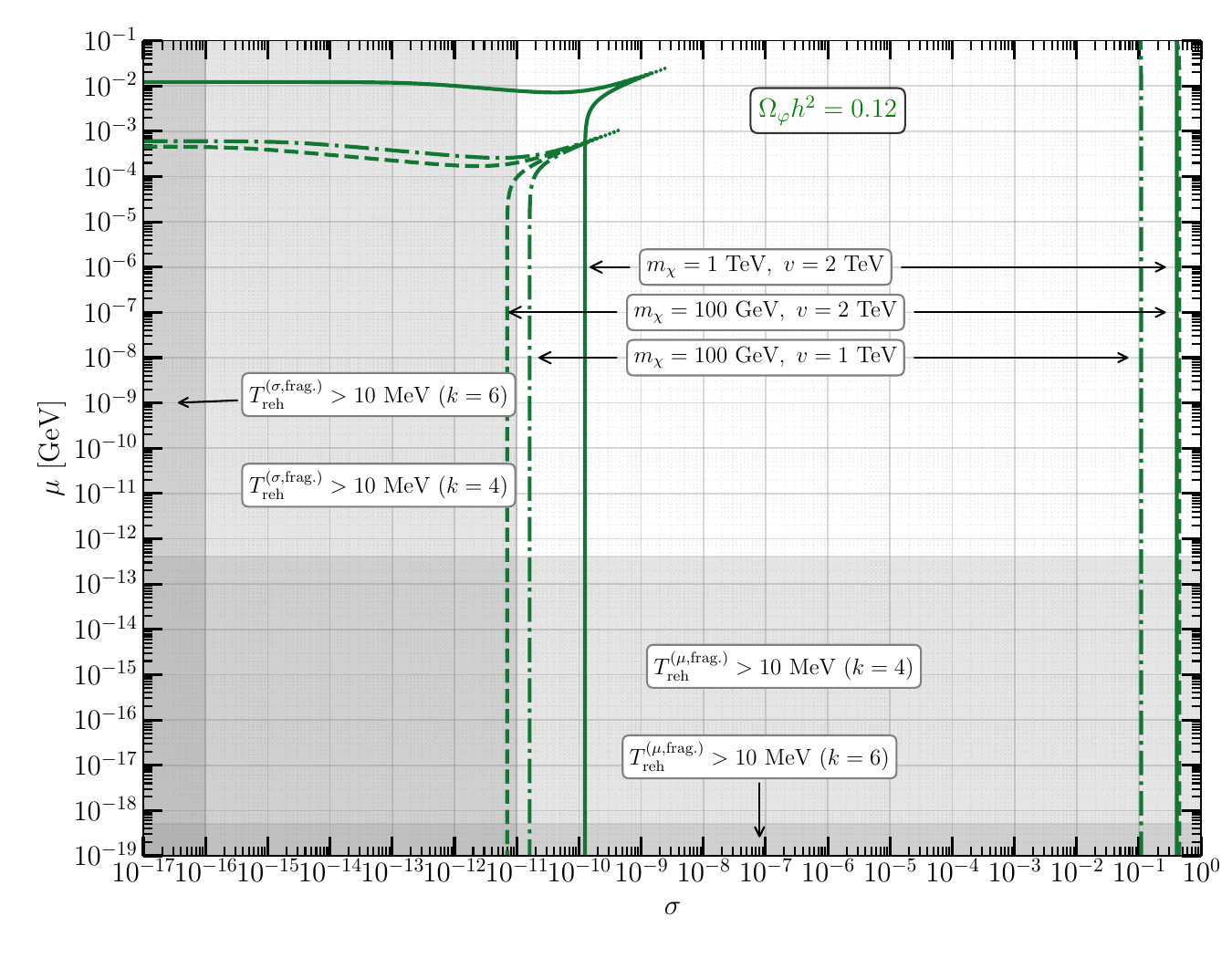}
   \caption{Relic abundance of $\varphi$ interacting with a generic scalar $\chi$. 
   The solid, dot-dashed, and dashed lines correspond to the cases for different values of $m_\chi$ and $v$ as shown in the figure. Regions between two lines for each type are excluded due the overproduction of $\varphi$. The shaded gray regions are disfavored due to the reheating temperature lower than 10 MeV (BBN bound) with the fragmentation effect.
   }
   \label{fig:relic_abundance_generic_scalar}
\end{figure}

In the similar manner, when keeping $g_{\chi\varphi\varphi}$ generic (including both $\sigma$ and $\mu$ couplings), the decay contribution is given by
\begin{align}
   \gamma_{\chi,\varphi\varphi}
   &\simeq
   2\times\frac{1}{12}m_\chi\Gamma_{\chi\to\varphi\varphi}T^2\,,
\end{align}
with 
\begin{align}
   \Gamma_{\chi\to\varphi\varphi}
   &\simeq
   \frac{g_{\chi\varphi\varphi}^2}{8\pi m_\chi}\sqrt{1-\frac{4m_\varphi^2}{m_\chi^2}},
\end{align}
yielding
\begin{align}
   Y_0^{\rm (decay)}
   &\simeq
   \frac{15\sqrt{10}}{4\pi^3g_*^{3/2}(T=m_\chi)}\frac{M_P\Gamma_{\chi\to\varphi\varphi}}{m_\chi^2} + Y_{\rm ini}.
   \label{eq:Y0_decay}
\end{align}
With the initial condition $Y_{\rm ini}=0$, it turns out that the decay contribution is dominant in the freeze-in regime, and thus we obtain
\begin{align}
   \Omega_\varphi h^2 
   &\simeq
   0.12\times \left(
      \frac{g_*}{106.75}
   \right)^{-3/2}\left(
      \frac{\sigma}{4.3\times10^{-11}}
   \right)^{5/2}\left(
      \frac{v/10~{\rm TeV}}{m_\chi/1~{\rm TeV}}
   \right)^3,
   \label{eq:Omega_phi_FIMP_sigma}
\end{align}
when $\mu$ is vanishing, which is the case shown in Fig.~\ref{fig:Oh2_sigma_generic_scalar}.
On the other hand, as will be shown in Fig.~\ref{fig:relic_abundance_generic_scalar}, for sufficiently small $\sigma$ and non-vanishing $\mu$, we have $m_\varphi\simeq \mu/4\pi$ and thus obtain
\begin{align}
   \Omega_\varphi h^2 
   &\simeq
   0.12\times \left(
      \frac{g_*}{106.75}
   \right)^{-3/2}
   \left(
      \frac{v}{10~{\rm TeV}}
   \right)^3
   \left(\frac{1~{\rm TeV}}{m_\chi}\right)^7
   \left(
      \frac{\mu}{10^{-2}~{\rm GeV}}
   \right)^5.
   \label{eq:Omega_phi_FIMP_mu}
\end{align}

Figure~\ref{fig:relic_abundance_generic_scalar} shows contours of $\Omega_\varphi h^2=0.12$, namely, the produced inflaton being the whole dark matter in the present Universe, where we have assumed that $Y$ is initially vanishing.
The region enclosed by the two lines of $\Omega_\varphi h^2=0.12$ is excluded because the abundance is too large.
This initial condition, $Y_{\rm ini} \approx 0$, is motivated by the rapid decay of fragmentation products, 
leaving thermal regeneration as the dominant production mechanism.
The solid, dot-dashed, and dashed lines in the figure correspond to different values of $m_\chi$ and $v$.
Each type of line has two viable parameter space to explain dark matter abundance.
The vertical lines at larger $\sigma$ correspond to the WIMP region, while the ones at smaller $\sigma$ correspond to the FIMP region.
Note that there is a narrow strip in the parameter space where $g_{\chi\varphi\varphi}$ is nearly zero, which is due to the cancellation between the $\sigma$ and $\mu$ contributions to $g_{\chi\varphi\varphi}$. This cancellation leads to a suppression of the decay and scattering rates, requiring larger $\sigma$ and $\mu$ to achieve the observed amount of relic abundance.
Although omitted in the figure, there are also resonant regions where $m_\varphi\simeq m_\chi/2$ is satisfied.

Note that this initial condition, $Y_{\rm ini}\approx 0$, can be justified quantitatively.
The fragmentation products are relativistic $\varphi$ particles whose depletion during reheating was computed in Ref.~\cite{Garcia:2023dyf}.
Extrapolating their results beyond $a_{\rm reh}$, and characterizing the mean energy of the fragments as in Ref.~\cite{Garcia:2023dyf}, we obtain a surviving abundance
\begin{align}
   Y^{\rm frag} &\simeq
   \begin{cases}
      \displaystyle
      \frac{1}{1728g_*^{1/4}c_s}\left[
         13-\left(
            \frac{a}{a_{\rm reh}}
         \right)^3
      \right]^3
      &
      \text{($\mu$-decay)} \\
      \displaystyle
      \frac{1}{g_*^{1/4}c_s}
      \left(
         \frac{4}{3+a/a_{\rm reh}}
      \right)^3
      &
      \text{($\sigma$-scattering)}
   \end{cases}
   ,
   \label{eq:Yfrag}
\end{align}
where $c_s=2\pi^4/45\zeta(3) \simeq 3.6$.
It can be readily seen that when reheating is dominated by $\mu$-decay, $Y^{\rm frag}$ is quickly depleted at $a/a_{\rm reh}=13^{1/3}\simeq 2.35$, and thus the initial abundance of $\varphi$ is negligible for the subsequent thermal production.
When reheating is dominated by $\sigma$-scattering, we may parametrize the $Y^{\rm frag}$ contribution by using Eq.~\eqref{eq:Y0_decay} as
\begin{align}
   Y_0^{(\rm decay)}
   &\simeq
   \frac{15\sqrt{10}}{4\pi^3g_*^{3/2}}\frac{M_P\Gamma_{\chi\to\varphi\varphi}}{m_\chi^2}\left(
      1+\Delta^{\rm frag}
   \right),
\end{align}
where
\begin{align}
   \Delta^{\rm frag}
   &\simeq
   \frac{2^{11}\pi^4}{15\sqrt{10}}\frac{g_*^{5/4}}{c_s \sigma^2}\frac{m_\chi^6}{M_P v^2 T_{\rm reh}^3}\nonumber\\
   &\simeq
   10^{-3}\left(
      \frac{g_*}{106.75}
   \right)^{5/4}
   \left(
      \frac{4.3\times10^{-11}}{\sigma}
   \right)^2
   \left(
      \frac{10~{\rm TeV}}{v}
   \right)^2
   \left(
      \frac{m_\chi}{1~{\rm TeV}}
   \right)^6
   \left(
      \frac{10^7~{\rm GeV}}{T_{\rm reh}}
   \right)^3.
\end{align}
The quantity $\Delta^{\rm frag}$ thus measures the fractional contribution of the fragmentation remnant to the freeze-in yield. 
Here we have fixed the value of $\sigma$,  corresponding to $\Omega_\varphi h^2 = 0.12$ as an illustrative example, and $T_{\rm reh}$ then follows from Eq.~\eqref{eq:T_sigma}.
Evaluating $\Delta^{\rm frag}$ at the benchmarks of Fig.~\ref{fig:relic_abundance_generic_scalar}, we find $\Delta^{\rm frag}\lesssim10^{-5}$ for $k\ge6$, so that the initial abundance produced by fragmentation is negligible and setting $Y_{\rm ini}\approx0$ is justified throughout the viable region. 
For $k=4$, by contrast, the coupling required to achieve $\Omega_{\varphi}h^2=0.12$ renders $T_{\rm reh}$ comparable to $m_\chi$, so that the assumption $T_{\rm reh}\gg m_\chi$ underlying the post-reheating freeze-in description is not satisfied.

Combining the freeze-out and the freeze-in production of $\varphi$, we find that the parameter space between the two regimes is excluded due to the overproduction of $\varphi$.
Furthermore, by incorporating the fragmentation effects, we require that the reheating temperature ($T^{(\mu,\text{frag.})}_{\rm reh}$ and $T^{(\sigma,\text{frag.})}_{\rm reh}$ achieved by either $\mu$- or $\sigma$-couplings, respectively) should be larger than 10 MeV so that the successful BBN is realized.
Disfavored regions are translated into $\sigma\lesssim 10^{-11}$ and $\mu\lesssim4\times10^{-13}~{\rm GeV}$ for $k=4$, and $\sigma\lesssim 10^{-16}$ and $\mu\lesssim 5\times10^{-19}~{\rm GeV}$ for $k=6$~\cite{Garcia:2023dyf}, which are depicted in gray region in Fig~\ref{fig:relic_abundance_generic_scalar}.

Before closing this section, we note that the results presented in Fig.~\ref{fig:relic_abundance_generic_scalar} assume the inflaton particle stable for simplicity.
However, inflaton is in general unstable since it mixes with $\chi$, and $\chi$ should interact with particles in the thermal bath including those in the SM.
The inflaton lifetime depends on the model of how $\chi$ couples to other matter particles.
For instance, three possibilities may be considered for how $\chi$ eventually thermalizes with the SM. (i) $\chi$ is identified with the SM Higgs boson itself, which we analyze in detail in the next section. (ii) $\chi$ is a distinct real singlet that mixes with the Higgs via a renormalizable portal coupling $\lambda_{\chi H}|\chi|^2 H^\dagger H$. The resulting experimental constraints are of the same type as those discussed in Sec.~\ref{subsec:constraints} for the Higgs-mixing case. (iii) $\chi$ is promoted to a complex scalar charged under a dark $U(1)'$ that thermalizes with the SM via kinetic mixing $\varepsilon F'_{\mu\nu}F^{\mu\nu}$ with hypercharge, without invoking Higgs mixing. In this case the relevant experimental constraints come from beam-dump and dark-photon searches in the $(\varepsilon, m_{A'})$ plane, independent of $\sigma$, $\mu$, $v$, and $m_\chi$.
In the next section, we will focus on the first possibility, where $\chi$ is identified with the SM Higgs boson, and discuss the phenomenological implications of the inflaton decay into SM particles.


\section{Case Study: The Standard Model Higgs Portal}
\label{sec:higgs_portal}

In the general framework discussed in the previous sections, the scalar field $\chi$ was treated as a generic singlet. We now turn to a specific and phenomenologically rich scenario where $\chi$ is identified with the SM Higgs doublet $H$.

\subsection{Interaction and Mixing}
\label{subsec:interaction}

In a scenario where the inflaton $\varphi$ interacts with the SM Higgs doublet, the relevant interaction Lagrangian is given by
\begin{align}
   \mathcal{L}_{\rm int} = \sigma |H|^2 \varphi^2 + 2\mu |H|^2 \varphi.
\end{align}
After electroweak symmetry breaking, the Higgs field develops a VEV $v \simeq 246$ GeV. By expanding the Higgs field as $H \sim (v+h)/\sqrt{2}$, we find that the linear coupling $\mu$ induces a mixing between the inflaton $\varphi$ and the physical Higgs boson $h$. 
Neglecting small tadpole contributions (assuming $\mu \ll v$), the mass matrix calculation in Eq.~\eqref{eq:mass_matrix} yields a mixing angle $\theta$:
\begin{align}
   \theta &\simeq \frac{2\mu v}{m_h^2},
\end{align}
where we have assumed $m_\varphi \ll m_h$. 
This mixing is crucial as it allows the inflaton to decay into SM particles, rendering it unstable, unlike a generic stable scalar case often assumed in simple dark matter models. 
The physical eigenstates are obtained by rotating the fields by $\theta$. 
This generates effective interaction terms as
\begin{align}
    {\cal L}_{\rm int}
    &=
    \frac{1}{2}\sigma \varphi^2 h^2
    +(\sigma v-2\theta \mu)\varphi^2 h + (\mu+2\theta\sigma v)\varphi h^2 + {\cal O}(\theta^2)\nonumber\\[8pt]
    &\equiv 
    \frac{1}{2}\sigma \varphi^2 h^2
    + g_{h\varphi\varphi} h\varphi^2
    + g_{hh\varphi}h^2\varphi
    + {\cal O}(\theta^2),
\end{align}
where
\begin{align}
    g_{h\varphi\varphi} &= \sigma v - 2\theta\mu \simeq \sigma v - 4\mu^2 v/m_h^2,\\[8pt]
    g_{hh\varphi} &= \mu + 2\theta\sigma v \simeq \mu + 4\sigma \mu v^2/m_h^2.
\end{align}
These effective couplings drive the production of inflaton quanta from the thermal bath.
Note that in the following analysis, we again consider the inflaton mass arising from both the tree- and loop-levels as given in Eq.~\eqref{eq:mass_phi}.

\subsection{Inflaton Production and Relic Abundance}
\label{subsec:production}

\begin{figure}[htbp]
   \centering
   \includegraphics[width=.3\textwidth]{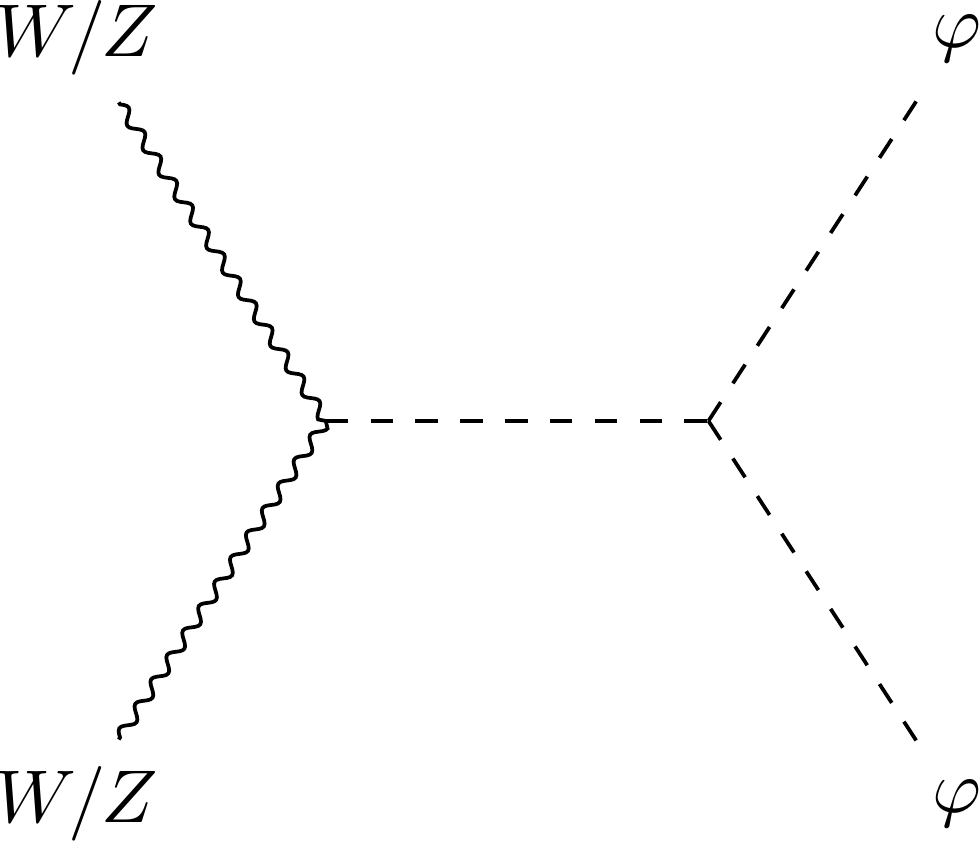}~~~~~~~~~~~
   \includegraphics[width=.3\textwidth]{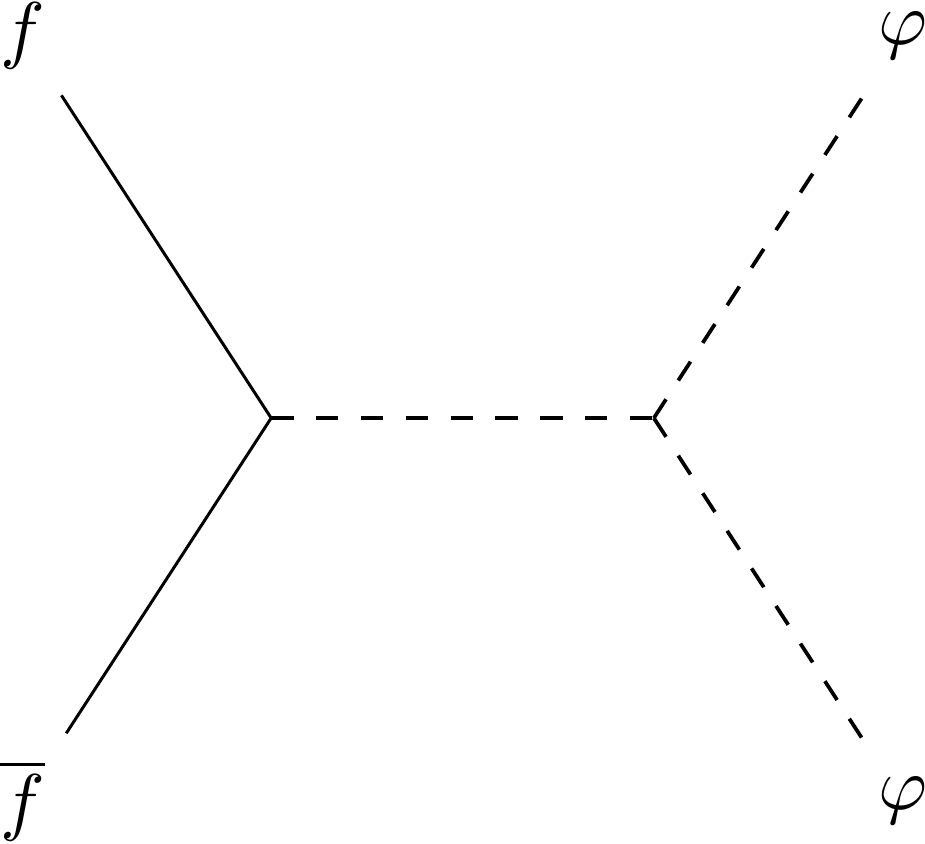}
   \caption{Diagrams of vector scattering ($W^+W^-/ZZ\to h \to \varphi\varphi$) and fermion scattering ($f\bar{f}\to h \to \varphi\varphi$) in the Higgs portal scenario.}
   \label{fig:diagram_VV_ff}
\end{figure}

\begin{figure}[htbp]
   \centering
   \includegraphics[width=.8\textwidth]{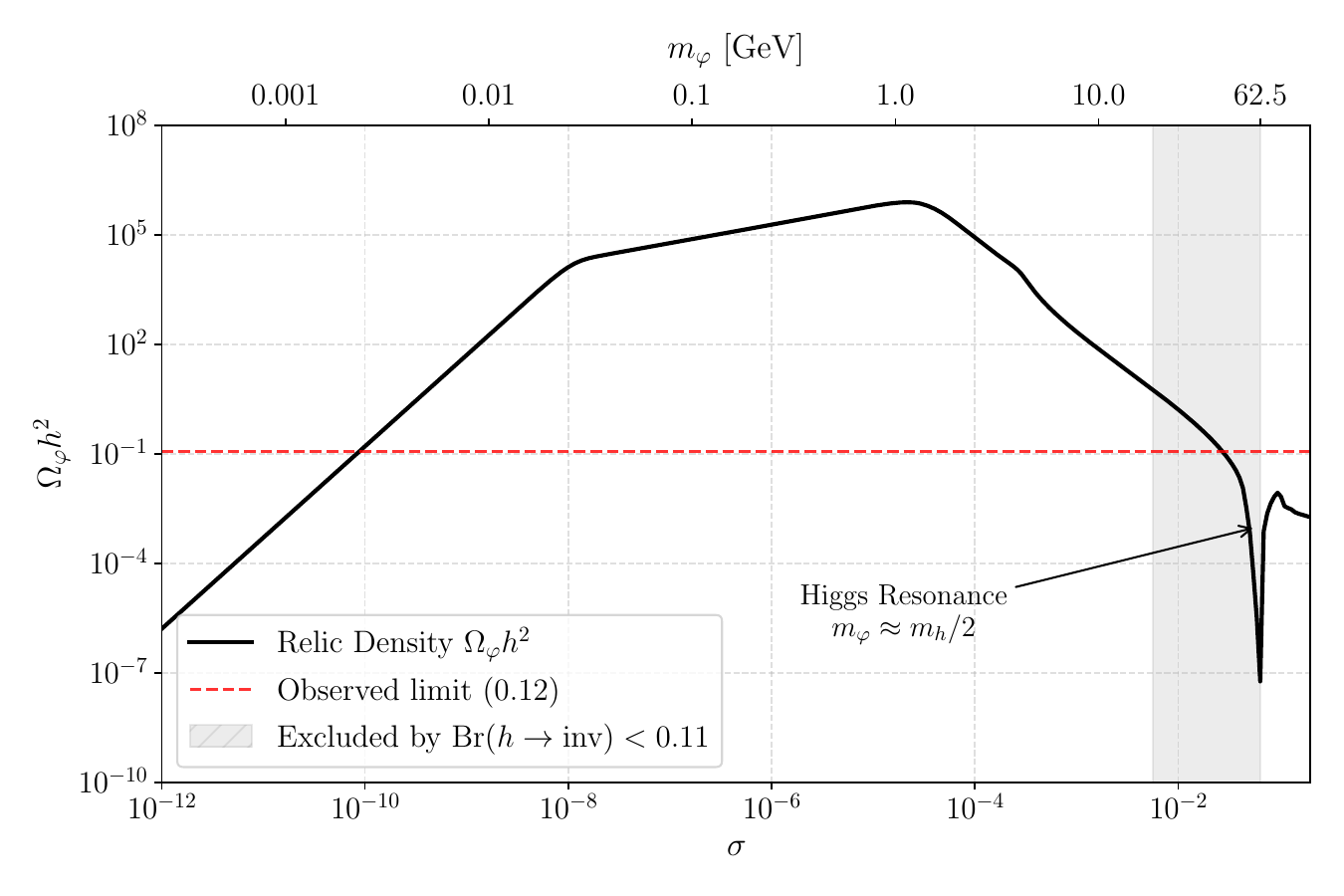}
   \caption{Relic abundance of $\varphi$ as a function of $\sigma$ when fixing $\mu=0$ in the Higgs portal scenario. 
   Gray shaded region is excluded by the invisible Higgs decay measurement, which includes the WIMP DM case. 
   }
   \label{fig:Oh2_sigma_higgs_portal}
\end{figure}

The production of inflaton particles proceeds through the decay of the Higgs boson ($h \rightarrow \varphi\varphi$) and 2-to-2 scattering processes involving the Higgs and weak gauge bosons ($hh \rightarrow \varphi\varphi$, $WW/ZZ \rightarrow \varphi\varphi$). 
The diagrams of $hh\to\varphi\varphi$ are the same as shown in Fig.~\ref{fig:scattering} with identifying $\chi$ as $h$, and those of $W^+W^-/ZZ\to h \to \varphi\varphi$ and $f\overline f\to h \to \varphi\varphi$ are shown in Fig.~\ref{fig:diagram_VV_ff}.
The scattering kernels for these processes, including the interferences between contact, $s\text{-}$, $t\text{-}$, and $u$-channel diagrams, are detailed in Appendix~\ref{appendix:kernels}.
Solving the Boltzmann equation numerically, we determine the relic abundance of the inflaton particle.

To better understand which physical processes determine the relic abundance, we first consider the $\mu=0$ case (see Fig.~\ref{fig:Oh2_sigma_higgs_portal}).
In this scenario, the inflaton mass is solely determined by the $\sigma$-coupling, i.e., $m_\varphi = \sqrt{\sigma}v$.
For $\sigma \lesssim 2\times10^{-5}$, $\varphi$ never thermalizes with the bath, and thus the freeze-in mechanism determines its final abundance.
For $2\times 10^{-7}\lesssim \sigma \lesssim 2\times 10^{-5}$, the slope of the black solid line in Fig~\ref{fig:Oh2_sigma_higgs_portal} is shallower than that for $\sigma\lesssim 2\times 10^{-7}$.
This is because when $\sigma$ increases in the freeze-in regime, the backreaction of the produced $\varphi$ particles on the thermal bath becomes more significant, leading to a suppression of further production.
At $\sigma \simeq 2\times10^{-5}$, $\varphi$ becomes a member of the thermal bath, and thus its abundance is determined by the freeze-out mechanism for larger $\sigma$.
When $m_\varphi$ comes to close to $m_h/2$, the reaction rate of $\varphi\varphi\to h \to f\overline f$ is enhanced by the Higgs resonance.
As $\sigma$ increases further, the cross section decreases because the resonance loses efficiency, despite kinematically opening $WW/ZZ$ annihilation channels for $\varphi$.
For the region where the inflaton can play a role of DM, however, the measurement of the invisible Higgs decay channel places a stringent bound, and the viable inflaton DM scenario of $\Omega_\varphi h^2=0.12$ is excluded~\cite{ATLAS:2023tkt}.

\begin{figure}[htbp]
   \centering
   \includegraphics[width=\textwidth]{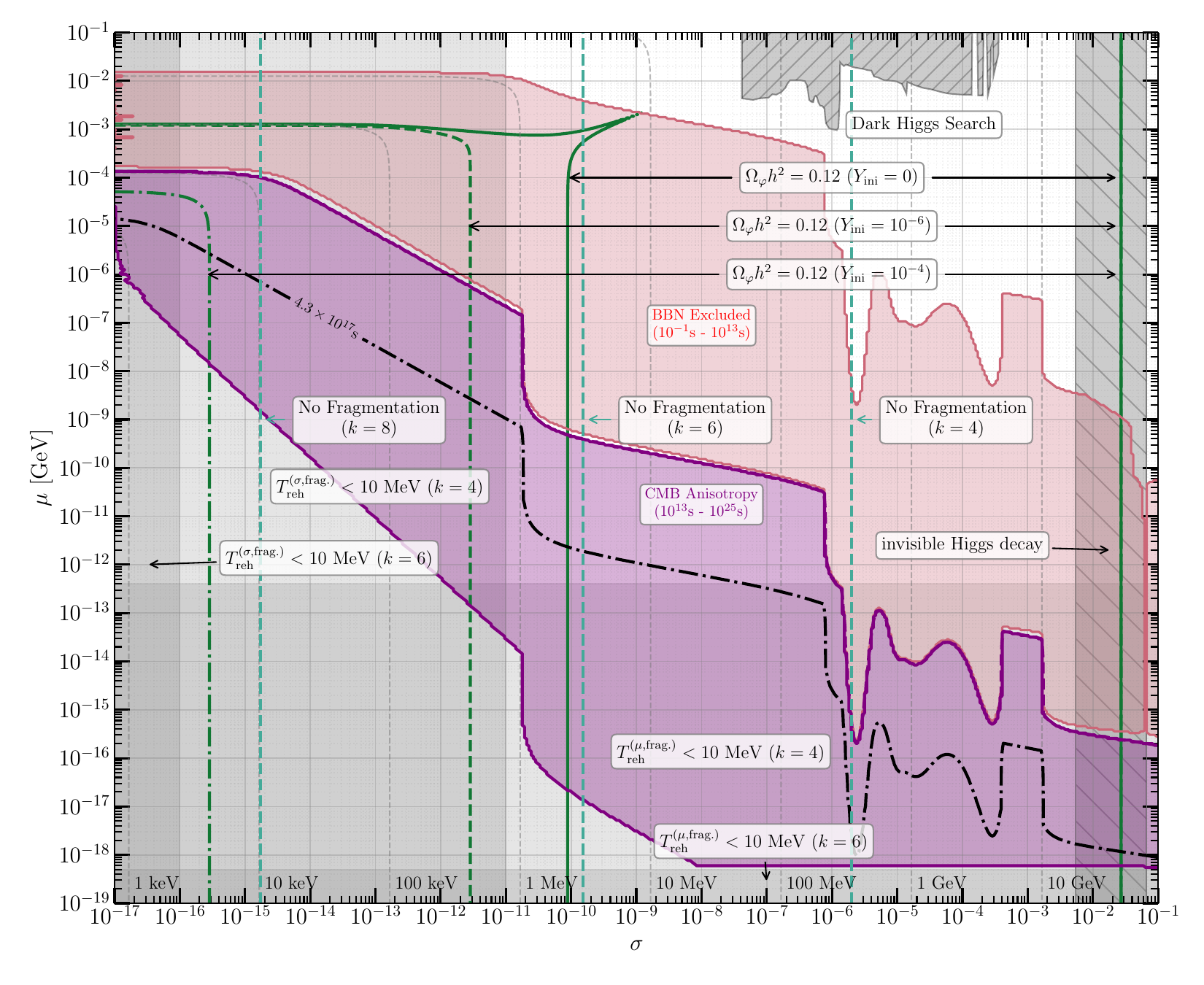}
   \caption{Various constraints on the Higgs portal scenario and the survival corridor. The gray dashed lines indicate the value of $m_\varphi$. Note that the constraints from the BBN ($\tau=10^{-1}~{\rm s}-10^{13}~{\rm s}$) and the CMB anisotropy ($\tau=10^{13}~{\rm s}-10^{25}~{\rm s}$) assume $Y_{\rm ini}=0$.}
   \label{fig:main_result_higgs_portal}
\end{figure}

In Fig.~\ref{fig:main_result_higgs_portal}, various constraints listed in Sec.~\ref{subsec:constraints} on the Higgs portal scenario in the $\sigma$--$\mu$ plane. 
As illustrated in  the figure, the abundance depends strongly on the couplings $\sigma$,  and $\mu$ when $\mu$ is large. 
As in the generic scalar case discussed in the previous section, we observe two distinct regimes:
\begin{itemize}
    \item Freeze-in: For small couplings, the abundance accumulates gradually. The production is dominated by the decay $h \rightarrow \varphi\varphi$ when kinematically allowed.
    \item Freeze-out: For larger couplings, the inflaton reaches thermal equilibrium with the SM bath before decoupling.
\end{itemize}
Contours of $\Omega_\varphi h^2 = 0.12$ are depicted by the green lines in Fig~\ref{fig:main_result_higgs_portal}, depending on the amount of pre-existing $\varphi$ number density, $Y_{\rm ini}$. 
In the freeze-out regime where $\sigma\simeq 0.1$, the resultant abundance does not depend on $Y_{\rm ini}$, as expected.
On the other hand, in the freeze-in regime where the required value of $\sigma$ is smaller, it is sensitive to the initial abundance.
As we have seen in the generic scalar case, there also exists a strip along which the coupling $g_{h\varphi\varphi}$ is nearly zero, due to the cancellation between the $\sigma$ and $\mu$ contributions to $g_{h\varphi\varphi}$.
This region, however, does not exist when $Y_{\rm ini}$ is nonzero since the initial abundance of $\varphi$ makes $\Omega_\varphi h^2$ irrelevant for the cancellation in the effective coupling.

In the freeze-in regime, we can readily estimate the abundance by using Eq.~\eqref{eq:Omega_phi_FIMP_sigma} (for sufficiently small $\mu$) and Eq.~\eqref{eq:Omega_phi_FIMP_mu} (for sufficiently small $\sigma$), respectively, as
\begin{align}
   \Omega_\varphi h^2 
   &\simeq
   \begin{cases}
      \displaystyle
      0.12\left(\frac{\sigma}{3.1\times10^{-10}}\right)^{5/2} + \left(\frac{\sigma}{3.1\times10^{-10}}\right)^{1/2}\left(
         \frac{Y_{\rm ini}}{8.1\times10^{-7}}
      \right) 
      \qquad (\textrm{for small}~ \mu),
      \\[14pt]
      \displaystyle
      0.12\left(\frac{\mu}{2.4~{\rm MeV}}\right)^5 + \left(\frac{\mu}{2.4~{\rm MeV}}\right)\left(\frac{Y_{\rm ini}}{1.8\times10^{-5}}\right)
      \qquad (\textrm{for small}~ \sigma),
   \end{cases}
\end{align}
where $Y_{\rm ini}$ parametrizes a pre-existing population of $\varphi$ particles, whatever its origin,
and we have taken $g_*=106.75$ and $m_h=125$ GeV.
As can be seen in Fig.~\ref{fig:main_result_higgs_portal}, when $\sigma$-coupling dominates in $m_\varphi$, the initial value of $Y_{\rm ini}\gtrsim10^{-6}$ affects the relic abundance, while when $\mu$-coupling dominates in the mass, the resultant abundance is affected for $Y_{\rm ini}\gtrsim10^{-5}$.
The fragmentation remnant estimated in Sec.~\ref{subsec:relic} carries over to the Higgs portal upon replacing $m_\chi\to m_h$ and $v\to246$~GeV, at the viable freeze-in point ($\sigma\simeq10^{-10}$) it amounts to $Y^{\rm frag}\simeq1.2\times 10^{-18}$ (with $T_{\rm reh}\simeq 2.1\times10^{8}$ GeV) for $k=6$ and is smaller for larger $k$.
The contours with $Y_{\rm ini}\neq0$ in Fig.~\ref{fig:main_result_higgs_portal} are therefore to be read as parametrizing an initial population of a different origin, such as the ultraviolet-dominated production discussed in Sec.~\ref{sec:conclusion}, rather than the fragmentation of the condensate.

Note that the parameter space between the freeze-out and the freeze-in regions is disfavored by the overproduction of $\varphi$, except for the case where $\varphi$ decays before BBN, as discussed in the next subsection.

\subsection{Inflaton Lifetime and Decay}
\label{subsec:lifetime}

\begin{figure}[htbp]
    \centering
    \includegraphics[width=.9\textwidth]{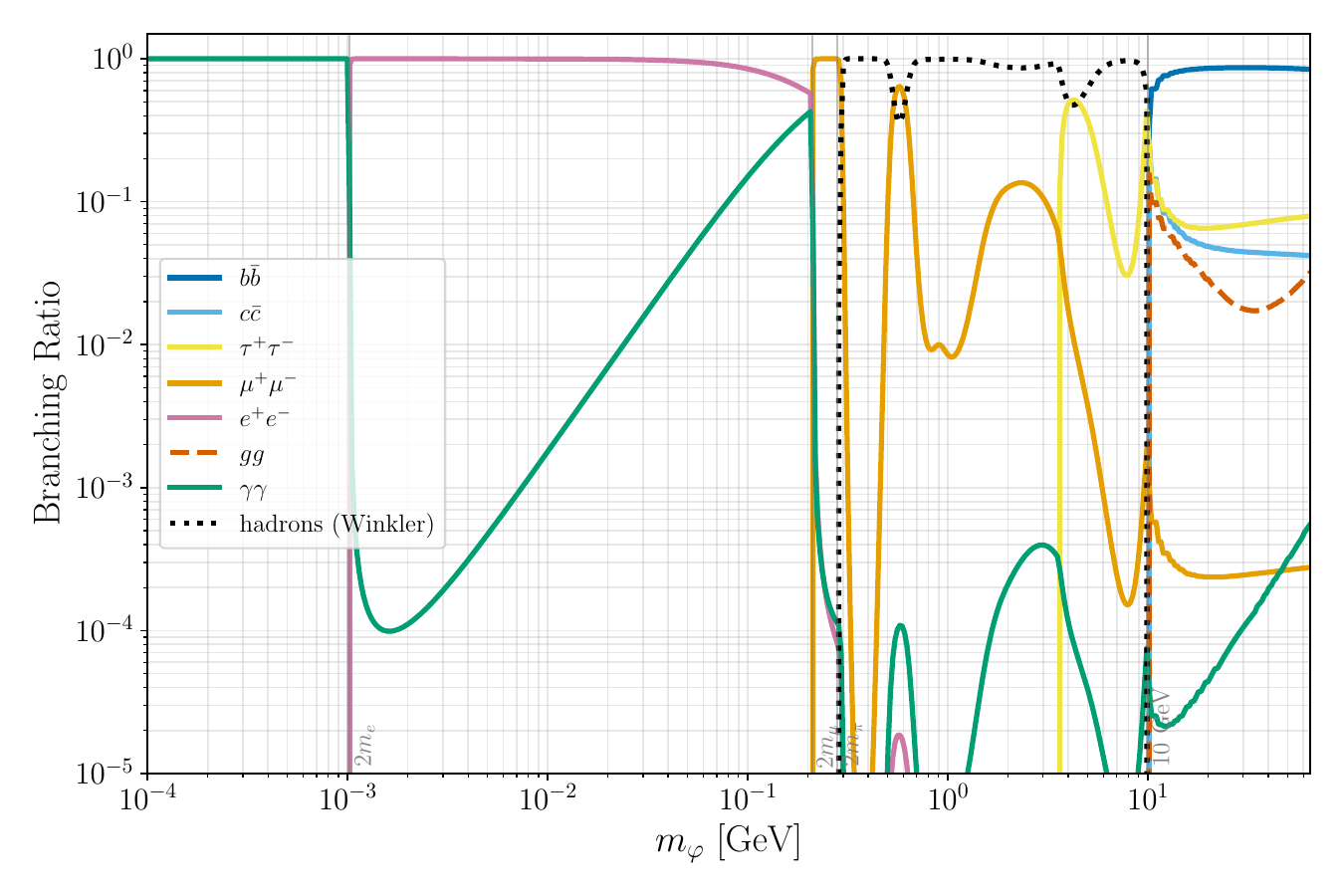}
    \caption{Branching ratio of inflaton decay through the mixing with the SM Higgs boson.}
    \label{fig:branching_ratio}
\end{figure}

The $\mu$-coupling induces a mixing between $\varphi$ and the SM Higgs boson, allowing $\varphi$ to decay into SM particles.
The decay width of $\varphi$ can be computed by using the Higgs decay width as
\begin{align}
    \Gamma_\varphi
    & \simeq
    \theta^2\times \Gamma_h(s = m_\varphi^2).
\end{align}
This approximation is good enough when $m_\varphi \ll m_h$.
Figure~\ref{fig:branching_ratio} shows the branching ratio of the decay of $\varphi$.
The precise estimate of $\Gamma_h$ is obtained by {\tt HDECAY}~\cite{Djouadi:2018xqq}, supplemented by the decay width of $h\to f\overline f$ and $h\to\gamma\gamma$ for the lower mass region, e.g., $m_\varphi < 10$ GeV~\cite{Djouadi:2005gi}.
The partial decay widths of the hadronic channel are taken from Ref.~\cite{Winkler:2018qyg}.

In the lower mass region, we used the well-known result\footnote{See, e.g., Ref.~\cite{Djouadi:2005gi} and references therein.}:
\begin{align}
    \Gamma_{h\to\bar f f}(s=m_\varphi^2) &=
    \frac{N_C m_\varphi m_f^2}{8\pi v^2}\beta_f^3,\\[8pt]
    \Gamma_{h\to\gamma\gamma}(s=m_\varphi^2) &=
    \frac{\alpha_{\rm em}^2m_\varphi^3}{256\pi^3v^2}
    \left|
        \sum_f N_C Q_f^2 A_{1/2}(\tau_f)+A_1(\tau_W)
    \right|^2,
\end{align}
where $N_C$ is the number of colors (3 for quarks, 1 for leptons), $Q_f$ is the electric charge of fermion~$f$, and $A_1(\tau_W)$ and $A_{1/2}(\tau_f)$ are loop functions given by
\begin{align}
   A_{1/2}(\tau_i) &=
   2\left[
      \tau_i + (\tau_i-1)\arcsin^2\sqrt{\tau_i}
   \right]\tau_i^{-2},\\[8pt]
   A_1(\tau_i) &=
   -\left[
      2\tau_i^2 + 3\tau_i (2\tau_i-1)\arcsin^2\sqrt{\tau_i}
   \right]\tau_i^{-2},
\end{align}
for $\tau_i\le 1$, with $\tau_i = m_\varphi^2/4m_i^2$.
For $m_\varphi\ll m_W$, we have
\begin{align}
    A_1(\tau_W) &\to 7,\\[8pt]
    N_C Q_t^2 A_{1/2}(\tau_t) &\to  -16/9
\end{align}
to draw Fig.~\ref{fig:branching_ratio} for the parameter space of $m_\varphi < 2m_\mu$, i.e., in an approximation incorporating only $W$ and $t$ contributions as they are dominant in the loop.

\subsection{Cosmological and Experimental Constraints}
\label{subsec:constraints}

The viable parameter space is carved out by several observational bounds, summarized in Fig.~\ref{fig:main_result_higgs_portal}:
\begin{itemize}
   \item \textbf{BBN Constraints:}
   \begin{itemize}
      \item \textit{Energy Injection:} Inflaton decays injecting energy between the lifetime $\tau_\varphi \sim 0.1$~s (neutrino decoupling) and $10^{13}$~s can disrupt the synthesis of light elements. The constraints on $m_\varphi Y$ as a function of $\tau_\varphi$~\cite{Kawasaki:2004qu,Kawasaki:2017bqm} excludes a significant band of the parameter space where the lifetime falls within this range.
      Note that the exclusion region in red assumes that $Y_{\rm ini}=0$ and that the decay of $\varphi$ is hadronic, so the constraint may become weaker in the lower mass region (decay being either leptonic or radiative).
      \item \textit{Fragmentation effect on $T_{\rm reh}$}: As discussed in Section~\ref{sec:production}, the requirement that the reheating temperature allows for successful BBN ($T^{(\mu,\text{frag.})}_{\rm reh}, T^{(\sigma,\text{frag.})}_{\rm reh} > 10$ MeV for reheating achieved by only $\mu$ or $\sigma$ coupling, respectively) places lower bounds on the couplings. The values of $\sigma$ and $\mu$ are taken from Ref.~\cite{Garcia:2023dyf}.
      The region inconsistent with the BBN corresponds to the gray overlapped region in both $\sigma$ and $\mu$ directions, which is the left-bottom corner of the figure.
   \end{itemize}
   \item \textbf{CMB Constraints:}
   \begin{itemize}
      \item \textit{Spectral Distortions:} Decays occurring between $10^4$ s and $10^6$ s result in $\mu$- and $y$-type distortions of the CMB blackbody spectrum, as the injected energy cannot be fully thermalized~\cite{Hu:1992dc,Chluba:2011hw}. These constraints are, however, overlapped with the BBN bound from the energy injection, and thus we suppressed them in the figure.
      \item \textit{Anisotropy:} 
      Late-time decays (after recombination, $\tau_\varphi \gtrsim 10^{13}$ s) deposit energy into the intergalactic medium (IGM), altering the ionization history and heating the IGM which affect CMB anisotropies.      
      Lifetimes up to $\tau_\varphi \sim 10^{25}$ s are excluded by Planck data~\cite{Slatyer:2016qyl,Cang:2020exa}.
      The constraint is given to the energy injection rate, which is proportional to $\Omega_{\rm DM}/\tau_{\rm DM}$ where the decaying particle is assumed to be 100\% DM.
      Using the result of Ref.~\cite{Cang:2020exa}, we place bounds on the DM lifetime.
      Denoting $\tau_{\rm DM}>\tau_{\rm lim}$ and considering the energy injection rate by $\varphi$ being proportional to $\Omega_\varphi/\tau_\varphi$, we find $\Omega_\varphi/\tau_\varphi = \Omega_{\rm DM}/\tau_{\rm DM} < \Omega_{\rm DM}/\tau_{\rm lim}\;\Rightarrow\;\Omega_{\rm DM}h^2\times(\tau_\varphi/\tau_{\rm lim}) > \Omega_\varphi h^2$. The corresponding constraint shown in the figure sets $\Omega_{\rm DM}h^2=0.12$.
   \end{itemize}
   \item \textbf{Collider and Direct Search Constraints:}
   \begin{itemize}
      \item \textit{Invisible Higgs Decay:} If $m_\varphi < m_h/2$, the Higgs boson can decay into a pair of inflatons ($h \to \varphi\varphi$). Current LHC limits constrain the branching ratio of invisible Higgs decays, ${\rm Br}(h \to \text{inv}) < 0.11$ at 95\% CL \cite{Arcadi:2019lka,ATLAS:2023tkt}. This translates into an upper bound on the coupling $\sigma$. In particular, the parameter space where $\varphi$ explain whole DM in the WIMP regime is in conflict with this limit.
      \item \textit{Dark Higgs Search:} For light inflatons ($50~{\rm MeV} < m_\varphi \lesssim 50$ GeV), various meson decay channels at fixed target and collider experiments place strong constraints on the scalar mass and its mixing with the SM Higgs boson~\cite{Ferber:2023iso}.
      \item \textit{Direct Detection:} If the inflaton constitutes a significant fraction of dark matter, direct detection experiments (e.g., XENONnT, LUX, PandaX) search for its elastic scattering off nuclei. The interaction strength is mediated by the Higgs portal coupling. Particularly, the WIMP region has already been excluded~\cite{EscuderoAbenza:2025cfj}.
   \end{itemize}
\end{itemize}

Finally, the vertical dashed lines in blue represent the constraints from the fragmentation of the inflaton condensate, which are taken from Ref.~\cite{Garcia:2023dyf}.
For $k=4$, when $\sigma$ is smaller than $2.0\times 10^{-6}$, inflaton fragments away before the completion of reheating, suggesting that there is a relic inflaton quanta through fragmentation resulting in nonzero $Y_{\rm ini}$.
For $k=6$ and $k=8$, such thresholds are $\sigma=1.5\times 10^{-10}$ and $1.7\times10^{-15}$.
The corresponding thresholds for $\mu$ ($k=4,6,8$) is greater than the mass scales shown in Fig~\ref{fig:main_result_higgs_portal}.
Therefore, since the viable freeze-in region with $Y_{\rm ini}=0$ lies at $\mu \lesssim 10^{-17}~{\rm GeV}$ and $\sigma \simeq 10^{-10}$, the $k=6$ threshold, $\sigma=1.5\times10^{-10}$, is within a factor of two of the viable region\footnote{
In this regime the freeze-in yield is dominated by $h\to\varphi\varphi$ decays at $T\sim m_h$, that is, just below the electroweak phase transition. A complete treatment would follow the temperature dependence of the Higgs VEV through the crossover and would add the contact process $HH^{\dagger}\to\varphi\varphi$ in the symmetric phase. Such a possible refinement is, however, beyond the scope of the paper.}, so this case sits close to the boundary between fragmentation occurring before and after the completion of reheating.
This proximity, however, does not affect our conclusions. 
Even if the condensate does fragment for $k=6$, the surviving population quoted above, $Y^{\rm frag}\simeq1.2\times10^{-18}$, remains negligible, and the freeze-in prediction is therefore unchanged.
For the corresponding value of $\sigma$, the reheating temperature is $T_{\rm reh}\simeq2.1\times10^{8}$~GeV (dominated by the $\sigma$-channel), which is well above $m_h$. Since the freeze-in abundance is evaluated deep in the IR tail of the thermal history, at $T\sim m_h$ or $m_\chi$, well after reheating completes, it is insensitive to (a few orders of magnitude) corrections to $T_{\rm reh}$ that could arise from non-perturbative effects not captured in Eqs.~\eqref{eq:T_mu} and \eqref{eq:T_sigma}.

\section{Discussion and Conclusion}
\label{sec:conclusion}

We have demonstrated that, for inflaton potentials $V(\phi) \propto \phi^k$ where $k \ge 4$, the effective inflaton mass decreases as the Universe expands and this ensures that even if the inflaton condensate is depleted during reheating, and inflaton quanta can be regenerated from the thermal bath once the temperature exceeds the effective mass.

Our analysis shows that this regeneration proceeds through two primary regimes:
\begin{itemize}
   \item WIMP Regime: Large couplings ($\sigma \simeq 10^{-1}$ to $10^{-2}$) where the inflaton particle reaches thermal equilibrium.
   \item FIMP Regime: Small couplings where the abundance is produced via freeze-in, dominated by a generic scalar (or Higgs) decays.
\end{itemize}

First we have investigated the case for a generic scalar and shown that the size of the coupling between $\phi$ and $\chi$ can be constrained via the inflaton regeneration, which serves as a unique probe of reheating scenarios. Specifically, requiring the regenerated inflaton particles to constitute the observed DM abundance divides the parameter space into two viable regimes: the freeze-out and freeze-in scenarios (WIMP and FIMP regions, respectively). We have demonstrated that the coupling values lying between these two regimes are excluded due to the overproduction of inflaton quanta. Furthermore, the requirement of a successful BBN (the reheating temperature higher than 10 MeV), combined with the fragmentation effects, imposes lower bounds on the portal couplings.

In particular, when identifying the mediator $\chi$ as the SM Higgs, the parameter space is significantly constrained: 
\begin{itemize}
   \item Collider Constraints: The WIMP region is largely excluded by current LHC limits on the invisible Higgs branching ratio, ${\rm Br}(h \to {\rm inv}) < 0.11$.
   \item Cosmological Bounds: Lifetimes that result in energy injection between $0.1$~s and $10^{13}$~s are disallowed by BBN. Late-time decays are further restricted by CMB anisotropy limit.
   \item The Survival Corridor: A narrow corridor remains in the freeze-in regime, particularly when the initial abundance $Y_{\rm ini}$ is small or the lifetime of $\varphi$ is short enough, indicating a larger $\mu$ that results in high reheating temperature.
\end{itemize}  

A critical point of discussion is the required smallness of the portal couplings $\mu$ and $\sigma$ to satisfy the FIMP dark matter abundance. 
Furthermore, the impact of initial conditions, such as the fragmentation of the inflaton condensate, remains a crucial factor in determining the final relic density.
While we argued that field-dependent decay suppresses $Y_{\rm ini}$ for large effective masses, suppression efficiency may vary with $k$.
As estimated in Sec.~\ref{subsec:relic}, the fragmentation remnant is negligible for $k\ge6$, whereas for $k=4$ the coupling required by the relic abundance drives $T_{\rm reh}$ down to values comparable to $m_\chi$, where the post-reheating freeze-in description, which assumes $T_{\rm reh}\gg m_\chi$, may no longer apply.
In this regard, $Y_{\rm ini}$ is not a free parameter as far as fragmentation is concerned.
What remains open is the initial population left by the non-perturbative processes of Sec.~\ref{subsec:non-perturbative} and by ultraviolet-dominated production.
Determining these contributions lies beyond the scope of this paper and is left for future work.

Although this study restricts the portal coupling to the renormalizable level, any higher dimensional terms could be incorporated.
For instance, if $\phi |H|^4/\Lambda$ term presents, then $\varphi$ may also be produced at high temperatures, contributing $Y_{\rm ini}$ in the freeze-in regime as such a contribution is ultraviolet dominated.

We also note that our analysis does not involve the thermal mass in $\varphi$, e.g., $m_\varphi\sim \sqrt{\sigma} T$, which could potentially affect the production rate at high temperatures.
On the other hand, in the FIMP regime, the production of $\varphi$ is dominated by the decay of $\chi$ (or $h$), and thus the thermal mass of $\varphi$ is not relevant for the production rate as $m_{\chi,h}\gg m_\varphi$.

A possible distinctive feature of the framework we propose is the stochastic gravitational wave background (SGWB), which might be detected in the near future.
For the monomial potentials considered ($k \ge 4$), the Universe undergoes an era of stiff expansion where the equation of state $w_{\phi} = (k-2)/(k+2)$ exceeds $1/3$. 
This ``kination-like" phase is known to induce blue-tilted GW power spectrum, amplifying high-frequency modes~\cite{Giovannini:1998bp,Peebles:1998qn}. 
Furthermore, the fragmentation of the inflaton condensate provides a secondary, non-perturbative source of GWs. 
These fragmentation-induced signals typically manifest as sharply peaked features in the GHz to PHz range.

Such signatures offer a ``fingerprint" of the reheating dynamics that is complementary to relic abundance constraints. 
While the survival ``corridors" are primarily shaped by portal couplings $\sigma$ and $\mu$, the GW spectrum provides a direct window into the value of $k$ and the fragmentation history, potentially detectable by future experiments such as DECIGO or ultra-high frequency GW detectors.

Our results highlight that the post-reheating history of the inflaton is more complex than typically assumed. 
Importantly, although the reheating process is usually considered to be challenging to test, through the inflaton regeneration mechanism, we can constrain or probe it.
When we specify the model such as the Higgs portal scenario, future observations, specifically improved measurements of the invisible Higgs width and upcoming CMB experiments sensitive to spectral distortions and anisotropy, will be crucial in probing the remaining open parameter space. 
This work underscores the necessity of considering the full dynamical history of the inflaton to understand its role in the late-time Universe.

\section*{Acknowledgements}
This work was partially supported by JSPS KAKENHI Grant Numbers 25K01004 (TT),  and MEXT KAKENHI 23H04515 (TT), 25H01543 (TT). 

\pagebreak

\appendix
\noindent {\LARGE \bf Appendix} 

\section{Thermally averaged cross sections}
\label{appendix:thermally_averaged_cross_sections}

The thermally averaged cross section $\langle \sigma v \rangle$\footnote{The cross section $\sigma$ and relative velocity $v$ are not to be confused with the $\sigma$-coupling and the VEV $v$ in the text.} is a crucial quantity in cosmology and particle physics, particularly in the study of dark matter annihilation and early Universe processes. It represents the average of the product of the cross section $\sigma$ and the relative velocity $v$ of interacting particles, weighted by their thermal distribution.
In this appendix, we review the definition and calculation of $\langle \sigma v \rangle$.

\subsection{2-to-2 processes}
\label{subsec:2_to_2_processes}

As a typical example, we consider the two-to-two process where two particles 1 and 2 annihilate into two particles 3 and 4.
The collision integral for this process is given by
\begin{align}
C[f_1]
&= 
\frac{1}{2E_1} \int d\Pi_2 d\Pi_3 d\Pi_4 (2\pi)^4 \delta^4(p_1 + p_2 - p_3 - p_4)\nonumber\\
&\times 
\left[ |\mathcal{M}_{34, 12}|^2 f_3 f_4 (1 \pm f_1)(1 \pm f_2) - |\mathcal{M}_{12, 34}|^2 f_1 f_2 (1 \pm f_3)(1 \pm f_4) \right],
\end{align}
where $d\Pi_i = \displaystyle\frac{d^3 p_i}{(2 \pi)^3 2E_i}$ is the Lorentz-invariant phase space measure for particle $i$, and $|\mathcal{M}_{ij,kl}|^2$ is the squared matrix element for the process of $ij\to kl$.
The partial reaction rate $\gamma_{12,34}$ for the annihilation process $1 + 2 \to 3 + 4$ can be expressed as
\begin{align}
\gamma_{12,34} 
&= 
\int d\Pi_1 d\Pi_2 d\Pi_3 d\Pi_4 (2\pi)^4 \delta^4(p_1 + p_2 - p_3 - p_4) 
~|\mathcal{M}_{12,34}|^2 f_1 f_2 (1 \pm f_3)(1 \pm f_4).
\end{align}
In the following, we work under the dilute gas approximation, i.e., $1 \pm f_i \simeq 1$.
Notice that
\begin{align}
    d{\rm LIPS}_{34}
    &\equiv
    d\Pi_3 d\Pi_4 (2\pi)^4 \delta^4(p_1 + p_2 - p_3 - p_4)
\end{align}
is the Lorentz-invariant phase space for the final state particles 3 and 4.
Since $|\mathcal{M}_{12,34}|^2$ is Lorentz invariant, we may introduce
\begin{align}
    w(s)
    &\equiv
    \int d{\rm LIPS}_{34} |\mathcal{M}_{12,34}|^2
\end{align}
as a Lorentz-invariant function of the Mandelstam variable $s = (p_1 + p_2)^2$.
In the plasma rest frame, we have
\begin{align}
    s &= m_1^2 + m_2^2 + 2E_1 E_2 - 2|\vec{p}_1||\vec{p}_2|\cos\theta_{12},
\end{align}
where $\theta_{12}$ is the angle between the momenta $\vec{p}_1$ and $\vec{p}_2$, and thus $w(s)$ depends on $E_1$, $E_2$, and $\cos\theta_{12}$.

The phase space integration for the initial state particles 1 and 2 can be expressed as
\begin{align}
    \int d\Pi_1 d\Pi_2
    &=
    \int \frac{d^3 p_1}{(2\pi)^3 2E_1} \frac{d^3 p_2}{(2\pi)^3 2E_2} \nonumber\\[8pt]
    &=
    \int \frac{4\pi |\vec{p}_1|^2 d|\vec{p}_1|}{(2\pi)^3 2E_1}
    \frac{4\pi |\vec{p}_2|^2 d|\vec{p}_2|}{(2\pi)^3 2E_2}
    \int_{-1}^{1} \frac{d\cos\theta_{12}}{2}.
\end{align}
Using $|\vec p_i| d|\vec{p}_i| = E_i dE_i$, we can rewrite this as
\begin{align}
    \int d\Pi_1 d\Pi_2
    &=
    \frac{1}{32\pi^4} \int_{m_1}^\infty dE_1 \int_{m_2}^\infty dE_2 |\vec{p}_1| |\vec{p}_2| \int_{-1}^{1} d\cos\theta_{12}.
\end{align}
Thus, the partial reaction rate can be expressed as
\begin{align}
    \gamma_{12,34}
    &=
    \frac{1}{32\pi^4} \int_{m_1}^\infty dE_1 \int_{m_2}^\infty dE_2 |\vec{p}_1| |\vec{p}_2| f_1 f_2 \int_{-1}^{1} d\cos\theta_{12} w(s).
\end{align}

The final state phase space in $w(s)$ should also be integrated at the plasma rest frame, which can be done in the following way:
\begin{align}
    \int d{\rm LIPS}_{34}
    &=
    \int \frac{d^3 p_3}{(2\pi)^3 2E_3} \frac{d^3 p_4}{(2\pi)^3 2E_4} (2\pi)^4 \delta^4(p_q  + p_{\bar{q}} - p_3 - p_4) \nonumber\\[8pt]
    &=
    \int \frac{d^3 p_3}{(2\pi)^3 2E_3} \frac{d^4 p_4}{(2\pi)^3} \delta(p_4^2 - m_4^2) \theta(E_4) (2\pi)^4 \delta^4(p_1 + p_2 - p_3 - p_4) \nonumber\\[8pt]
    &=
    \int \frac{d^3 p_3}{(2\pi)^3 2E_3} (2\pi) \delta((p_1 + p_2 - p_3)^2 - m_4^2) \theta(E_1 + E_2 - E_3).
\end{align}
Here, the argument of the delta function can be rewritten as
\begin{align}
    (p_1 + p_2 - p_3)^2 - m_4^2
    &=
    (p_1+p_2)^2+m_3^2-m_4^2-2(E_1+E_2)E_3+2|\vec p_1+\vec p_2||\vec p_3|\cos\theta\nonumber\\[8pt]
    &=
    2|\vec p_1+\vec p_2||\vec p_3|\left(
        \cos\theta - \frac{2(E_1+E_2)E_3-s-m_3^2+m_4^2}{2|\vec p_1+\vec p_2||\vec p_3|}
    \right),
\end{align}
and hence the delta function can be expressed as
\begin{align}
    \delta((p_1 + p_2 - p_3)^2 - m_4^2)
    &=
    \frac{1}{2|\vec p_1+\vec p_2||\vec p_3|} \delta\left(
        \cos\theta - \frac{2(E_1+E_2)E_3-s-m_3^2+m_4^2}{2|\vec p_1+\vec p_2||\vec p_3|}
    \right).
\end{align}
Using this, we can perform the angular integration over $\cos\theta$ as
\begin{align}
    \int d{\rm LIPS}_{34}
    &=
    \int \frac{2\pi |\vec p_3|^2 d|\vec p_3|}{(2\pi)^3 2E_3} (2\pi) \frac{1}{2|\vec p_1+\vec p_2||\vec p_3|} \nonumber\\[8pt]
    &\times
    \theta(E_1 + E_2 - E_3)
    \theta\left(1 - \left| \frac{2(E_1+E_2)E_3 - s - m_3^2 + m_4^2}{2|\vec p_1+\vec p_2||\vec p_3|} \right| \right).
\end{align}
The second theta function here imposes the kinematic constraint on $E_3$ as
\begin{align}
    E_3^{\pm}
    &=
    \frac{1}{2s} \left[
        (E_1 + E_2)(s + m_3^2 - m_4^2) \pm |\vec p_1 + \vec p_2| \sqrt{\lambda(s, m_3^2, m_4^2)}
    \right],
\end{align}
where $\lambda(a,b,c) = a^2 + b^2 + c^2 - 2(ab + bc + ca)$ is the Källén function.
It is often useful to write as
\begin{align}
    \sqrt{\lambda(s, m_3^2, m_4^2)}
    &=
    s\beta\left(\frac{m_3}{\sqrt{s}},\frac{m_4}{\sqrt{s}}\right),\\[8pt]
    \beta\left(\frac{m_3}{\sqrt{s}},\frac{m_4}{\sqrt{s}}\right)
    &\equiv
    \sqrt{\left(
        1-\frac{(m_3+m_4)^2}{s}
    \right)\left(
        1-\frac{(m_3-m_4)}{s}
    \right)},
\end{align}
with which we can rewrite $E_3^\pm $ as
\begin{align}
    E_3^\pm = \frac{E_1+E_2}{2}\left(
        1+\frac{m_3^2-m_4^2}{s}
    \right)
    \pm
    \frac{|\vec p_1+\vec p_2|}{2}\beta\left(\frac{m_3}{\sqrt{s}},\frac{m_4}{\sqrt{s}}\right).
\end{align}
Thus, the final state phase space integration can be written as
\begin{align}
    \int d{\rm LIPS}_{34}
    &=
    \frac{1}{8\pi |\vec p_1 + \vec p_2|} \int_{E_3^-}^{E_3^+} dE_3.
    \label{eq:LIPS_34_integration}
\end{align}
Using this result, we can express $w(s)$ as
\begin{align}
    w(s)
    &=
    \frac{1}{8\pi |\vec p_1 + \vec p_2|} \int_{E_3^-}^{E_3^+} dE_3 |\mathcal{M}_{12,34}|^2.
\end{align}
Substituting this into the expression for $\gamma_{12,34}$, we obtain
\begin{align}
    \gamma_{12,34}
    &=
    \frac{1}{256\pi^5} \int_{m_1}^\infty dE_1 \int_{m_2}^\infty dE_2 \int_{-1}^{1}d\cos\theta_{12}\frac{|\vec p_1| |\vec p_2|}{|\vec p_1 + \vec p_2|} f_1 f_2 \int_{E_3^-}^{E_3^+} dE_3 |\mathcal{M}_{12,34}|^2.
\end{align}
The thermally averaged cross section $\langle \sigma v \rangle$ is defined through
\begin{align}
    \gamma_{12,34}
    &=
    n_1 n_2 \langle \sigma_{12,34} v \rangle,
\end{align}
where $n_i$ is the number density of particle $i$.
From this, we can express $\langle \sigma_{12,34} v \rangle$ as
\begin{align}
    \langle \sigma_{12,34} v \rangle
    &=
    \frac{1}{n_1 n_2} \frac{1}{256\pi^5} \int_{m_1}^\infty dE_1 \int_{m_2}^\infty dE_2 \int_{-1}^{1}d\cos\theta_{12} \frac{|\vec p_1| |\vec p_2|}{|\vec p_1 + \vec p_2|} f_1 f_2 \int_{E_3^-}^{E_3^+} dE_3 |\mathcal{M}_{12,34}|^2.
\end{align}

Specifically, if ${\cal M}_{12,34}$ is a constant, we find
\begin{align}
    \langle \sigma_{12,34} v \rangle
    &=
    \frac{|\mathcal{M}_{12,34}|^2}{n_1 n_2} \frac{1}{256\pi^5} \int_{m_1}^\infty dE_1 \int_{m_2}^\infty dE_2 \int_{-1}^1d\cos\theta_{12} \frac{|\vec p_1| |\vec p_2|}{|\vec p_1 + \vec p_2|} f_1 f_2 (E_3^+ - E_3^-)\nonumber\\[8pt]
    &=
    \frac{|\mathcal{M}_{12,34}|^2}{n_1 n_2} \frac{1}{256\pi^5} \int_{m_1}^\infty dE_1 \int_{m_2}^\infty dE_2 \int_{-1}^1d\cos\theta_{12} |\vec p_1| |\vec p_2|\, f_1 f_2 \,\beta\left(\frac{m_3}{\sqrt{s}},\frac{m_4}{\sqrt{s}}\right).
\end{align}
If we further assume that $m_3, m_4 \ll \sqrt{s}$, we have $\beta(m_3/\sqrt{s}, m_4/\sqrt{s}) \simeq 1$, and thus
\begin{align}
    \langle \sigma_{12,34} v \rangle
    &=
    \frac{|\mathcal{M}_{12,34}|^2}{n_1 n_2} \frac{1}{128\pi^5} \int_{m_1}^\infty dE_1 \int_{m_2}^\infty dE_2 |\vec p_1| |\vec p_2| f_1 f_2.
\end{align}
This expression can be further simplified depending on the specific forms of the distribution functions $f_1$ and $f_2$.
When both particles follow the Boltzmann distribution, i.e., $f_i = e^{-E_i/T}$, we find
\begin{align}
    \langle \sigma_{12,34} v \rangle
    &=
    \frac{|\mathcal{M}_{12,34}|^2}{n_1 n_2} \frac{1}{128\pi^5} \int_{m_1}^\infty dE_1 |\vec p_1| e^{-E_1/T} \int_{m_2}^\infty dE_2 |\vec p_2| e^{-E_2/T}\nonumber\\[8pt]
    &=
    \frac{|\mathcal{M}_{12,34}|^2}{n_1 n_2} \frac{1}{128\pi^5} m_1 T K_1\left(\frac{m_1}{T}\right) m_2 T K_1\left(\frac{m_2}{T}\right),
\end{align}
where $K_1$ is the modified Bessel function of the second kind.

\subsection{1-to-2 processes}
\label{subsec:1_to_2_processes}

Consider the decay process where a particle 1 decays into two particles 2 and 3.
The collision integral for this process is given by
\begin{align}
C[f_1]
&= 
\frac{1}{2E_1} \int d\Pi_2 d\Pi_3 (2\pi)^4 \delta^4(p_1 - p_2 - p_3)\nonumber\\[8pt]
& \qquad\qquad \times 
\left[ |\mathcal{M}_{23, 1}|^2 f_2 f_3 (1 \pm f_1) - |\mathcal{M}_{1, 23}|^2 f_1 (1 \pm f_2)(1 \pm f_3) \right],
\end{align}
where
$|\mathcal{M}_{ij,k}|^2$ is the squared matrix element for the process of $ij\to k$.
The partial decay rate $\gamma_{1,23}$ for the decay process $1 \to 2 + 3$ can be expressed as
\begin{align}
\gamma_{1,23} 
&= 
\int d\Pi_1 d\Pi_2 d\Pi_3 (2\pi)^4 \delta^4(p_1 - p_2 - p_3) 
~|\mathcal{M}_{1,23}|^2 f_1 (1 \pm f_2)(1 \pm f_3).
\end{align}
As mentioned above, we work under the dilute gas approximation, i.e., $1 \pm f_i \simeq 1$, and
notice that
\begin{align}
    d{\rm LIPS}_{23}
    &\equiv
    d\Pi_2 d\Pi_3 (2\pi)^4 \delta^4(p_1 - p_2 - p_3)
\end{align}
is the Lorentz-invariant phase space for the final state particles 2 and 3.
Since $|\mathcal{M}_{1,23}|^2$ is Lorentz invariant, we may introduce
\begin{align}
    w_{1,23}(s)
    &\equiv
    \int d{\rm LIPS}_{23} |\mathcal{M}_{1,23}|^2
\end{align}
as a Lorentz-invariant function of the Mandelstam variable $s = p_1^2 = (p_2 + p_3)^2$, by which we may write
\begin{align}
    \gamma_{1,23}
    &=
    \int d\Pi_1 f_1 w_{1,23}(s)
    =
    \int \frac{|\vec p_1|dE_1}{4\pi^2} f_1 w_{1,23}(s).
\end{align}

The previously calculated two-body phase space integration \eqref{eq:LIPS_34_integration} can be directly applied here, yielding
\begin{align}
    \int d{\rm LIPS}_{23}
    &=
    \frac{1}{8\pi |\vec p_1|} \int_{E_2^-}^{E_2^+} dE_2,
\end{align}
where
\begin{align}
    E_2^{\pm}
    &=
    \frac{E_1}{2}\left(
        1+\frac{m_2^2 - m_3^2}{m_1^2}
    \right)
    \pm
    \frac{|\vec p_1|}{2}\beta\left(\frac{m_2}{m_1},\frac{m_3}{m_1}\right).
\end{align}
Thus, the partial decay rate can be expressed as
\begin{align}
    \gamma_{1,23}
    &=
    \int \frac{|\vec p_1|dE_1}{4\pi^2} f_1 \frac{1}{8\pi |\vec p_1|} \int_{E_2^-}^{E_2^+} dE_2 |\mathcal{M}_{1,23}|^2\nonumber\\[8pt]
    &=
    \frac{1}{32\pi^3}\int_{m_1}^\infty dE_1 f_1 \int_{E_2^-}^{E_2^+} dE_2 |\mathcal{M}_{1,23}|^2.
\end{align}

Suppose again that $|\mathcal{M}_{1,23}|^2$ is a constant.
Then, we find
\begin{align}
    \gamma_{1,23}
    &=
    \frac{|\mathcal{M}_{1,23}|^2}{32\pi^3}\int_{m_1}^\infty dE_1 f_1 (E_2^+ - E_2^-)\nonumber\\[8pt]
    &=
    \frac{|\mathcal{M}_{1,23}|^2}{32\pi^3}\int_{m_1}^\infty dE_1 f_1 |\vec p_1| \beta\left(\frac{m_2}{m_1},\frac{m_3}{m_1}\right).
\end{align}
Recall that the decay width $\Gamma_{1\to 23}$ in the rest frame of particle 1 is given by
\begin{align}
    \Gamma_{1\to 23}
    &=
    \frac{1}{2m_1} \int d{\rm LIPS}_{23} |\mathcal{M}_{1,23}|^2
    =
    \frac{|\mathcal{M}_{1,23}|^2}{16\pi m_1} \beta\left(\frac{m_2}{m_1},\frac{m_3}{m_1}\right).
\end{align}
Using this, we can rewrite the partial decay rate as
\begin{align}
    \gamma_{1,23}
    &=
    \frac{m_1\Gamma_{1\to 23}}{2\pi^2} \int_{m_1}^\infty dE_1|\vec p_1| f_1.
\end{align}
When $f_1$ is given by the Boltzmann distribution, i.e., $f_1 = e^{-E_1/T}$, we have
\begin{align}
    \gamma_{1,23}
    &=
    \frac{m_1\Gamma_{1\to 23}}{2\pi^2} \int_{m_1}^\infty dE_1|\vec p_1| e^{-E_1/T}
    =
    \frac{\Gamma_{1\to 23}}{2\pi^2} m_1^2 T K_1\left(\frac{m_1}{T}\right).
\end{align}

Next we consider the inverse decay process where two particles 2 and 3 annihilate into one particle 1.
The partial reaction rate $\gamma_{23,1}$ for the inverse decay process $2 + 3 \to 1$ can be expressed as
\begin{align}
\gamma_{23,1} 
&= 
\int d\Pi_2 d\Pi_3 d\Pi_1 (2\pi)^4 \delta^4(p_2 + p_3 - p_1) 
|\mathcal{M}_{23,1}|^2 f_2 f_3 (1 \pm f_1).
\end{align}
In the dilute gas approximation, this becomes
\begin{align}
    \gamma_{23,1}
    &=
    \int d\Pi_2 d\Pi_3 f_2 f_3 w_{23,1}(s),
\end{align}
where $w_{23,1}(s)$ is the same as defined before, with $s = (p_2 + p_3)^2$, leading to
\begin{align}
    w_{23,1}(s)
    &=
    \int d{\rm LIPS}_{1} |\mathcal{M}_{23,1}|^2.
\end{align}
Here, the one-body phase space integration is straightforward:
\begin{align}
    \int d\Pi_1 (2\pi)^4 \delta^4(p_2 + p_3 - p_1)
    &=
    \int \frac{d^4p_1}{(2\pi)^3}\delta(p_1^2 - m_1^2) \theta(E_1) (2\pi)^4 \delta^4(p_2 + p_3 - p_1)\nonumber\\[8pt]
    &=
    2\pi \delta(s - m_1^2) \theta(E_2 + E_3),
\end{align}
and thus we have
\begin{align}
    w_{23,1}(s)
    &=
    2\pi \delta(s - m_1^2) |\mathcal{M}_{23,1}|^2.
\end{align}
Using
\begin{align}
    s &= m_2^2 + m_3^2 + 2E_2 E_3 - 2|\vec p_2||\vec p_3|\cos\theta_{23} 
    \,,
\end{align}
the delta function can be recast into the following form:
\begin{align}
 \delta(s - m_1^2)
    =
    \frac{1}{2|\vec p_2||\vec p_3|} \delta\left(
        \cos\theta_{23} - \frac{2E_2 E_3 - m_1^2 + m_2^2 + m_3^2}{2|\vec p_2||\vec p_3|}
    \right) \,.   
\end{align}
Then we can perform the angular integration over $\cos\theta_{23}$ in $\displaystyle\int d\Pi_2 d\Pi_3$.
The condition for the argument of the delta function to be in the range $[-1, 1]$ leads to the kinematic constraint on~$E_3$.
In principle, for a given range of $E_2$ and $E_3$, we can compute $\gamma_{23,1}$.
In practice, however, it is convenient to redefine the integration variables to $E_+ = E_2 + E_3$ and $E_- = E_2 - E_3$ to simplify the integration.
The condition $| \cos\theta_{23} | \leq 1$ then translates into constraints on $E_+$ and $E_-$, which can be used to determine the limits of integration:
\begin{align}
    E_-^{(\pm)}
    &=
    \frac{m_2^2-m_3^2}{m_1^2}E_+ \pm \sqrt{E_+^2 - m_1^2} \cdot \beta\left(\frac{m_2}{m_1},\frac{m_3}{m_1}\right).
\end{align}
Thus, we can express the partial reaction rate for the inverse decay as
\begin{align}
    \gamma_{23,1}
    &=
    \int \frac{|\vec p_2|E_2dE_2}{2\pi^2\cdot 2E_2}\frac{|\vec p_3|E_3dE_3}{2\pi^2\cdot 2E_3}\frac{d\cos\theta_{23}}{2}\cdot
    2\pi\delta(s-m_1^2)|\mathcal{M}_{23,1}|^2 f_2 f_3\nonumber\\[8pt]
    &=
    \frac{1}{64\pi^3}\int_{m_2 + m_3}^\infty dE_+ \int_{E_-^{(-)}}^{E_-^{(+)}} dE_- f_2 f_3 |\mathcal{M}_{23,1}|^2.
\end{align}
From unitarity, we have $|\mathcal{M}_{23,1}|^2 = |\mathcal{M}_{1,23}|^2$.
Thus, using the previously defined decay width $\Gamma_{1\to 23}$, we can rewrite this as
\begin{align}
    \gamma_{23,1}
    &=
    \frac{m_1 \Gamma_{1\to 23}}{4\pi^2\beta(m_2/m_1,m_3/m_1)}\int_{\text{max}(m_1,m_2 + m_3)}^\infty dE_+ \int_{E_-^{(-)}}^{E_-^{(+)}} dE_- f_2 f_3.
\end{align}
When we further assume that both particles 2 and 3 follow the Boltzmann distribution, i.e., $f_i = e^{-E_i/T}$, we find
\begin{align}
    \gamma_{23,1}
    &=
    \frac{m_1 \Gamma_{1\to 23}}{4\pi^2\beta(m_2/m_1,m_3/m_1)}\int_{\text{max}(m_1,m_2 + m_3)}^\infty dE_+ \int_{E_-^{(-)}}^{E_-^{(+)}} dE_- e^{-E_+/T}\nonumber\\
    &=
    \frac{\Gamma_{1\to 23}}{2\pi^2} m_1^2 T K_1\left(\frac{m_1}{T}\right),
\end{align}
where in the last line we have assumed $m_1 > m_2 + m_3$.
Notice that we obtain the relation
\begin{align}
    \gamma_{23,1}
    &=
    \gamma_{1,23},
\end{align}
which is a manifestation of the principle of detailed balance.

\section{Transformation between Laboratory and CM frames\label{appendix:lab_to_cm_and_back}}

In practice, it is sometimes convenient to expand the reaction rate in the non-relativistic limit.
To do so, a more suitable frame to compute the Lorentz-invariant function $w(s)$ is the center-of-mass (CM) frame, rather than the laboratory (lab) frame, which corresponds to the plasma rest frame.
In this section, we summarize the relations between the two frames.

Let us again consider the two-to-two process where two particles 1 and 2 annihilate into two particles 3 and 4.
We write the initial total momentum as
\begin{align}
    P^{\rm Lab}_{\rm{tot},\mu} &= p_{1,\mu}+p_{2,\mu} \equiv (E,\vec P),
\end{align}
and the same quantity in the CM frame as
\begin{align}
    P^{\rm CM}_{\rm{tot},\mu} &= p^*_{1,\mu}+p^*_{2,\mu} \equiv (\sqrt{s},\vec 0).
\end{align}
Recalling that a four-momentum of a particle of mass $m$ and velocity $\vec v$ may be written as $p_\mu = m\gamma(1,\vec v)$ with $\gamma = 1/\sqrt{1-v^2}$, we can express the total momentum in the lab frame as
\begin{align}
    P^{\rm Lab}_{\rm{tot},\mu}
    &=
    \sqrt{s}\gamma(1,\beta\vec n),
\end{align}
where $\vec n$ is a unit vector in the direction of $\vec P$, and
\begin{align}
    \sqrt{s}\gamma=E
    &\;\;\;\Rightarrow\;\;\;
    \gamma = \frac{E}{\sqrt{s}},\\
    \sqrt{s}\gamma\beta = |\vec P| 
    &\;\;\;\Rightarrow\;\;\;
    \beta = \frac{|\vec P|}{E}.
\end{align}
Without loss of generality, we may take $\vec n = (0,0,1)$.
The Lorentz transformation from the CM frame to the lab frame is then given by
\begin{align}
    \Lambda^\mu_{\;\nu}
    &=
    \begin{pmatrix}
        \gamma & 0 & 0 & \beta\gamma\\
        0 & 1 & 0 & 0\\
        0 & 0 & 1 & 0\\
        \beta\gamma & 0 & 0 & \gamma
    \end{pmatrix}.
\end{align}

Now, we compute the two-body phase space integration in the CM frame:
\begin{align}
   \int d{\rm LIPS}_{34}
    &=
    \int \frac{d^3 p^*_3}{(2\pi)^3 2E^*_3} \frac{d^3 p^*_4}{(2\pi)^3 2E^*_4} (2\pi)^4 \delta^4(p^*_1 + p^*_2 - p^*_3 - p^*_4)\nonumber\\[8pt]
    &=
    \int \frac{d^3 p^*_3}{(2\pi)^3 2E^*_3} \frac{d^4 p^*_4}{(2\pi)^3} \delta(p_4^{*2} - m_4^2) \theta(E^*_4) (2\pi)^4 \delta^4(p^*_1 + p^*_2 - p^*_3 - p^*_4)\nonumber\\[8pt]
    &=
    \int \frac{d^3 p^*_3}{(2\pi)^3 2E^*_3} (2\pi) \delta((p^*_1 + p^*_2 - p^*_3)^2 - m_4^2) \theta(\sqrt{s} - E^*_3).
\end{align}
Here, the argument of the delta function can be rewritten as
\begin{align}
    (p^*_1 + p^*_2 - p^*_3)^2 - m_4^2
    &=
    s + m_3^2 - m_4^2 - 2\sqrt{s} E^*_3\nonumber\\[8pt]
    &=
    (-2\sqrt{s})\left(
        E_3^* - \frac{s+m_3^2-m_4^2}{2\sqrt{s}}
    \right),
\end{align}
and hence the delta function is expressed as
\begin{align}
    \delta((p^*_1 + p^*_2 - p^*_3)^2 - m_4^2)
    &=
    \frac{1}{2\sqrt{s}} \delta\left(
        E_3^* - \frac{s+m_3^2-m_4^2}{2\sqrt{s}}
    \right).
\end{align}
Using this, we can perform the energy integration over $E_3^*$ as
\begin{align}
\label{eq:LIPS34}
    \int d{\rm LIPS}_{34}
    &=
    \int \frac{|\vec p_3^{\mkern6mu *}| E_3^* dE_3^* d\Omega^*}{(2\pi)^3 2E_3^*} (2\pi) \frac{1}{2\sqrt{s}} \delta\left(
        E_3^* - \frac{s+m_3^2-m_4^2}{2\sqrt{s}}
    \right)\nonumber\\
    &=
    \int\frac{1}{16\pi^2} \frac{|\vec p_3^{\mkern6mu *}|}{\sqrt{s}} d\Omega^*.
\end{align}
Substituting
\begin{align}
    |\vec p_3^{\mkern6mu *}|
    &=
    \frac{\sqrt{\lambda(s, m_3^2, m_4^2)}}{2\sqrt{s}}
    =
    \frac{\sqrt{s}}{2}\beta\left(\frac{m_3}{\sqrt{s}},\frac{m_4}{\sqrt{s}}\right)
    \label{eq:p3*}
\end{align}
into Eq.~\eqref{eq:LIPS34}, 
we obtain
\begin{align}
    \int d{\rm LIPS}_{34}
    &=
    \frac{1}{32\pi^2} \beta\left(\frac{m_3}{\sqrt{s}},\frac{m_4}{\sqrt{s}}\right) \int d\Omega^*.
\end{align}
In most cases, the integration over the azimuthal angle $\phi^*$ is trivial, and we have
\begin{align}
    \int d{\rm LIPS}_{34}
    &=
    \frac{1}{16\pi} \beta\left(\frac{m_3}{\sqrt{s}},\frac{m_4}{\sqrt{s}}\right) \int d\cos\theta^*.
\end{align}
In the following, by performing the Lorentz transformation, we will show that this result can also be obtained by the result in the lab frame \eqref{eq:LIPS_34_integration}.

A key relation is that the energy of particle 3 in the CM frame is related to the variables in the lab frame as
\begin{align}
    E_3 &= \gamma(E_3^* + \beta |\vec p_3^{\mkern6mu *}| \cos\theta^*).
    \label{eq:E3_lab_cm}
\end{align}
In the CM frame, from $\vec P_{\rm{tot}}^* = \vec 0$ and $E^*_{\rm{tot}} = \sqrt{s}$, we have
\begin{align}
    E_3^*
    &=
    \frac{s + m_3^2 - m_4^2}{2\sqrt{s}},\label{eq:E3*}\\[8pt]
\end{align}
with $|\vec p_3^{\mkern6mu *}|$ given by Eq.~\eqref{eq:p3*}.
Together with
\begin{align}
    \gamma &= \frac{E_1+E_2}{\sqrt{s}},\\[8pt]
    \gamma\beta &= \frac{|\vec p_1 + \vec p_2|}{\sqrt{s}},
\end{align}
we find
\begin{align}
    E_3
    &=
    \frac{E_1 + E_2}{2}\left(
        1 + \frac{m_3^2 - m_4^2}{s}
    \right)
    + \frac{|\vec p_1 + \vec p_2|}{2} \beta\left(
        \frac{m_3}{\sqrt{s}},\frac{m_4}{\sqrt{s}}
    \right) \cos\theta^*.
\end{align}
From this relation, we obtain
\begin{align}
    E_3^\pm
    &=
    \frac{E_1 + E_2}{2}\left(
        1 + \frac{m_3^2 - m_4^2}{s}
    \right)
    \pm \frac{|\vec p_1 + \vec p_2|}{2} \beta\left(
        \frac{m_3}{\sqrt{s}},\frac{m_4}{\sqrt{s}}
    \right),
\end{align}
for $\cos\theta^* = \pm 1$, which matches the upper and lower limits of the $E_3$ integration in the lab frame \eqref{eq:LIPS_34_integration}.

From \eqref{eq:E3_lab_cm}, the transformation of the integration measure is given by
\begin{align}
    dE_3
    &=
    \frac{|\vec p_1 + \vec p_2|}{2} \beta\left(
        \frac{m_3}{\sqrt{s}},\frac{m_4}{\sqrt{s}}
    \right) d\cos\theta^*.
\end{align}
Using this, we can rewrite the lab frame phase space integration \eqref{eq:LIPS_34_integration} as
\begin{align}
    \int d{\rm LIPS}_{34}
    &=
    \frac{1}{8\pi |\vec p_1 + \vec p_2|} \int_{E_3^-}^{E_3^+} dE_3\nonumber\\[8pt]
    &=
    \frac{1}{8\pi |\vec p_1 + \vec p_2|} \int_{-1}^{1} \frac{|\vec p_1 + \vec p_2|}{2} \beta\left(
        \frac{m_3}{\sqrt{s}},\frac{m_4}{\sqrt{s}}
    \right) d\cos\theta^*\nonumber\\[8pt]
    &=
    \frac{1}{16\pi} \beta\left(
        \frac{m_3}{\sqrt{s}},\frac{m_4}{\sqrt{s}}
    \right) \int_{-1}^{1} d\cos\theta^*,
\end{align}
which matches the CM frame result.

Given that the squared amplitude is also Lorentz invariant, we can compute it in the CM frame as a function of $s$ and $\cos\theta^*$: $|\mathcal{M}_{12,34}|^2 = |\mathcal{M}_{12,34}(s,\cos\theta^*)|^2$~\footnote{
The squared amplitude is in general a function of $s$, $t$, and $u$. The Mandelstam variables on the other hand satisfy $s+t+u=\sum_i m_i^2$ by which we may eliminate, e.g., $u$ in the squared amplitude. Furthermore, $t$ is a function of $s$ and the scattering angle $\theta^*$ in the CM frame, and thus $|\mathcal{M}_{12,34}(s,\cos\theta^*)|^2$ follows.
}.
Then, the Lorentz-invariant function $w(s)$ can be expressed as
\begin{align}
    w(s)
    &=
    \int d{\rm LIPS}_{34} |\mathcal{M}_{12,34}|^2\nonumber\\[8pt]
    &=
    \frac{1}{16\pi} \beta\left(
        \frac{m_3}{\sqrt{s}},\frac{m_4}{\sqrt{s}}
    \right) \int_{-1}^{1} d\cos\theta^* |\mathcal{M}_{12,34}(s,\cos\theta^*)|^2.
\end{align}
Again, $w(s)$ is a Lorentz-invariant function, and thus can be used in any frame, including the lab frame, to compute the reaction rates.
Thus, it is often convenient to compute $w(s)$ in the CM frame, i.e., integrating over scattering angle, and then use it to perform the phase space integration in the lab frame as shown in Secs.~\ref{subsec:2_to_2_processes} and \ref{subsec:1_to_2_processes} with $s=(p_1 + p_2)^2=m_1^2+m_2^2+2(E_1E_2-|\vec p_1||\vec p_2|\cos\theta_{12})$.

For later convenience, we rewrite $w(s)$ in terms of the integration over $t$, instead of $\cos\theta^*$.
This can be done by using
\begin{align}
    t &= (p_3-p_1)^2 = m_1^2+m_3^2-2(E_1^* E_3^* - |\vec p_1^{\mkern6mu *}||\vec p_3^{\mkern6mu *}|\cos\theta^*)\\[8pt]
    \Rightarrow \;\;\;\;&
    d\cos\theta^* = \frac{1}{2|\vec p_1^{\mkern6mu *}||\vec p_3^{\mkern6mu *}|} dt. \label{eq:dcost_dt}
\end{align}
The integration region becomes
\begin{align}
    t_\pm
    &= m_1^2+m_3^2-2(E_1^* E_3^* \mp |\vec p_1^{\mkern6mu *}||\vec p_3^{\mkern6mu *}|).
\end{align}
Recall that the function obtained after integrating over $t$ should be Lorentz invariant, and thus it is convenient to write everything in terms of Lorentz invariant quantities, namely, $s$ in this case.
Using
\begin{align}
    |\vec p_1^{\mkern6mu *}|
    &=
    \frac{\sqrt{s}}{2}\beta\left(\frac{m_1}{\sqrt{s}},\frac{m_2}{\sqrt{s}}\right),\\[12pt]
    E_1^*
    &=
    \sqrt{|\vec p_1^{\mkern6mu *}|^2 + m_1^2}
    = \frac{s+m_1^2-m_2^2}{2\sqrt{s}},
\end{align}
together with $|\vec p_3^{\mkern6mu *}|$ and $E_3^*$ given by Eqs.~\eqref{eq:p3*} and \eqref{eq:E3*}, respectively,
we have
\begin{align}
    w(s)
    &=
    \frac{1}{8\pi s \beta(m_1/\sqrt{s},m_2/\sqrt{s})}\int_{t_-}^{t_+}dt |{\cal M}_{12,34}(s,t)|^2,\label{eq:w_s_integrating_over_t}\\[12pt]
    t_{\pm}
    &=
    m_1^2 + m_3^2 -\frac{s}{2}
    \left[
        \left(
            1+\frac{m_1^2-m_2^2}{s}
        \right)
        \left(
            1+\frac{m_3^2-m_4^2}{s}
        \right)
        \mp
        \beta\left(\frac{m_1}{\sqrt{s}},\frac{m_2}{\sqrt{s}}\right)
        \beta\left(\frac{m_3}{\sqrt{s}},\frac{m_4}{\sqrt{s}}\right)
    \right].\label{eq:t_plus_minus}
\end{align}

The $t$-integration can be readily recast into the integral over $u$ by using
\begin{align}
    u
    &=
    \sum_i m_i^2 -s-t
    \;\;\;\Rightarrow\;\;\;
    dt = -du.
\end{align}
Notice that the integration region is also flipped:
\begin{align}
    u_{\rm max} &= \sum_i m_i^2 - s - t_{\rm min},\\[8pt]
    u_{\rm min} &= \sum_i m_i^2 - s - t_{\rm max},
\end{align}
namely,
\begin{align}
    u_\pm &=
    m_1^2 + m_4^2 - \frac{s}{2}
    \left[
        \left(
            1+\frac{m_1^2-m_2^2}{s}
        \right)
        \left(
            1+\frac{m_4^2-m_3^2}{s}
        \right)
        \mp
        \beta\left(\frac{m_1}{\sqrt{s}},\frac{m_2}{\sqrt{s}}\right)
        \beta\left(\frac{m_3}{\sqrt{s}},\frac{m_4}{\sqrt{s}}\right)
    \right].
\end{align}
Thus, we have
\begin{align}
    w(s)
    &=
    \frac{1}{8\pi s \beta(m_1/\sqrt{s},m_2/\sqrt{s})}\int_{u_-}^{u_+}du |{\cal M}_{12,34}(s,u)|^2.
\end{align}
Notice that when $m_1=m_2$ and $m_3=m_4$, we have $t_{\pm}=u_{\pm}$, so the integration limits are the same if the integrand is symmetric under $t \leftrightarrow u$.

\section{Gondolo-Gelmini formula}
\label{appendix:gondolo_gelmini_formula}

In the two-to-two process discussed in Sec.~\ref{subsec:2_to_2_processes}, if we further assume that both particles 1 and 2 follow the Boltzmann distribution, i.e., $f_i = e^{-E_i/T}$, we can simplify the expression \`{a}~la Gondolo and Gelmini~\cite{Gondolo:1990dk}.
A strategy is to express the reaction rate as a single integral over the Mandelstam variable $s$.
To do so, we apply the following identities 
\begin{align}
    1 &= \int ds \,\,\delta(s - (p_1 + p_2)^2),\\
    1 &= \int d^4 P_{\rm tot} \,\, \delta^4(P_{\rm tot} - p_1 - p_2),
\end{align}
to rewrite the reaction rate as
\begin{align}
    \gamma_{12,34}
    &=
    \int d\Pi_1 d\Pi_2 \,\, w(s) \,\,e^{-E_{\rm tot}/T}
    \equiv \frac{1}{(2\pi)^6}\int ds \,\, w(s) J(s),
\end{align}
where we have defined
\begin{align}
    J(s)
    &\equiv
    \int d^4 P_{\rm tot} \frac{d^3p_1}{2E_1}\frac{d^3p_2}{2E_2} e^{-E_{\rm tot}/T} \delta^4(P_{\rm tot} - p_1 - p_2) \delta(s - (p_1 + p_2)^2).
\end{align}
In terms of the Lorentz-invariant two-body phase space, $J(s)$ can be expressed as
\begin{align}
    J(s)
    &=
    (2\pi)^2\int d^4 P_{\rm tot}d{\rm LIPS}_{12} e^{-E_{\rm tot}/T} \delta(s - P_{\rm tot}^2).
\end{align}
Since the integrand does not depend on the ``scattering angles" of particles 1 and 2, we can perform the angle integration in $d{\rm LIPS}_{12}$ in either the lab frame or the CM frame, yielding
\begin{align}
    \int d{\rm LIPS}_{12}
    &=
    \frac{1}{8\pi} \beta\left(\frac{m_1}{\sqrt{s}},\frac{m_2}{\sqrt{s}}\right),
\end{align}
regardless of the frame choice.
We can further perform the integration over $P_{\rm tot}$ as
\begin{align}
    J(s)
    &=
    \frac{(2\pi)^2}{8\pi}\beta\left(\frac{m_1}{\sqrt{s}},\frac{m_2}{\sqrt{s}}\right) \int d^4 P_{\rm tot} e^{-E_{\rm tot}/T} \delta(s - P_{\rm tot}^2)\nonumber\\[8pt]
    &=
    \frac{(2\pi)^3}{8\pi^2}\beta\left(\frac{m_1}{\sqrt{s}},\frac{m_2}{\sqrt{s}}\right) T \sqrt{s} K_1\left(\frac{\sqrt{s}}{T}\right).
\end{align}
Here, in the last line, we have used the integral representation of the modified Bessel function of the first kind.
Putting all together, we obtain the Gondolo-Gelmini formula:
\begin{align}
    \gamma_{12,34}
    &=
    \frac{T}{64\pi^4} \int_{(m_1 + m_2)^2}^\infty ds \sqrt{s} \beta\left(\frac{m_1}{\sqrt{s}},\frac{m_2}{\sqrt{s}}\right) w(s) K_1\left(\frac{\sqrt{s}}{T}\right).
\end{align}

\section{Collision term kernels}
\label{appendix:kernels}

In this appendix, we compute the squared amplitudes of the three processes in the Higgs portal scenario: 1. Decay, 2. Scattering (Higgs), 3. Scattering (Vector), and 4. Scattering (Fermion).
From these results, we can readily obtain the generic scalar case as well by replacing the Higgs boson $h$ with the generic scalar $\chi$ in the relevant expressions, i.e., those of decay and scattering (Higgs).

\begin{enumerate}
    \item Decay: $h \to \varphi\varphi$\\
    Unlike the case where there is only the $\sigma$-coupling, the current setup allows the inflaton production from the decay of the Higgs boson.
    The amplitude can be readily obtained as
   \begin{align}
      {\cal M}_{h,\varphi\varphi}
      &=
      2g_{h\varphi\varphi},
   \end{align}
   and thus
   \begin{align}
      |{\cal M}_{h,\varphi\varphi}|^2
      &=
      4g_{h,\varphi\varphi}^2
      =
      4\left(
         \sigma - \frac{4\mu^2}{m_h^2}
      \right)^2 v^2.
   \end{align}
   Integrating over the final state phase space, we obtain
   \begin{align}
      w_{h,\varphi\varphi}(s)
      &=
      \frac{1}{8\pi} |{\cal M}_{h,\varphi\varphi}|^2 \beta\left(\frac{m_\varphi}{\sqrt{s}},\frac{m_\varphi}{\sqrt{s}}\right)
      =
      \frac{v^2}{2\pi} \left(
         \sigma-\frac{4\mu^2}{m_h^2}
      \right)^2\sqrt{1-\frac{4m_\varphi^2}{s}}.
   \end{align}

   \item Scattering (Higgs): $hh\to\varphi\varphi$\\
   The relevant vertices are given by
   \begin{itemize}
      \item $h$-$\varphi$-$\varphi:\,\,\,-2ig_{h\varphi\varphi}$
      \item $h$-$h$-$\varphi$-$\varphi:\,\,\,-2i\sigma$
      \item $h$-$h$-$\varphi:\,\,\,-2ig_{hh\varphi}$
      \item $h$-$h$-$h:\,\,\,-3i m_h^2/v$
   \end{itemize}
   There are 6 diagrams in total: contact, $s$-channel ($h$-exchange), $t$-channel ($\varphi$ and $h$ exchange), and $u$-channel ($\varphi$ and $h$ exchange).
   The amplitudes can be computed as
   \begin{align}
      i{\cal M}_{\rm contact}
      &=
      -2i\sigma,\\
      i{\cal M}_s
      &=
      \left(
         -3i\frac{m_h^2}{v}
      \right)\cdot\frac{i}{s-m_h^2}\cdot
      \left(-2ig_{h\varphi\varphi}\right)
      =
      -i\frac{6(g_{h\varphi\varphi}/v) m_h^2}{s-m_h^2},\\
      i{\cal M}^{(\varphi)}_t
      &=
      (-2ig_{h\varphi\varphi})\cdot\frac{i}{t-m_\varphi^2}\cdot (-2ig_{h\varphi\varphi})
      =-i\frac{4g_{h\varphi\varphi}^2}{t-m_\varphi^2},\\
      i{\cal M}^{(h)}_t
      &=
      (-2ig_{hh\varphi})\cdot\frac{i}{t-m_h^2}\cdot (-2ig_{hh\varphi})
      =-i\frac{4g_{hh\varphi}^2}{t-m_h^2},\\
      i{\cal M}^{(\varphi)}_{u}
      &=
      -i\frac{4g_{h\varphi\varphi}^2}{u-m_\varphi^2},\\
      i{\cal M}^{(h)}_{u}
      &=
      -i\frac{4g_{hh\varphi}^2}{u-m_h^2}.
   \end{align}
   Defining
   \begin{align}
      C_\varphi(t,u)
      &\equiv
      \frac{1}{t-m_\varphi^2} + \frac{1}{u-m_\varphi^2},\\
      C_h(t,u)
      &\equiv
      \frac{1}{t-m_h^2} + \frac{1}{u-m_h^2},
   \end{align}
   we obtain
   \begin{align}
      &
      {\cal M}_{hh,\varphi\varphi}
      =
      2\sigma
      + \frac{6(g_{h\varphi\varphi}/v)m_h^2}{s-m_h^2}
      + 4g_{h\varphi\varphi}^2 C_\varphi(t,u)
      + 4g_{hh\varphi}^2 C_h(t,u) \,,
    \end{align}      
and hence 
   \begin{align}
      \overline{|{\cal M}_{hh,\varphi\varphi}|}^2
      \equiv
      A + B + C + D + E + F + G + H + I + J,
    \end{align}
    where 
    \begin{align}
      &
      A = 2\sigma^2,\;\;\;
      B = \frac{12\sigma(g_{h\varphi\varphi}/v) m_h^2}{s-m_h^2},\;\;\;
      C = 8\sigma g_{h\varphi\varphi}^2 C_\varphi(t,u),\;\;\;
      D = 8\sigma g_{hh\varphi}^2 C_h(t,u),\nonumber\\[8pt]
      &
      E = \frac{18(g_{h\varphi\varphi}/v)^2m_h^4}{(s-m_h^2)^2},\;\;\;
      F = \frac{24(g_{h\varphi\varphi}^3/v)m_h^2}{s-m_h^2}C_\varphi(t,u),\;\;\;
      G = \frac{24(g_{h\varphi\varphi}g^2_{hh\varphi}/v)m_h^2}{s-m_h^2}C_h(t,u),\nonumber\\[8pt]
      &
      H = 8g_{h\varphi\varphi}^4C_\varphi^2(t,u),\;\;\;
      I = 16g_{h\varphi\varphi}^2g_{hh\varphi}^2C_\varphi(t,u)C_h(t,u),\;\;\;
      J = 8g_{hh\varphi}^4C_h^2(t,u),
   \end{align}
 and,  in $\overline{|{\cal M}_{hh,\varphi\varphi}|}^2$,  we have multiplied by a symmetry factor 1/2 for the identical incoming particles.
   Thus, we find
   \begin{align}
      w_{hh,\varphi\varphi}^A(s)
      &=
      \frac{1}{8\pi}A\beta\left(\frac{m_\varphi}{\sqrt{s}},\frac{m_\varphi}{\sqrt{s}}\right)
      = \frac{\sigma^2}{4\pi}\beta_\varphi,\\[8pt]
      w_{hh,\varphi\varphi}^B(s)
      &=
      \frac{1}{8\pi}B\beta\left(\frac{m_\varphi}{\sqrt{s}},\frac{m_\varphi}{\sqrt{s}}\right)
      = \frac{3\sigma(g_{h\varphi\varphi}/v)m_h^2}{2\pi s}\left(
         1-\frac{m_h^2}{s}
      \right)^{-1}\beta_\varphi,
   \end{align}
   where $\beta_i \equiv \sqrt{1-4m_i^2/s}$.
   The other terms involve either $t$- or $u$-integrations.
   For example, for the term $C$, we have
   \begin{align}
      \int_{t_-}^{t_+}dt C_\varphi(t,u)
      &=
      \int_{t_-}^{t_+} 
         \frac{dt}{t-m_\varphi^2} + 
      \int_{u_-}^{u_+}
         \frac{du}{u-m_\varphi^2}.
   \end{align}
   Given that
   \begin{align}
      t_{\pm}
      &=
      u_{\pm} 
      =
      m_h^2 + m_\varphi^2 - \frac{s}{2}(1\mp\beta_h\beta_\varphi),
   \end{align}
   we find
   \begin{align}
      \int_{t_-}^{t_+}dt C_\varphi(t,u)
      &=
      2\ln\left(
         \frac{1-\beta_h\beta_\varphi-2m_h^2/s}{1+\beta_h\beta_\varphi-2m_h^2/s}
      \right).
   \end{align}
   For later convenience, we define
   \begin{align}
      L(\beta)
      &\equiv
      \ln\left(
         \frac{1+\beta^2+2\beta_h\beta_\varphi}{1+\beta^2-2\beta_h\beta_\varphi}
      \right).
   \end{align}
   Using
   \begin{align}
      \frac{1-\beta_h\beta_\varphi-2m_h^2/s}{1+\beta_h\beta_\varphi-2m_h^2/s}
      &=
      \frac{1+\beta_h^2-2\beta_h\beta_\varphi}{1+\beta_h^2+2\beta_h\beta_\varphi},
   \end{align}
   we have
   \begin{align}
      \int_{t_-}^{t_+}dt C_\varphi(t,u)
      &=
      -2L(\beta_h),
   \end{align}
   and thus
   \begin{align}
      w_{hh,\varphi\varphi}^C(s)
      &=
      \frac{16\sigma g_{h\varphi\varphi}^2}{2\cdot8\pi \beta_h s}\left[
         \int_{t_-}^{t_+}\frac{dt}{t-m_\varphi^2} + \int_{u_-}^{u_+}\frac{du}{u-m_\varphi^2}
      \right]
      \nonumber\\[8pt]
      &=
      -\frac{2\sigma g_{h\varphi\varphi}^2}{\pi\beta_h s}L(\beta_h).
   \end{align}
   In the same manner, we obtain
   \begin{align}
      w_{hh,\varphi\varphi}^D(s)
      &=
      \frac{16\sigma g_{hh\varphi}^2}{2\cdot8\pi \beta_h s}\left[
         \int_{t_-}^{t_+}\frac{dt}{t-m_h^2} + \int_{u_-}^{u_+}\frac{du}{u-m_h^2}
      \right]\nonumber\\[8pt]
      &=
      -\frac{2\sigma g_{hh\varphi}^2}{\pi\beta_h s}L(\beta_\varphi).
   \end{align}

   It is straightforward to obtain
   \begin{align}
      w_{hh,\varphi\varphi}^E(s)
      &=
      \frac{9 g_{h\varphi\varphi}^2m_h^4}{4\pi v^2 s^2}\beta_\varphi \left(1-\frac{m_h^2}{s}\right)^{-2},\\[8pt]
      w_{hh,\varphi\varphi}^F(s)
      &=
      -\frac{6g_{h\varphi\varphi}^3m_h^2}{\pi \beta_h v s^2}\left(1-\frac{m_h^2}{s}\right)^{-1}L(\beta_h),\\[8pt]
      w_{hh,\varphi\varphi}^G(s)
      &=
      -\frac{6g_{h\varphi\varphi}g_{hh\varphi}^2m_h^2}{\pi \beta_h v s^2}\left(
         1-\frac{m_h^2}{s}
      \right)^{-1}L(\beta_\varphi).
   \end{align}

   To integrate over the remaining terms needs some care.
   For example, we have
   \begin{align}
      \int_{t_-}^{t_+}dt C_\varphi^2(t,u)
      &=
      \int dt\left(
         \frac{1}{t-m_\varphi^2} + \frac{1}{u-m_\varphi^2}
      \right)^2.    
   \end{align}
   In this integral, it is straightforward to perform
   \begin{align}
      \int_{t_-}^{t_+}\frac{dt}{(t-m_\varphi^2)^2}
      &=
      \left[
         -\frac{1}{t-m_\varphi^2}
      \right]_{t_-}^{t_+}
      =
      \frac{s\beta_h\beta_\varphi}{m_h^4+m_\varphi^2s\beta_h}
      = \int_{u_-}^{u^+}\frac{du}{(u-m_\varphi^2)^2}.
   \end{align}
   The cross term can be computed as
   \begin{align}
      \int \frac{dt}{(t-m_\varphi^2)(u-m_\varphi^2)}
      &=
      \frac{1}{2m_h^2-s}\left[
         \int_{t_-}^{t_+}\frac{dt}{t-m_\varphi^2} + \int_{u_-}^{u_+}\frac{du}{u-m_\varphi^2}
      \right]\nonumber\\[8pt]
      &=
      \frac{2}{2m_h^2-s}\ln\left(
         \frac{1-\beta_h\beta_\varphi-2m_h^2/s}{1+\beta_h\beta_\varphi-2m_h^2/s}
      \right).
   \end{align}
   After further simplification, we find
   \begin{align}
      \int_{t_-}^{t_+}dt C_\varphi^2(t,u)
      &=
      \frac{8}{s}\left[
         \frac{4\beta_h\beta_\varphi}{(1+\beta_h^2)^2-4\beta_h^2\beta_\varphi^2}
         +
         \frac{1}{1+\beta_h^2}L(\beta_h)
      \right].
   \end{align}
   Therefore, we obtain
   \begin{align}
      w_{hh,\varphi\varphi}^H(s)
      &=
      \frac{8g_{h\varphi\varphi}^4}{\pi\beta_h s^2}
      \left[
         \frac{4\beta_h\beta_\varphi}{(1+\beta_h^2)^2-4\beta_h^2\beta_\varphi^2}
         +
         \frac{1}{1+\beta_h^2}L(\beta_h)
      \right]
   \end{align}
   In the same manner, we find
   \begin{align}
      w_{hh,\varphi\varphi}^I(s)
      &=
      \frac{32g_{h\varphi\varphi}^2g_{hh\varphi}^2}{\pi s^2\beta_h(2\beta_h^2+\beta_\varphi^2)(\beta_h^2-\beta_\varphi^2)}\left[
         (1+\beta_h^2)L(\beta_\varphi)-(1+\beta_\varphi^2)L(\beta_h)
      \right],\\[8pt]
      w_{hh,\varphi\varphi}^J(s)
      &=
      \frac{8g_{hh\varphi}^4}{\pi\beta_h s^2}
      \left[
         \frac{4\beta_h\beta_\varphi}{(1+\beta_\varphi^2)^2-4\beta_h^2\beta_\varphi^2}
         +
         \frac{1}{1+\beta_\varphi^2}L(\beta_\varphi)
      \right].
   \end{align}
   \item Scattering (Vector): $W^+W^-\to\varphi\varphi, ZZ\to \varphi\varphi$\\
   The relevant vertices are given by
   \begin{itemize}
      \item $V$-$V$-$h:\,\,\, 2i m_V^2/v$ ($V=W^\pm, Z$)
      \item $h$-$\varphi$-$\varphi:\,\,\, 2i g_{h\varphi\varphi}$
   \end{itemize}
   The amplitude becomes
   \begin{align}
      i{\cal M}_{VV,\varphi\varphi}
      &=
      \left(
         2i\frac{m_V^2}{v}\epsilon^*_\mu\epsilon^{*\mu}
      \right)\cdot\frac{i}{s-m_h^2}\cdot
      \left(
         2ig_{h\varphi\varphi}
      \right)\nonumber\\[8pt]
      \Rightarrow\;\;\;
      &
      |{\cal M}_{VV,\varphi\varphi}|^2
      =
      \frac{16(g_{h\varphi\varphi}/v)^2m_V^4}{(s-m_h^2)^2}
      \left[
         2+\frac{(s-2m_V^2)^2}{4m_V^4}
      \right].
   \end{align}
   From the straightforward calculation, we obtain
   \begin{align}
      w_{VV,\varphi\varphi}(s)
      &=
      \frac{\beta_\varphi}{8\pi}\times \frac{16(g_{h\varphi\varphi}/v)^2m_V^4}{(s-m_h^2)^2}
      \left[
         2+\frac{(s-2m_V^2)^2}{4m_V^4}
      \right]\nonumber\\[8pt]
      &=
      \frac{2(g_{h\varphi\varphi}/v)^2m_V^4\beta_\varphi}{\pi(s-m_h^2)^2}\left[
         2+\frac{(s-2m_V^2)^2}{4m_V^4}
      \right].
   \end{align}
   In the relativistic limit, we have
   \begin{align}
      w_{VV,\varphi\varphi}(s)
      &\to
      \frac{\sigma^2}{2\pi}
   \end{align}
   for $\mu=0$.
   When $V=Z$, a factor of 1/2 should be multiplied since the initial state particles are identical.
   A final remark is that in the relativistic limit we have
   \begin{align}
      w_{hh,\varphi\varphi}(s) &\to \frac{\sigma^2}{2\pi},
   \end{align}
   so we find
   \begin{align}
      w_{hh,\varphi\varphi} + w_{WW,\varphi\varphi} + w_{ZZ,\varphi\varphi}
      &\to
      \frac{1}{2}\frac{\sigma^2}{2\pi} + \frac{\sigma^2}{2\pi} + \frac{1}{2}\frac{\sigma^2}{2\pi} = \frac{\sigma^2}{\pi},
   \end{align}
   which coincides with the expression
   \begin{align}
      \sum_{i=1}^4w_{h_ih_i,\varphi\varphi}(s) = \frac{\sigma^2}{\pi}
   \end{align}
   where we take the four real scalar degrees of freedom in the Higgs doublet as the dynamical ones.
   This is the manifestation of the Nambu-Goldstone boson equivalence theorem.
   \item Scattering (Fermion): 
   Writing the total decay width of $h$ as $\Gamma_h$, we obtain
   \begin{align}
      w_{f\overline f,\varphi\varphi}(s)
      &=
      \frac{N_C}{\pi}\frac{g_{hh\varphi}^2y_f^2}{(s-m_h^2)^2+m_h^2\Gamma_h^2}s\beta_f^3,
   \end{align}
   where $N_C$ is the number of colors (3 for quarks, 1 for leptons).
   In our analysis, we used $\Gamma_h=4.1$ MeV~\cite{LHCHiggsCrossSectionWorkingGroup:2016ypw}.
\end{enumerate}

We list the kernels for the collision terms obtained above.
The following includes the symmetry factor of the initial state particles as well as a factor of 2 due to two $\varphi$'s produced per reaction, which is eventually cancelled by a symmetry factor of 1/2 for $\varphi$ in the final state:
\begin{align}
   w_{h,\varphi\varphi}(s)
   &=
   \frac{g_{h\varphi\varphi}^2}{2\pi}\beta_\varphi,\\[8pt]
   w_{hh,\varphi\varphi}(s)
   &=
   w_{hh,\varphi\varphi}^A(s) + w_{hh,\varphi\varphi}^B(s) + w_{hh,\varphi\varphi}^C(s) + w_{hh,\varphi\varphi}^D(s) + w_{hh,\varphi\varphi}^E(s)\nonumber\\[8pt]
   &+ w_{hh,\varphi\varphi}^F(s) + w_{hh,\varphi\varphi}^G(s) + w_{hh,\varphi\varphi}^H(s) + w_{hh,\varphi\varphi}^I(s) + w_{hh,\varphi\varphi}^J(s),\label{eq:w_scalar_scattering}
\end{align}
where each term is given as follows:
\begin{align}
      w_{hh,\varphi\varphi}^A(s)
      &= 
      \displaystyle\frac{\sigma^2}{4\pi}\beta_\varphi,\\[8pt]
      w_{hh,\varphi\varphi}^B(s)
      &=
      \displaystyle\frac{3\sigma(g_{h\varphi\varphi}/v)m_h^2}{2\pi s}\left(
         1-\frac{m_h^2}{s}
      \right)^{-1}\beta_\varphi,\\[8pt]
      w_{hh,\varphi\varphi}^C(s)
      &=
      \displaystyle-\frac{2\sigma g_{h\varphi\varphi}^2}{\pi\beta_h s}L(\beta_h),\\[8pt]
      w_{hh,\varphi\varphi}^D(s)
      &=
      \displaystyle-\frac{2\sigma g_{hh\varphi}^2}{\pi\beta_h s}L(\beta_\varphi),\\[8pt]
      w_{hh,\varphi\varphi}^E(s)
      &=
      \displaystyle\frac{9 g_{h\varphi\varphi}^2m_h^4}{4\pi v^2 s^2}\beta_\varphi \left(1-\frac{m_h^2}{s}\right)^{-2},\\[8pt]
      w_{hh,\varphi\varphi}^F(s)
      &=
      -\displaystyle\frac{6g_{h\varphi\varphi}^3m_h^2}{\pi \beta_h v s^2}\left(1-\frac{m_h^2}{s}\right)^{-1}L(\beta_h),\\[8pt]
      w_{hh,\varphi\varphi}^G(s)
      &=
      -\displaystyle\frac{6g_{h\varphi\varphi}g_{hh\varphi}^2m_h^2}{\pi \beta_h v s^2}\left(
         1-\frac{m_h^2}{s}
      \right)^{-1}L(\beta_\varphi),\\[8pt]
      w_{hh,\varphi\varphi}^H(s)
      &=
      \displaystyle\frac{8g_{h\varphi\varphi}^4}{\pi\beta_h s^2}
      \left[
         \frac{4\beta_h\beta_\varphi}{(1+\beta_h^2)^2-4\beta_h^2\beta_\varphi^2}
         +
         \frac{1}{1+\beta_h^2}L(\beta_h)
      \right],\\[8pt]
      w_{hh,\varphi\varphi}^I(s)
      &=
      \displaystyle\frac{32g_{h\varphi\varphi}^2g_{hh\varphi}^2}{\pi s^2\beta_h(2\beta_h^2+\beta_\varphi^2)(\beta_h^2-\beta_\varphi^2)}\left[
         (1+\beta_h^2)L(\beta_\varphi)-(1+\beta_\varphi^2)L(\beta_h)
      \right],\\[8pt]
      w_{hh,\varphi\varphi}^J(s)
      &=
      \displaystyle\frac{g_{hh\varphi}^4}{\pi\beta_h s^2}
      \left[
         \frac{4\beta_h\beta_\varphi}{(1+\beta_\varphi^2)^2-4\beta_h^2\beta_\varphi^2}
         +
         \frac{1}{1+\beta_\varphi^2}L(\beta_\varphi)
      \right].
   \label{eq:w_scalar_scattering_from_A_to_J}
   \end{align}
Furthermore, we also have
   \begin{align}
   w_{WW,hh}(s)
      &=
      \frac{16(g_{h\varphi\varphi}/v)^2m_W^4\beta_\varphi}{\pi(s-m_h^2)^2}\left[
         2+\frac{(s-2m_W^2)^2}{4m_W^4}
      \right],\label{eq:w_vector_scattering_W}\\[8pt]
   w_{ZZ,hh}(s)
   &=
   \frac{8(g_{h\varphi\varphi}/v)^2m_Z^4\beta_\varphi}{\pi(s-m_h^2)^2}\left[
      2+\frac{(s-2m_Z^2)^2}{4m_Z^4}
   \right],\label{eq:w_vector_scattering_Z}\\[8pt]
   w_{f\overline f,\varphi\varphi}(s)
      &=
      \frac{N_C}{\pi}\frac{g_{hh\varphi}^2y_f^2}{(s-m_h^2)^2+m_h^2\Gamma_h^2}s\beta_f^3
      \label{eq:w_fermion_scattering},
\end{align}
where
\begin{align}
    \beta_i 
    &=
    \sqrt{1-4m_i^2/s},\\[8pt]
    L(\beta)
    &=
    \ln\left(
        \frac{1+\beta^2+2\beta_h\beta_\varphi}{1+\beta^2-2\beta_h\beta_\varphi}
    \right),\\[8pt]
    g_{h\varphi\varphi} &= \sigma v - 2\theta\mu \simeq \sigma v - 4\mu^2 v/m_h^2,\\[8pt]
    g_{hh\varphi} &= \mu + 2\theta\sigma v \simeq \mu + 4\sigma \mu v^2/m_h^2.
\end{align}

\clearpage 
\bibliography{refs}

@article{Planck:2018vyg,
    author = "Aghanim, N. and others",
    collaboration = "Planck",
    title = "{Planck 2018 results. VI. Cosmological parameters}",
    eprint = "1807.06209",
    archivePrefix = "arXiv",
    primaryClass = "astro-ph.CO",
    doi = "10.1051/0004-6361/201833910",
    journal = "Astron. Astrophys.",
    volume = "641",
    pages = "A6",
    year = "2020",
    note = "[Erratum: Astron.Astrophys. 652, C4 (2021)]"
}

@article{AtacamaCosmologyTelescope:2025nti,
    author = "Calabrese, Erminia and others",
    collaboration = "Atacama Cosmology Telescope",
    title = "{The Atacama Cosmology Telescope: DR6 constraints on extended cosmological models}",
    eprint = "2503.14454",
    archivePrefix = "arXiv",
    primaryClass = "astro-ph.CO",
    reportNumber = "FERMILAB-PUB-25-0157-PPD",
    doi = "10.1088/1475-7516/2025/11/063",
    journal = "JCAP",
    volume = "11",
    pages = "063",
    year = "2025"
}

@article{Turner:1983he,
    author = "Turner, Michael S.",
    title = "{Coherent Scalar Field Oscillations in an Expanding Universe}",
    reportNumber = "EFI-83-29-CHICAGO",
    doi = "10.1103/PhysRevD.28.1243",
    journal = "Phys. Rev. D",
    volume = "28",
    pages = "1243",
    year = "1983"
}

@article{Gondolo:1990dk,
    author = "Gondolo, Paolo and Gelmini, Graciela",
    title = "{Cosmic abundances of stable particles: Improved analysis}",
    reportNumber = "UCLA-90-TEP-68",
    doi = "10.1016/0550-3213(91)90438-4",
    journal = "Nucl. Phys. B",
    volume = "360",
    pages = "145--179",
    year = "1991"
}

@article{Djouadi:2018xqq,
    author = "Djouadi, Abdelhak and Kalinowski, Jan and Muehlleitner, Margarete and Spira, Michael",
    collaboration = "HDECAY",
    title = "{HDECAY: Twenty$_{++}$ years after}",
    eprint = "1801.09506",
    archivePrefix = "arXiv",
    primaryClass = "hep-ph",
    reportNumber = "LPT-ORSAY-18-04, CERN-TH-2017-262, LPT-Orsay-18-04, KA-TP-03-2018, PSI-PR-18-02",
    doi = "10.1016/j.cpc.2018.12.010",
    journal = "Comput. Phys. Commun.",
    volume = "238",
    pages = "214--231",
    year = "2019"
}

@article{Djouadi:2005gi,
    author = "Djouadi, Abdelhak",
    title = "{The Anatomy of electro-weak symmetry breaking. I: The Higgs boson in the standard model}",
    eprint = "hep-ph/0503172",
    archivePrefix = "arXiv",
    reportNumber = "LPT-ORSAY-05-17",
    doi = "10.1016/j.physrep.2007.10.004",
    journal = "Phys. Rept.",
    volume = "457",
    pages = "1--216",
    year = "2008"
}

@article{Winkler:2018qyg,
    author = "Winkler, Martin Wolfgang",
    title = "{Decay and detection of a light scalar boson mixing with the Higgs boson}",
    eprint = "1809.01876",
    archivePrefix = "arXiv",
    primaryClass = "hep-ph",
    reportNumber = "NORDITA-2018-087",
    doi = "10.1103/PhysRevD.99.015018",
    journal = "Phys. Rev. D",
    volume = "99",
    number = "1",
    pages = "015018",
    year = "2019"
}

@article{Arcadi:2019lka,
    author = "Arcadi, Giorgio and Djouadi, Abdelhak and Raidal, Martti",
    title = "{Dark Matter through the Higgs portal}",
    eprint = "1903.03616",
    archivePrefix = "arXiv",
    primaryClass = "hep-ph",
    reportNumber = "LAPTH-010/19",
    doi = "10.1016/j.physrep.2019.11.003",
    journal = "Phys. Rept.",
    volume = "842",
    pages = "1--180",
    year = "2020"
}

@article{ATLAS:2023tkt,
    author = "Aad, Georges and others",
    collaboration = "ATLAS",
    title = "{Combination of searches for invisible decays of the Higgs boson using 139 fb{\ensuremath{-}}1 of proton-proton collision data at s=13 TeV collected with the ATLAS experiment}",
    eprint = "2301.10731",
    archivePrefix = "arXiv",
    primaryClass = "hep-ex",
    reportNumber = "CERN-EP-2022-289",
    doi = "10.1016/j.physletb.2023.137963",
    journal = "Phys. Lett. B",
    volume = "842",
    pages = "137963",
    year = "2023"
}

@article{Ferber:2023iso,
    author = "Ferber, Torben and Grohsjean, Alexander and Kahlhoefer, Felix",
    title = "{Dark Higgs bosons at colliders}",
    eprint = "2305.16169",
    archivePrefix = "arXiv",
    primaryClass = "hep-ph",
    reportNumber = "P3H-23-034, TTP23-018",
    doi = "10.1016/j.ppnp.2024.104105",
    journal = "Prog. Part. Nucl. Phys.",
    volume = "136",
    pages = "104105",
    year = "2024"
}

@article{EscuderoAbenza:2025cfj,
    author = "Escudero Abenza, Miguel and Hambye, Thomas",
    title = "{The simplest dark matter model at the edge of perturbativity}",
    eprint = "2505.02408",
    archivePrefix = "arXiv",
    primaryClass = "hep-ph",
    reportNumber = "CERN-TH-2025-087, ULB-TH/25-04",
    doi = "10.1016/j.physletb.2025.139696",
    journal = "Phys. Lett. B",
    volume = "868",
    pages = "139696",
    year = "2025"
}

@article{Garcia:2023dyf,
    author = "Garcia, Marcos A. G. and Gross, Mathieu and Mambrini, Yann and Olive, Keith A. and Pierre, Mathias and Yoon, Jong-Hyun",
    title = "{Effects of fragmentation on post-inflationary reheating}",
    eprint = "2308.16231",
    archivePrefix = "arXiv",
    primaryClass = "hep-ph",
    reportNumber = "UMN--TH--4223/23, FTPI--MINN--23/15, DESY-23-122",
    doi = "10.1088/1475-7516/2023/12/028",
    journal = "JCAP",
    volume = "12",
    pages = "028",
    year = "2023"
}

@article{Planck:2018jri,
    author = "Akrami, Y. and others",
    collaboration = "Planck",
    title = "{Planck 2018 results. X. Constraints on inflation}",
    eprint = "1807.06211",
    archivePrefix = "arXiv",
    primaryClass = "astro-ph.CO",
    doi = "10.1051/0004-6361/201833887",
    journal = "Astron. Astrophys.",
    volume = "641",
    pages = "A10",
    year = "2020"
}

@article{BICEP:2021xfz,
    author = "Ade, P. A. R. and others",
    collaboration = "BICEP, Keck",
    title = "{Improved Constraints on Primordial Gravitational Waves using Planck, WMAP, and BICEP/Keck Observations through the 2018 Observing Season}",
    eprint = "2110.00483",
    archivePrefix = "arXiv",
    primaryClass = "astro-ph.CO",
    doi = "10.1103/PhysRevLett.127.151301",
    journal = "Phys. Rev. Lett.",
    volume = "127",
    number = "15",
    pages = "151301",
    year = "2021"
}

@article{Kallosh:2013hoa,
    author = "Kallosh, Renata and Linde, Andrei",
    title = "{Universality Class in Conformal Inflation}",
    eprint = "1306.5220",
    archivePrefix = "arXiv",
    primaryClass = "hep-th",
    doi = "10.1088/1475-7516/2013/07/002",
    journal = "JCAP",
    volume = "07",
    pages = "002",
    year = "2013"
}

@article{Kallosh:2013pga,
    author = "Kallosh, Renata and Linde, Andrei and Roest, Diederik",
    title = "{Superconformal Inflationary Attractors}",
    eprint = "1311.0472",
    archivePrefix = "arXiv",
    primaryClass = "hep-th",
    doi = "10.1007/JHEP11(2013)198",
    journal = "JHEP",
    volume = "11",
    pages = "198",
    year = "2013"
}

@article{Ellis:2015pla,
    author = "Ellis, John and Garcia, Marcos A. G. and Nanopoulos, Dimitri V. and Olive, Keith A.",
    title = "{Calculations of Inflaton Decays and Reheating: with Applications to No-Scale Inflation Models}",
    eprint = "1505.06986",
    archivePrefix = "arXiv",
    primaryClass = "hep-ph",
    reportNumber = "KCL-PH-TH-2015-23, LCTS-2015-13, CERN-PH-TH-2015-122, ACT-04-15, UMN-TH-3438-15, FTPI-MINN-15-26, MI-TH-1513",
    doi = "10.1088/1475-7516/2015/07/050",
    journal = "JCAP",
    volume = "07",
    pages = "050",
    year = "2015"
}

@article{Garcia:2020wiy,
    author = "Garcia, Marcos A. G. and Kaneta, Kunio and Mambrini, Yann and Olive, Keith A.",
    title = "{Inflaton Oscillations and Post-Inflationary Reheating}",
    eprint = "2012.10756",
    archivePrefix = "arXiv",
    primaryClass = "hep-ph",
    reportNumber = "UMN-TH-4006/20, FTPI-MINN-20/37, IFT-UAM/CSIC-20-185, KIAS-P20071",
    doi = "10.1088/1475-7516/2021/04/012",
    journal = "JCAP",
    volume = "04",
    pages = "012",
    year = "2021"
}

@article{Kaneta:2025xuo,
    author = "Kaneta, Kunio and Takahashi, Tomo and Watanabe, Natsumi",
    title = "{Post-Reheating Inflaton Production as a Probe of Reheating Dynamics}",
    eprint = "2508.20402",
    archivePrefix = "arXiv",
    primaryClass = "hep-ph",
    month = "8",
    year = "2025"
}

@article{Kawasaki:2004qu,
    author = "Kawasaki, Masahiro and Kohri, Kazunori and Moroi, Takeo",
    title = "{Big-Bang nucleosynthesis and hadronic decay of long-lived massive particles}",
    eprint = "astro-ph/0408426",
    archivePrefix = "arXiv",
    reportNumber = "ICRR-REPORT-508-2004-6, OU-TAP-234, TU-727",
    doi = "10.1103/PhysRevD.71.083502",
    journal = "Phys. Rev. D",
    volume = "71",
    pages = "083502",
    year = "2005"
}

@article{Hu:1992dc,
    author = "Hu, Wayne and Silk, Joseph",
    title = "{Thermalization and spectral distortions of the cosmic background radiation}",
    reportNumber = "CFPA-TH-92-36",
    doi = "10.1103/PhysRevD.48.485",
    journal = "Phys. Rev. D",
    volume = "48",
    pages = "485--502",
    year = "1993"
}

@article{Chluba:2011hw,
    author = "Chluba, J. and Sunyaev, R. A.",
    title = "{The evolution of CMB spectral distortions in the early Universe}",
    eprint = "1109.6552",
    archivePrefix = "arXiv",
    primaryClass = "astro-ph.CO",
    doi = "10.1111/j.1365-2966.2011.19786.x",
    journal = "Mon. Not. Roy. Astron. Soc.",
    volume = "419",
    pages = "1294--1314",
    year = "2012"
}

@article{Slatyer:2016qyl,
    author = "Slatyer, Tracy R. and Wu, Chih-Liang",
    title = "{General Constraints on Dark Matter Decay from the Cosmic Microwave Background}",
    eprint = "1610.06933",
    archivePrefix = "arXiv",
    primaryClass = "astro-ph.CO",
    reportNumber = "MIT-CTP-4842",
    doi = "10.1103/PhysRevD.95.023010",
    journal = "Phys. Rev. D",
    volume = "95",
    number = "2",
    pages = "023010",
    year = "2017"
}

@article{Cang:2020exa,
    author = "Cang, Junsong and Gao, Yu and Ma, Yin-Zhe",
    title = "{Probing dark matter with future CMB measurements}",
    eprint = "2002.03380",
    archivePrefix = "arXiv",
    primaryClass = "astro-ph.CO",
    doi = "10.1103/PhysRevD.102.103005",
    journal = "Phys. Rev. D",
    volume = "102",
    number = "10",
    pages = "103005",
    year = "2020"
}

@article{LHCHiggsCrossSectionWorkingGroup:2016ypw,
    author = "de Florian, D. and others",
    collaboration = "LHC Higgs Cross Section Working Group",
    title = "{Handbook of LHC Higgs Cross Sections: 4. Deciphering the Nature of the Higgs Sector}",
    eprint = "1610.07922",
    archivePrefix = "arXiv",
    primaryClass = "hep-ph",
    reportNumber = "CERN-2017-002-M, CERN-2017-002",
    doi = "10.23731/CYRM-2017-002",
    journal = "CERN Yellow Rep. Monogr.",
    volume = "2",
    pages = "1--869",
    year = "2017"
}

@article{Hooper:2018buz,
    author = "Hooper, Dan and Krnjaic, Gordan and Long, Andrew J. and Mcdermott, Samuel D.",
    title = "{Can the Inflaton Also Be a Weakly Interacting Massive Particle?}",
    eprint = "1807.03308",
    archivePrefix = "arXiv",
    primaryClass = "hep-ph",
    reportNumber = "FERMILAB-PUB-18-309-A",
    doi = "10.1103/PhysRevLett.122.091802",
    journal = "Phys. Rev. Lett.",
    volume = "122",
    number = "9",
    pages = "091802",
    year = "2019"
}

@article{Garcia:2021gsy,
    author = "Garcia, Marcos A. G. and Mambrini, Yann and Olive, Keith A. and Verner, Sarunas",
    title = "{On the Realization of WIMPflation}",
    eprint = "2107.07472",
    archivePrefix = "arXiv",
    primaryClass = "hep-ph",
    reportNumber = "UMN--TH--4019/21, FTPI--MINN--21/12, IFT-UAM/CSIC-21-83",
    doi = "10.1088/1475-7516/2021/10/061",
    journal = "JCAP",
    volume = "10",
    pages = "061",
    year = "2021"
}

@article{Kawasaki:2017bqm,
    author = "Kawasaki, Masahiro and Kohri, Kazunori and Moroi, Takeo and Takaesu, Yoshitaro",
    title = "{Revisiting Big-Bang Nucleosynthesis Constraints on Long-Lived Decaying Particles}",
    eprint = "1709.01211",
    archivePrefix = "arXiv",
    primaryClass = "hep-ph",
    reportNumber = "KEK-COSMO-211, IPMU17-0117, UT-17-29, KEK-Cosmo-211, KEK-TH-1998",
    doi = "10.1103/PhysRevD.97.023502",
    journal = "Phys. Rev. D",
    volume = "97",
    number = "2",
    pages = "023502",
    year = "2018"
}

@article{Guth:1980zm,
    author = "Guth, Alan H.",
    editor = "Fang, Li-Zhi and Ruffini, R.",
    title = "{The Inflationary Universe: A Possible Solution to the Horizon and Flatness Problems}",
    reportNumber = "SLAC-PUB-2576",
    doi = "10.1103/PhysRevD.23.347",
    journal = "Phys. Rev. D",
    volume = "23",
    pages = "347--356",
    year = "1981"
}

@article{Sato:1981qmu,
    author = "Sato, Katsuhiko",
    title = "{First-order phase transition of a vacuum and the expansion of the Universe}",
    doi = "10.1093/mnras/195.3.467",
    journal = "Mon. Not. Roy. Astron. Soc.",
    volume = "195",
    number = "3",
    pages = "467--479",
    year = "1981"
}

@article{Linde:1981mu,
    author = "Linde, Andrei D.",
    editor = "Fang, Li-Zhi and Ruffini, R.",
    title = "{A New Inflationary Universe Scenario: A Possible Solution of the Horizon, Flatness, Homogeneity, Isotropy and Primordial Monopole Problems}",
    reportNumber = "LEBEDEV-81-229",
    doi = "10.1016/0370-2693(82)91219-9",
    journal = "Phys. Lett. B",
    volume = "108",
    pages = "389--393",
    year = "1982"
}

@article{Albrecht:1982wi,
    author = "Albrecht, Andreas and Steinhardt, Paul J.",
    editor = "Fang, Li-Zhi and Ruffini, R.",
    title = "{Cosmology for Grand Unified Theories with Radiatively Induced Symmetry Breaking}",
    reportNumber = "UPR-0185T",
    doi = "10.1103/PhysRevLett.48.1220",
    journal = "Phys. Rev. Lett.",
    volume = "48",
    pages = "1220--1223",
    year = "1982"
}

@article{Bardeen:1980kt,
    author = "Bardeen, James M.",
    title = "{Gauge Invariant Cosmological Perturbations}",
    doi = "10.1103/PhysRevD.22.1882",
    journal = "Phys. Rev. D",
    volume = "22",
    pages = "1882--1905",
    year = "1980"
}

@article{Mukhanov:1981xt,
    author = "Mukhanov, Viatcheslav F. and Chibisov, G. V.",
    title = "{Quantum Fluctuations and a Nonsingular Universe}",
    journal = "JETP Lett.",
    volume = "33",
    pages = "532--535",
    year = "1981"
}

@article{Hawking:1982cz,
    author = "Hawking, S. W.",
    title = "{The Development of Irregularities in a Single Bubble Inflationary Universe}",
    reportNumber = "Print-83-0015 (CAMBRIDGE)",
    doi = "10.1016/0370-2693(82)90373-2",
    journal = "Phys. Lett. B",
    volume = "115",
    pages = "295",
    year = "1982"
}

@article{Bardeen:1983qw,
    author = "Bardeen, James M. and Steinhardt, Paul J. and Turner, Michael S.",
    title = "{Spontaneous Creation of Almost Scale - Free Density Perturbations in an Inflationary Universe}",
    reportNumber = "UPR-0202T, EFI-83-13-CHICAGO",
    doi = "10.1103/PhysRevD.28.679",
    journal = "Phys. Rev. D",
    volume = "28",
    pages = "679",
    year = "1983"
}

@article{Abbott:1982hn,
    author = "Abbott, L. F. and Farhi, Edward and Wise, Mark B.",
    title = "{Particle Production in the New Inflationary Cosmology}",
    reportNumber = "MIT-CTP-983",
    doi = "10.1016/0370-2693(82)90867-X",
    journal = "Phys. Lett. B",
    volume = "117",
    pages = "29",
    year = "1982"
}

@article{Dolgov:1982th,
    author = "Dolgov, A. D. and Linde, Andrei D.",
    title = "{Baryon Asymmetry in Inflationary Universe}",
    reportNumber = "ITEP-78-1982",
    doi = "10.1016/0370-2693(82)90292-1",
    journal = "Phys. Lett. B",
    volume = "116",
    pages = "329",
    year = "1982"
}

@article{Albrecht:1982mp,
    author = "Albrecht, Andreas and Steinhardt, Paul J. and Turner, Michael S. and Wilczek, Frank",
    title = "{Reheating an Inflationary Universe}",
    reportNumber = "UPR-0189T, EFI-82-09-CHICAGO",
    doi = "10.1103/PhysRevLett.48.1437",
    journal = "Phys. Rev. Lett.",
    volume = "48",
    pages = "1437",
    year = "1982"
}

@article{Alpher:1948ve,
    author = "Alpher, R. A. and Bethe, H. and Gamow, G.",
    title = "{The origin of chemical elements}",
    doi = "10.1103/PhysRev.73.803",
    journal = "Phys. Rev.",
    volume = "73",
    pages = "803--804",
    year = "1948"
}

@article{Alpher:1948srz,
    author = "Alpher, Ralph A. and Herman, Robert",
    title = "{Evolution of the Universe}",
    doi = "10.1038/162774b0",
    journal = "Nature",
    volume = "162",
    number = "4124",
    pages = "774--775",
    year = "1948"
}

@article{Wagoner:1966pv,
    author = "Wagoner, Robert V. and Fowler, William A. and Hoyle, Fred",
    title = "{On the Synthesis of elements at very high temperatures}",
    doi = "10.1086/149126",
    journal = "Astrophys. J.",
    volume = "148",
    pages = "3--49",
    year = "1967"
}

@article{Walker:1991ap,
    author = "Walker, Terry P. and Steigman, Gary and Schramm, David N. and Olive, Keith A. and Kang, Ho-Shik",
    title = "{Primordial Nucleosynthesis Redux}",
    reportNumber = "OSU-TA-10-90, UMN-TH-826-90, FERMILAB-PUB-91-036-A",
    doi = "10.1086/170255",
    journal = "Astrophys. J.",
    volume = "376",
    pages = "51--69",
    year = "1991"
}

@article{Linde:1983gd,
    author = "Linde, Andrei D.",
    title = "{Chaotic Inflation}",
    doi = "10.1016/0370-2693(83)90837-7",
    journal = "Phys. Lett. B",
    volume = "129",
    pages = "177--181",
    year = "1983"
}

@article{Starobinsky:1980te,
    author = "Starobinsky, Alexei A.",
    editor = "Khalatnikov, I. M. and Mineev, V. P.",
    title = "{A New Type of Isotropic Cosmological Models Without Singularity}",
    doi = "10.1016/0370-2693(80)90670-X",
    journal = "Phys. Lett. B",
    volume = "91",
    pages = "99--102",
    year = "1980"
}

@article{Bezrukov:2007ep,
    author = "Bezrukov, Fedor L. and Shaposhnikov, Mikhail",
    title = "{The Standard Model Higgs boson as the inflaton}",
    eprint = "0710.3755",
    archivePrefix = "arXiv",
    primaryClass = "hep-th",
    doi = "10.1016/j.physletb.2007.11.072",
    journal = "Phys. Lett. B",
    volume = "659",
    pages = "703--706",
    year = "2008"
}

@article{Kallosh:2013yoa,
    author = "Kallosh, Renata and Linde, Andrei and Roest, Diederik",
    title = "{Superconformal Inflationary $\alpha$-Attractors}",
    eprint = "1311.0472",
    archivePrefix = "arXiv",
    primaryClass = "hep-th",
    doi = "10.1007/JHEP11(2013)198",
    journal = "JHEP",
    volume = "11",
    pages = "198",
    year = "2013"
}

@article{Roest:2013fha,
    author = "Roest, Diederik",
    title = "{Universality classes of inflation}",
    eprint = "1309.1285",
    archivePrefix = "arXiv",
    primaryClass = "hep-th",
    doi = "10.1088/1475-7516/2014/01/007",
    journal = "JCAP",
    volume = "01",
    pages = "007",
    year = "2014"
}

@article{Kallosh:2014rga,
    author = "Kallosh, Renata and Linde, Andrei and Roest, Diederik",
    title = "{Large field inflation and double $\alpha$-attractors}",
    eprint = "1405.3646",
    archivePrefix = "arXiv",
    primaryClass = "hep-th",
    doi = "10.1007/JHEP08(2014)052",
    journal = "JHEP",
    volume = "08",
    pages = "052",
    year = "2014"
}

@article{Liddle:2003as,
    author = "Liddle, Andrew R and Leach, Samuel M",
    title = "{How long before the end of inflation were observable perturbations produced?}",
    eprint = "astro-ph/0305263",
    archivePrefix = "arXiv",
    doi = "10.1103/PhysRevD.68.103503",
    journal = "Phys. Rev. D",
    volume = "68",
    pages = "103503",
    year = "2003"
}

@article{Martin:2013tda,
    author = "Martin, Jerome and Ringeval, Christophe and Vennin, Vincent",
    title = "{Encyclop{\ae}dia Inflationaris}: {Opiparous Edition}",
    eprint = "1303.3787",
    archivePrefix = "arXiv",
    primaryClass = "astro-ph.CO",
    doi = "10.1016/j.dark.2024.101653",
    journal = "Phys. Dark Univ.",
    volume = "5-6",
    pages = "75--235",
    year = "2014"
}

@article{Balkenhol:2025wms,
    author = "Balkenhol, L. and others",
    title = "{Inflation at the End of 2025: Constraints on $r$ and $n_s$ Using the Latest CMB and BAO Data}",
    eprint = "2512.10613",
    archivePrefix = "arXiv",
    primaryClass = "astro-ph.CO",
    month = "12",
    year = "2025"
}

@article{AtacamaCosmologyTelescope:2025blo,
    author = "Louis, Thibaut and others",
    collaboration = "Atacama Cosmology Telescope",
    title = "{The Atacama Cosmology Telescope: DR6 power spectra, likelihoods and {\ensuremath{\Lambda}}CDM parameters}",
    eprint = "2503.14452",
    archivePrefix = "arXiv",
    primaryClass = "astro-ph.CO",
    reportNumber = "FERMILAB-PUB-25-0071-PPD",
    doi = "10.1088/1475-7516/2025/11/062",
    journal = "JCAP",
    volume = "11",
    pages = "062",
    year = "2025"
}

@article{SPT-3G:2025bzu,
    author = "Camphuis, E. and others",
    collaboration = "SPT-3G",
    title = "{SPT-3G D1: CMB temperature and polarization power spectra and cosmology from 2019 and 2020 observations of the SPT-3G main field}",
    eprint = "2506.20707",
    archivePrefix = "arXiv",
    primaryClass = "astro-ph.CO",
    reportNumber = "FERMILAB-PUB-25-0144-PPD",
    doi = "10.1103/7wt3-9v2y",
    journal = "Phys. Rev. D",
    volume = "113",
    number = "8",
    pages = "083504",
    year = "2026"
}

@article{Ellis:2025zrf,
    author = "Ellis, John and Garcia, Marcos A. G. and Olive, Keith A. and Verner, Sarunas",
    title = "{Constraints on attractor models of inflation and reheating from Planck, BICEP/Keck, ACT DR6, and SPT-3G data}",
    eprint = "2510.18656",
    archivePrefix = "arXiv",
    primaryClass = "hep-ph",
    reportNumber = "UMN-TH-4512/25, FTPI-MINN-25/14, KCL-PH-TH/2025-42, CERN-TH-2025-199",
    doi = "10.1103/d35r-7bn8",
    journal = "Phys. Rev. D",
    volume = "113",
    number = "6",
    pages = "063571",
    year = "2026"
}

@article{Giovannini:1998bp,
    author = "Giovannini, Massimo",
    title = "{Gravitational waves constraints on postinflationary phases stiffer than radiation}",
    eprint = "hep-ph/9806329",
    archivePrefix = "arXiv",
    doi = "10.1103/PhysRevD.58.083504",
    journal = "Phys. Rev. D",
    volume = "58",
    pages = "083504",
    year = "1998"
}

@article{Peebles:1998qn,
    author = "Peebles, P. J. E. and Vilenkin, A.",
    title = "{Quintessential inflation}",
    eprint = "astro-ph/9810509",
    archivePrefix = "arXiv",
    doi = "10.1103/PhysRevD.59.063505",
    journal = "Phys. Rev. D",
    volume = "59",
    pages = "063505",
    year = "1999"
}

@article{Griest:1990kh,
    author = "Griest, Kim and Seckel, David",
    title = "{Three exceptions in the calculation of relic abundances}",
    reportNumber = "CFPA-TH-90-001A, BA-90-79",
    doi = "10.1103/PhysRevD.43.3191",
    journal = "Phys. Rev. D",
    volume = "43",
    pages = "3191--3203",
    year = "1991"
}

@article{DAgnolo:2015ujb,
    author = "D'Agnolo, Raffaele Tito and Ruderman, Joshua T.",
    title = "{Light Dark Matter from Forbidden Channels}",
    eprint = "1505.07107",
    archivePrefix = "arXiv",
    primaryClass = "hep-ph",
    doi = "10.1103/PhysRevLett.115.061301",
    journal = "Phys. Rev. Lett.",
    volume = "115",
    number = "6",
    pages = "061301",
    year = "2015"
}

@article{Henrich:2025sli,
    author = "Henrich, Stephen E. and Mambrini, Yann and Olive, Keith A.",
    title = "{Ultrarelativistic Freeze-Out: A Bridge from WIMPs to FIMPs}",
    eprint = "2511.02117",
    archivePrefix = "arXiv",
    primaryClass = "hep-ph",
    reportNumber = "UMN-TH-4513/25, FTPI-MINN-25/15",
    doi = "10.1103/zk9k-nbpj",
    journal = "Phys. Rev. Lett.",
    volume = "135",
    number = "22",
    pages = "221002",
    year = "2025"
}

@article{Henrich:2025gsd,
    author = "Henrich, Stephen E. and Gross, Mathieu and Mambrini, Yann and Olive, Keith A.",
    title = "{Ultrarelativistic freeze-out during reheating}",
    eprint = "2505.04703",
    archivePrefix = "arXiv",
    primaryClass = "hep-ph",
    reportNumber = "UMN-TH-4422/25, FTPI-MINN-25/04",
    doi = "10.1103/6yrm-g8t2",
    journal = "Phys. Rev. D",
    volume = "112",
    number = "10",
    pages = "103538",
    year = "2025"
}

@article{Henrich:2025pca,
    author = "Henrich, Stephen E. and Mambrini, Yann and Olive, Keith A.",
    title = "{Z' portal dark matter from post-inflationary reheating: WIMPs, FIMPs, and UFOs}",
    eprint = "2512.04229",
    archivePrefix = "arXiv",
    primaryClass = "hep-ph",
    reportNumber = "UMN-TH-4515/25, FTPI-MINN-25/16",
    month = "12",
    year = "2025"
}

@article{Lerner:2009xg,
    author = "Lerner, Rose Natalie and McDonald, John",
    title = "{Gauge singlet scalar as inflaton and thermal relic dark matter}",
    eprint = "0909.0520",
    archivePrefix = "arXiv",
    primaryClass = "hep-ph",
    doi = "10.1103/PhysRevD.80.123507",
    journal = "Phys. Rev. D",
    volume = "80",
    pages = "123507",
    year = "2009"
}

@article{Okada:2010jd,
    author = "Okada, Nobuchika and Shafi, Qaisar",
    title = "{WIMP Dark Matter Inflation with Observable Gravity Waves}",
    eprint = "1007.1672",
    archivePrefix = "arXiv",
    primaryClass = "hep-ph",
    doi = "10.1103/PhysRevD.84.043533",
    journal = "Phys. Rev. D",
    volume = "84",
    pages = "043533",
    year = "2011"
}

@article{Mukaida:2014kpa,
    author = "Mukaida, Kyohei and Nakayama, Kazunori",
    title = "{Dark Matter Chaotic Inflation in Light of BICEP2}",
    eprint = "1404.1880",
    archivePrefix = "arXiv",
    primaryClass = "hep-ph",
    reportNumber = "UT-14-15",
    doi = "10.1088/1475-7516/2014/08/062",
    journal = "JCAP",
    volume = "08",
    pages = "062",
    year = "2014"
}

@article{Sfakianakis:2018lzf,
    author = "Sfakianakis, Evangelos I. and van de Vis, Jorinde",
    title = "{Preheating after Higgs Inflation: Self-Resonance and Gauge boson production}",
    eprint = "1810.01304", archivePrefix = "arXiv", primaryClass = "hep-ph",
    doi = "10.1103/PhysRevD.99.083519", journal = "Phys. Rev. D", volume = "99", number = "8", pages = "083519", year = "2019"
}

@article{DeCross:2015uza,
    author = "DeCross, Matthew P. and Kaiser, David I. and Prabhu, Anirudh and Prescod-Weinstein, C. and Sfakianakis, Evangelos I.",
    title = "{Preheating after Multifield Inflation with Nonminimal Couplings, I: Covariant Formalism and Attractor Behavior}",
    eprint = "1510.08553", archivePrefix = "arXiv", primaryClass = "astro-ph.CO",
    doi = "10.1103/PhysRevD.97.023526", journal = "Phys. Rev. D", volume = "97", number = "2", pages = "023526", year = "2018"
}

@article{DeCross:2016fdz,
    author = "DeCross, Matthew P. and Kaiser, David I. and Prabhu, Anirudh and Prescod-Weinstein, Chanda and Sfakianakis, Evangelos I.",
    title = "{Preheating after multifield inflation with nonminimal couplings, II: Resonance Structure}",
    eprint = "1610.08868", archivePrefix = "arXiv", primaryClass = "astro-ph.CO",
    doi = "10.1103/PhysRevD.97.023527", journal = "Phys. Rev. D", volume = "97", number = "2", pages = "023527", year = "2018"
}

@article{DeCross:2016cbs,
    author = "DeCross, Matthew P. and Kaiser, David I. and Prabhu, Anirudh and Prescod-Weinstein, Chanda and Sfakianakis, Evangelos I.",
    title = "{Preheating after multifield inflation with nonminimal couplings, III: Dynamical spacetime results}",
    eprint = "1610.08916", archivePrefix = "arXiv", primaryClass = "astro-ph.CO",
    doi = "10.1103/PhysRevD.97.023528", journal = "Phys. Rev. D", volume = "97", number = "2", pages = "023528", year = "2018"
}

@article{Ema:2016dny,
    author = "Ema, Yohei and Jinno, Ryusuke and Mukaida, Kyohei and Nakayama, Kazunori",
    title = "{Violent Preheating in Inflation with Nonminimal Coupling}",
    eprint = "1609.05209",
    archivePrefix = "arXiv",
    primaryClass = "hep-ph",
    doi = "10.1088/1475-7516/2017/02/045",
    journal = "JCAP",
    volume = "02",
    pages = "045",
    year = "2017"
}

@article{Lozanov:2017hjm,
    author = "Lozanov, Kaloian D. and Amin, Mustafa A.",
    title = "{Self-resonance after inflation: oscillons, transients and radiation domination}",
    eprint = "1710.06851",
    archivePrefix = "arXiv",
    primaryClass = "astro-ph.CO",
    doi = "10.1103/PhysRevD.97.023533",
    journal = "Phys. Rev. D",
    volume = "97",
    number = "2",
    pages = "023533",
    year = "2018"
}

\end{document}